# Wolfgang Pauli 1900 to 1930: His Early Physics in Jungian Perspective

A Dissertation

Submitted to the Faculty of the Graduate School

of the University of Minnesota

by

## John Richard Gustafson

In Partial Fulfillment of the Requirements

for the Degree of Doctor of Philosophy

Advisor: Roger H. Stuewer

Minneapolis, Minnesota

July 2004

To my father and mother

Rudy and Aune Gustafson

# Abstract

Wolfgang Pauli's philosophy and physics were intertwined.  His philosophy was a variety of Platonism, in which Pauli's affiliation with Carl Jung formed an integral part, but Pauli's philosophical explorations in physics appeared before he met Jung.  Jung validated Pauli's psycho-philosophical perspective.  Thus, the roots of Pauli's physics and philosophy are important in the history of modern physics.  In his early physics, Pauli attempted to ground his theoretical physics in positivism.  He then began instead to trust his intuitive visualizations of entities that formed an underlying reality to the sensible physical world.  These visualizations included holistic kernels of mathematical-physical entities that later became for him synonymous with Jung's mandalas.  I have connected Pauli's visualization patterns in physics during the period 1900 to 1930 to the psychological philosophy of Jung and displayed some examples of Pauli's creativity in the development of quantum mechanics.  By looking at Pauli's early physics and philosophy, we gain insight into Pauli's contributions to quantum mechanics.  His exclusion principle, his influence on Werner Heisenberg in the formulation of matrix mechanics, his emphasis on firm logical and empirical foundations, his creativity in formulating electron spinors, his neutrino hypothesis, and his dialogues with other quantum physicists, all point to Pauli being the dominant genius in the development of quantum theory.  Because Pauli was in a difficult individuation process during his early years, his own writings on philosophy tend to be sparse and often contradictory.  My analysis of Pauli's physics and philosophy is based upon published and unpublished sources, and Pauli's later reflections.  A pattern has emerged.  Pauli changed his mind from relying on high rationality and empiricism, to valuing intuitive metaphysical visualizations.  This coupled with disturbing events in his life precipitated a breakdown and led Pauli to seek treatment at the Jung Clinic.  Pauli's psychological tension diminished after 1932.  His physics consistently involved symmetry and invariants.  His philosophy allied with Jung's resembled a Platonism of combined psyche and physics.  Pauli sought a rational unification and foundation for his philosophy, but that goal was cut short by his untimely death at the age of 58.

## Acknowledgements

This dissertation is the product of a long and arduous journey. I could not have written it without the contributions of many people. Dennis Gustafson encouraged me to love life and live it, and his spirit has found its way into this dissertaion. Daniel Gustafson was a wonderful role model for me, and awakened in me my love of physics. The American Philosophical Society kindly granted me permission to include significant excerpts from Thomas S. Kuhn's 1963 interview of Werner Heisenberg. The staff of the Niels Bohr Library assisted me in locating several important documents. Erika Eberhardt helped me with translations of German articles, and also introduced me to Jungian psychology. Karl von Meyenn provided me important information about Wolfgang Pauli's early education. The faculty and staff of the University of Minnesota's Program in the History of Science and Technology made this whole adventure possible. Members of my committee-- Benjamin F. Bayman, John M. Eyler, James H. Fetzer, Michel H.P. Janssen, Alan E. Shapiro, and Roger H. Stuewer--provided numerous important suggestions. My friends and family were understanding of my numerous absences from important activities and they continued to support my work. I am deeply grateful to the above people and organizations. Any errors and shortcomings, of course, are all mine.

My advisor, Roger H. Stuewer, needs to be thanked and recognized in a special way. He inspired me to see joy in the history of physics. He provided me with countless insights and clarifications of important concepts. He has a special way of demanding intellectual rigor while encouraging creativity. Using his editorial skill, he is directly responsible for transforming my ramblings into cogent paragraphs. He continued with me long after he had formally retired from the University of Minnesota. I am indebted to Roger. I thank you.

My wife Karen Johnson Gustafson knew I needed to do this and gave me the love and encouragement to bring this dissertation, and me along with it, into the light. I continue beyond measure to be indebted, in awe, and in love with this incredible woman. Thank you, Karen.

# Table of Contents

# Chapter I: Introduction

**Wolfgang Pauli: Physics and Psychology**

To understand the history of quantum mechanics, it is essential for the historian to understand Wolfgang Pauli's role in that history through his unique contributions to its physics. That requires the historian to understand Pauli's personality and philosophy. That, in turn, requires the historian to understand Pauli's receptivity to Carl Jung's psychological philosophy. This dissertation is my attempt to provide such a multifaceted understanding. As a heuristic model, I will portray Pauli as having a dual personality type that involved a strong rational side--the side that was public to his physicist colleagues, and a strong intuitive side--the side he largely kept shielded from view except to his Jungian colleagues. In Pauli, we see a dynamic playing out internally within his psyche between his two personality sides, and leading to the core of the radically new and important theory of quantum mechanics. In the course of my study of Pauli, his attraction to Jungian psychology, problematic to physicists, has become less spooky; Pauli received therapeutic help from Jung and naturally then became interested in his psychology. Jung did hit the mark in Pauli, and thus Pauli's philosophical interests that drew him to the core of quantum mechanics are deeply serious and important to a full understanding of the history of this enigmatic physical theory. Let me expand on why I find Pauli's story so fascinating.

Wolfgang Pauli's name appears repeatedly in the history of modern physics. His name, attached to the exclusion principle, permeates texts in chemistry and atomic physics. His principle also has a mysterious air, coming seemingly out of nowhere to explain atomic phenomena with little further call for justification. How did it arise? Where in a rational trail of history does it fit? What were the circumstances that led Pauli to discover it? Is the principle a numerological recipe from the old quantum theory, or is it a mathematical insight whose roots lie in the new quantum mechanics? If the exclusion principle arose from Pauli's discovery of a strange, classically nondescribable two-valuedness of the electron, then why did Pauli resist the idea of electron spin? When the names of the founders of matrix quantum mechanics are mentioned, Werner Heisenberg,

Max Born, and Niels Bohr come to mind with an associated accomplishment. Heisenberg created matrix mechanics, Born refined matrix mathematics and proposed a statistical interpretation of quantum mechanics, and Bohr argued for his philosophical perspective of complementarity. Pauli's name, by contrast, is inseparable from the history of matrix mechanics, but no clearly unique role for Pauli comes easily to mind beyond that of his confusing exclusion principle. In the literature that surrounds the history of quantum mechanics, the inner circle of quantum theorists, the cognoscenti, rave about Pauli, yet his name is seldom mentioned in more public assessments. Why was he so indispensable to the formation of quantum mechanics yet so mysterious? Why did it take until 1945 for Pauli to be recognized with a Nobel Prize?

Outside of modern physics, Wolfgang Pauli's name appears again, this time in discussions of the mystical psychology of Carl Jung. Pauli later was a colleague of Jung's in Zurich. When did this relationship start? What was the nature of their collaboration? How did the highly rational physicist Pauli come to be associated with the mystical psychologist Jung? Can one see in Pauli's physics any relationship to Jung's psychology? Did Jung influence Pauli's physics, and if so what are examples of that influence? Why do physicists seem not to know much about Pauli's relationship to Jung?

These questions came early to my mind as I became attracted to the personality of Wolfgang Pauli. In some initial reading of the literature where Pauli's name appears, stories surfaced of emotional breakdown, divorce, his mother's suicide, his sex life, his excessive drinking, his biting sarcasm, the secretive nature of his philosophical interests, his spouse's refusal to authorize his biographies, and so on. Here was too much temptation for me as a historian of modern physics to resist. The first startling and affirming break in my research on Pauli occurred when I read Thomas S. Kuhn's 1963 interview of Werner Heisenberg. I discovered here pointed quotations by Heisenberg, a member of the quantum-theoretical cognoscenti, and a respected firsthand observer and friend of Pauli. Heisenberg spoke of Pauli's early philosophy and physics. He verified the strong nature of Pauli's mystical side and interest in Jung, and my own suspicions of a connection between Pauli's philosophy and his physics.

Heisenberg's recollections, however, complicated my hope for an easy explanation. Pauli did not meet Jung until the early 1930s, after quantum mechanics had been formulated, and after Pauli's recognized contributions to modern physics largely had been made. Pauli's physics thus came first and his philosophical connection to Jung later. My research then focused on seeking out in Pauli's life prior to 1930, especially in his physics, situations and ideas that might have led Pauli to become receptive to the Jung school of psychological philosophy. Jung plays no role in this period of Pauli's life, which is a welcome conclusion when one is attempting to clarify the stimuli for Pauli's physics. While this helped in simplification, interpreting Pauli's role in physics became more complex because of the vacillating philosophical positions Pauli exhibited during the 1930s. Pauli's story became even more fascinating with the realization that Pauli's personal philosophy, his personality type, his religiosity, his courage and creativity in physics, all were in a complex process of flux while the roots of quantum mechanics were being developed.

A pattern did emerge from Pauli's early physics, however, that served to strengthen my perception of his key importance to the formulation of quantum theory. Pauli often was the hidden genius behind the discoveries of others. He saw in quantum physics many forms and processes that went beyond a rational description. He carefully screened out those forms and processes of quantum physics that could be rationally supported, and publicly voiced his perspectives on them. But there remained a few forms and processes that he could not derive rationally. Owing to what one might describe as a split personality, Pauli had difficulty voicing his intuitions that could not be completely rationalized. With reluctance and trepidation, he publicly released his idea of the exclusion principle, of spin matrices applied to the electron, and his hypothesis of the neutrino, all three contributions being of the highest creativity in quantum theory. Through a personally difficult process, Pauli found his roots in quantum theory, and those roots later served to ground his personality and his philosophy.

Initially, to explain Pauli's connection to Jung, I felt that Pauli reluctantly became a Platonist.[1] Pauli's brand of philosophy of his later years is invested with the archetypes

---

[1] Karl von Meyenn, "Pauli's Belief in Exact Symmetries," in Manuel Doncel, Armin Hermann, Louis Michel, and Abraham Pais, ed., *Symmetries in Physics (1600-1980)* (Barcelona: Universitat Autònoma Barcelona, 1987), p. 332. Von Meyenn characterizes Pauli as being a positivist in his early years and later following "Platonic-Pythagorean idealism" owing to Jung's influence. I was unaware of von Meyenn's characterization until well along in my own writing, but this characterization coming from a renowned

of Carl Jung, akin to Platonism.[2]  Through a difficult process of introspection and confrontation within his own psyche, Pauli came to terms with his internal voices and adopted what I will call a Platonic perspective.  Jung validated Pauli's own perspective.  That account of Pauli's individuation process comes close to describing what I see as his story, but it leaves too much out.  It tends to assume that Pauli's path to understanding quanta and to personal individuation were linear processes that can be neatly categorized.  Now, after several years of research, I am unsure if Pauli ever reached an equilibrium position of psychic individuation and emotional health, or if his physics perspective was ever one of certainty.   Regarding his personal philosophy, I feel Pauli should not be viewed as a disciple of Jung's, but instead as a colleague.  In several areas of the history of psychology, Pauli helped Jung more than is generally known.  Pauli is far too complex a historical figure, far too flexible and dramatic in his mood swings in physics and philosophy, for a neat categorical summary.  For now, I can only offer some metaphysical analysis and historical musings that seem to me to make Pauli's path more understandable.

I have concentrated on Pauli's early years, 1900 to 1930, to simplify the historical analysis but still survey a wealth of topics.  During this period in his life, Pauli participated in the transition from the old quantum theory to the new quantum mechanics.  He became his own person, separating from his parents and his mentors.  He underwent periods of psychological stress and crisis.   He created several far-reaching concepts in theoretical physics, but released only some to the public.  He married.  He divorced.  He had not yet met Carl Jung, nor his second wife.  He had not yet focused on or published his thoughts on his personal philosophy.

The following extract from Kuhn's 1963 interview of Heisenberg, which so delighted me in my early research, vividly displays the complexity of Pauli's story.  Heisenberg, an intimate friend of Pauli, offers historical insights that I will place into context in later chapters of my dissertation.

---

Pauli expert was welcome.  My dissertation addresses Pauli's process of conversion, which I feel came before Jung's influence.

[2] Pauli followed numerological clues in his physics, implying his philosophical orientation was toward the "all is number" of Pythagoras, but Pauli's philosophy also incorporated the archetypes of Jung as the base forms to reality.  Thus, I feel Pauli's confusing philosophy is better described as a variety of Platonism.

*Heisenberg*:  I might make one remark.  This doubling of states which Pauli first had called the unmechanical doubling, was actually connected with the Lorentz group.  But later on, as you know, one had found doublings which had nothing to do with the Lorentz group, say the iso-spin doubling, neutrons and protons.  This doubling itself was something which Pauli liked.  Therefore he was not too happy about the electronic spin.  The idea  that one should be forced, in such a discontinuous theory as quantum theory is, simply to double, to confront an alternative, either this or that, appealed to a very fundamental feature in Pauli's philosophy.

Now I wonder, did I ever show you this letter of Pauli in connection with our elementary particle business where he was for some time extremely enthusiastic about the whole thing?  Then there come a few sentences where he says, "Verdoppelung und Symmetrieverminderung.  'Das ist des Pudels Kern'."  That is, "The fundamental principle from which all nature is produced is doubling of states and then, later on, reduction of symmetries."  He adds, at this point, "Verdoppelung ist ein alter Zug des Teufels."  In the whole medieval philosophy of the Alchemist the devil is, of course, the one who would double things.  Then he adds that the devil is, of course, the one who makes doubts, hesitations, and the word "Teufel" has to do with "Zweifel," which, in the old time, meant doubling, i.e. you can do either this or that.  So Pauli says that to be put in front of an alternative and to double the possibilities is an old and most fundamental feature of the devil.  In this way, the devil has created the world.   Pauli loved to talk about these things.

Well, if I have not shown you this letter of Pauli, I really should show it to you.  It's very interesting for the psychology of Pauli.  I'm sure that this side of his philosophy must have played its role already in '24 when he wrote this paper on the "unmechanische Zwang."  Therefore, he wasn't too happy this dissolved into a rather trivial angular momentum of an electron.  In so far, he also approved of the doubling which I then tried in the iso-spin case and therefore also the doubling which occurred in the theory of elementary particles.

*Kuhn*: That letter I have not seen. Well, in some way it's appropriate that I haven't because it obviously gets into much more recent development. But you're clearly also right that it must reflect back on attitudes at these earlier points.

*Heisenberg*: Yes. It's so funny that Pauli in some way had some especially good relation to the devil. I would say, of course, also to God. I did probably tell you about [Paul A. M.] Dirac's philosophy--there is no God. Did I tell you that story? I think it reflects also the philosophy of Pauli. Since I have just spoken about his attitude with respect to the devil, I must tell another story.

This happened during the Solvay Conference in '27. There we lived in the same hotel and the younger people of the group sat one evening together drinking of wine or so. Somehow the problem had come up about religion and natural science. Dirac was a very eager defender of the view that religion was just nonsense, was opium for the people, it was just made to make people foolish, and so on. He argued rather strongly. Well, Dirac was a very young man and in some way he was interested in Communistic ideas, which, of course, was perfectly all right at that time. Pauli listened to it, and while Dirac became very angry about religion, he never said a word. He just sat there, you know his way, smiling a bit maliciously. Then finally somebody said, "Well, Pauli, you never say a word to this discussion. What is your opinion of it?" Then Pauli said, with a very malicious smile, "Yes, you know this Mr. Dirac has a religion. This religion is that there is no God and Dirac is his prophet!"

I do remember long discussions with Pauli, especially once when we took a boat from Langelinie, in Copenhagen, to the harbor and had a nice time. All of a sudden, I don't know why, Pauli came to the problem of religion and discussed the existence of God. He really was deeply interested in the question of how far one conveys meaning in using such words as "God." He, of course, at once would admit that a language never is suited for discussing these things, and so on. I remember another sentence in one of his letters. I had told him about the discussion I had had with some theologians. Then he said, "Ja, über Deine Theologen, zu denen ich ja in der archetypischen Relation der feindlichen Brüder stehe." [Yes, to theologians I stand in the archetypal relation of a hostile brother.]

That's very typical of him. He did think about these fundamental problems in terms of devil and God. At the same time he knew, of course, that these were very vague symbols by which one could not really express what one meant. Still he used it.

*Kuhn*: Had that interest in that way of talking gone back a long way with him? Does that go back to the period when you first knew him, or does that come only later?

*Heisenberg*: Well, I remember that I once made, being a student, a short trip on bicycle with him and [Otto] Laporte--the three of us. It was only for a few days. Actually, we came through Urfeld, if I remember right, we came to Garmisch and (???), and had a few days in the mountains. There I had some discussions with him on these problems. But I remember that when one really started to come into these problems, I would say the atmosphere became so tense that it was disagreeable to continue. I could see that this man was so engaged in problems of that kind that it was really better if one did not touch it. So we started a discussion, and he could see that I could understand him in this plane, and from this moment on he had a strong confidence in me. But also in some way, it was agreed that we should not talk about it. So I think for at least ten years more we never took up this discussion again. We only knew from each other that we both were also interested in this side of the world, not only in mathematics and physics.

So that made my relation to Pauli always different from the relation to many other students, because we had in this short trip just once discussed this point. Then we could see at once, "Well, here things become serious, better not talk about it." So, we never touched it again. Well, only then, of course, in this problem in Brussels, when he said that to Dirac, I could at once recognize what he meant. Then we had this discussion on the boat in the harbor of Copenhagen. Again, this side of Pauli comes out extremely strongly in these enthusiastic letters about the theory of the elementary particles. He was, I would say, for one or two months in a completely euphoric state. "Now all problems are solved." Later on he just--. There came an opposite extreme. Disappointment. Still, it was so clear when he was so much engaged in physics that it was in connection with this philosophical side of the world.

*Kuhn*: How far back does the influence of [Carl] Jung go?

*Heisenberg*: Well, I do not know from Pauli himself, but I think just from the time very soon after he came to Zurich. But how close the connection was I don't know. That I couldn't tell. But I would say his interest in philosophical problems has certainly been earlier than his encounter with Jung. Only Jung did hit the point, you know, in Pauli. Pauli was inclined in this direction, and therefore he could listen to Jung and hear what this man Jung actually meant, which many other people just didn't see.

*Kuhn*: I think most of the people who know that side of Pauli, many of whom would have described aspects of it as highly mystical, would also say that that was not something that you had. They would not expect to find that same sort of appreciation of a mystical approach to nature in you also. Would that be a mistake?

*Heisenberg*: Well, I find it difficult to know what other people think about me.

*Kuhn*: Well, what I really meant was with regard to the later Pauli attitude on points of this sort, would you feel comfortable and at home with them yourself? I think many physicists did not.

*Heisenberg*: You mean many physicists would disagree with Pauli on this side of the world, or what would you say?

*Kuhn*: I think they would indeed disagree, but that's a free privilege for anybody. I think to some extent they would be uncomfortable about the fact that anybody half as able and as critical as Pauli should himself have adventures which seem to them not only not physics, but almost a denial of physics.

*Heisenberg*: Yes, yes. Well, did I tell you the following thing? When Pauli died, I was asked to write this memorial volume. [Victor] Weisskopf had asked me. Then, actually, originally I had written an article on Pauli's philosophical views, but this article was not accepted. Weisskopf said, "Well, this article is very nice, but you know we don't like to discuss this side of Pauli so much. We want to see

Pauli as a physicist." So actually I was a bit angry about Weisskopf, but, well, I had to take his opinion, and apparently other people agreed.

Afterwards I did publish my article of Pauli's philosophical views. I first published it in German. May I give you a copy? Later on it appeared in a rather obscure periodical in the United States because there were still people in America who were still interested in it, but not the physicists. These were people of a different structure. Still, I like this article on Pauli's philosophical views. I think that I had succeeded in describing very accurately how Pauli's mind was constructed. I also hoped that I had made it clear to many people that I liked this kind of mind, and that my own mind is not so very different from that of Pauli. I may just have it here.

*Kuhn*: I would be very grateful if you would. Well, that I'm delighted to have. I should have known that this existed, but I didn't . ...

*Heisenberg*: Well, I hope that I have characterized this side of Pauli correctly. I discussed this paper with [B.L.] van der Waerden who knew Pauli well and he agreed and said, "Well, that is exactly like Pauli was."

Actually, I did quote very many things from Pauli, partly in his papers, partly in his letters, so that I think it is quite a correct picture. But I know that many physicists don't like this side of Pauli. I would say Pauli would never have made such ingenious physics as he has done if he had not had this side, you know. In order to invent the exclusion principle and all these things, one must be more that just a formal physicist. So I always loved this side of Pauli. This side of Pauli was really the first basis of an entire understanding between Pauli and myself. Although, as I said, we practically never talked about it until rather late and we were both rather old people. I just mentioned it because, in this first paper on the two-valuedness of the electron, undoubtedly this side has played some role for Pauli. It also fitted this role very nicely that he became Mephistopheles in our Faust.[3] That's absolutely right to the point.

*Kuhn*: It's almost a perfect role, including the Pauli effect.

---

[3] Pauli played the role of Mephistopheles in a skit at Bohr's institute in 1932; see, for example, Charles Enz, *No Time to be Brief: A Scientific Biography of Wolfgang Pauli* (New York: Oxford University Press, 2002), p. 226.

*Heisenberg*: But I was also surprised to see that even a man like Weisskopf, who first of all is a brilliant physicist and then also is a man who knew [Niels] Bohr and [Paul] Ehrenfest and all these people so well, would rather not like to speak about Pauli in these terms. I wonder how that is? I mean, why is it something one should not talk about, or why is it a feature which, so to say, spoils the picture of Pauli? Not for me, but for many physicists.

*Kuhn*: I don't know. I've talked further to Weisskopf about it. In what little conversation I've had with him that would have any reference to this, I do rather think that Mrs. Pauli is not eager to have this side--. I know because I've talked about microfilming the letters. Weisskopf has said that he thought there might be a problem with some of the more recent letters and that he thought Mrs. Pauli would rather have the more philosophical side, for the time being, withheld. ...

*Heisenberg*: Well, of course, I sent Mrs. Pauli both this thing and also several other things which I had written about Pauli. She wrote very nicely back, but I could not see from her reaction whether she approved or did not approve of it. I would think if she did not approve the story, I would have felt it from the letter that she sent to me, from some remark in that direction. So apparently she was quite happy about the way it was discussed.

Quite aside from Mrs. Pauli, you know physicists really do very serious things; they think about the structure of the world. After all, that's what we do. So then why is it that so many physicists are in disagreement with that way of thinking, also with this side of a man who took these things very seriously? It was not for Pauli a kind of funny game. It was certainly not meant as opium; it was the contrary of opium for Pauli. Pauli was so extremely skeptical that he very soon reached that point where he becomes skeptical against sceptics--where it turns round. That is a point which is unavoidable for everybody who wants to be consistent. That is apparently a point which very few people like to reach. It's very disagreeable to reach that point. It's very important if one is consistent and then, of course, one sees that rational thinking is only a limited approach to the world. Well, why not take it as it is? Pauli certainly tried, wherever he could, to do things rationally. He was a rationalist of the purest flavor. Still, at the same time, especially when one is so rational, one must see where the limits are, because there are limits. That can't be helped. Well, now we come to different things and you want to go back to physics again.[4]

**Pauli Studies**

First and foremost, the attraction of Wolfgang Pauli is his story. Historians are attracted to a good yarn, and Pauli's life and work offers a wealth of yet-to-be-mined material. He lived from the dawn to the middle of the twentieth century when world serenity was shattered by the unleashing of nuclear energy. Pauli's work in probing the mysteries of quanta, a pristine effort of the highest intellectualism, helped others to create a nuclear weapon. Pauli did not participate in the making of the atomic bomb, nor in trying to stop its use. There is a dark side to physics, in addition to its purity in endeavoring to probe the secrets of the universe. Pauli recognized this two-sidedness of physics, in nature, and in his own being. Herein lies his story. Two-sidedness characterizes quantum physics, where rationalism led to an unparalleled ability to calculate and predict measurements of physical quantities with extreme precision, while its foundational structure is based on uncertainty. Fresh attempts to reexamine the Copenhagen interpretation of quantum mechanics have revealed its unsatisfying philosophical foundation, but philosophers have been unable to replace it with one of satisfying consistency.

Moreover, aspects of mind in matter are left out. With the advent of seventeenth-century European science, mind was separated from matter, and the two have never been reunited. Pauli recognized this missing aspect, and sought to do something about it in his philosophical explorations, mostly during his mature years. He saw in the psychological metaphysics of Carl Jung the potential for a holistic philosophy of matter and mind. What generated his fascination for this intertwined view of mind and matter, which was at odds with his early philosophical leanings toward positivism? Two-sidedness was in Pauli's psyche. He was a rationalist of the highest order and attempted to move physics away from nonrational, intuitive, visualization methods. Secretly, he used them. By day he was a heady intellectual, a young professional academic respected by his colleagues. At night, he was a psychological mess. Pauli's mind when applied to physics appeared to be led by experimental facts, and his job was to generate mathematical predictive formulae. In his heart, the formulae had to be beautiful and meaningfully congruent to an aesthetic symmetry. The two-sidedness of physics is exhibited in its experimental and theoretical endeavors; the two-sidedness of the universe is reflected philosophically in the necessity to unite matter with mind; the two-sidedness of a person's psyche is reflected in its rational and nonrational character. Pauli's story is intertwined with all of these aspects of two-sidedness. In few other instances can the historian find in one person such an array of dualities, of polar opposites in a person's life and work, in personality type, in philosophical interests**,** in the drama of personal life.

---

[4] Thomas S. Kuhn interview of Werner Heisenberg, Session Nine, February 27, 1963, Archive for History of Quantum Physics, Niels Bohr Library, the University of Minnesota, and other repositories, pp. 16-20.

I was attracted to Pauli's story for several reasons. First, there is the mystic side of Pauli. Fellow graduate student in the Program in History of Science and Technology at the University of Minnesota, Charles Atchley, and I shared common interests and talked about Pauli. Atchley wrote his dissertation on the history of the neutrino.[5] I was amazed to learn that he was unaware of Pauli's relationship to Carl Jung. Atchley had analyzed the complex history of the neutrino, from Pauli's proposal of it to Frederick Reines and Clyde Cowan's verification of it a quarter century later. Nowhere, however, did he treat Pauli's interest in mysticism. Pauli's philosophical and mystical motivations in creating the neutrino concept seemed to me to be as important to a complete history of this elementary particle as were the methods used to verify its existence. Physics as a discipline tends to discount philosophy and mysticism as elements in its conduct, so the history of physics tends to exclude them as well as motivations of the physicist. Pauli, however, did have what one would call philosophical and mystical motivations in doing physics. What was this all about?

As noted above, Victor Weisskopf did not permit Werner Heisenberg to discuss Pauli's mystical side in his memorial paper for Pauli. Later, however, Weisskopf himself wrote candidly and at some length about Pauli's mystical side in his book, *The Joy of Insight,* and also reported that Pauli did not share that side of his personality with many of his physicist colleagues.[6] I had an opportunity to ask Weisskopf more about this when he visited the University of Minnesota and gave a physics colloquium in the 1990s. In

---

front of an audience of many professional physicists, Weisskopf replied to my inquiry by reemphasizing what he had written in his book, noting that Pauli had shared that side of his personality with Weisskopf only sparingly. Most striking about Weisskopf's reply was the reaction of his audience. Professional physicists were aghast! Pauli a mystic and colleague of Jung's? It was as if Pauli had been defrocked from the priesthood of physicists. If one aspires to write about a physicist and one encounters sensitivity--surely a fascinating story lurks there.

Second, I became convinced that Pauli's role in the formation of the Copenhagen hegemony needs clarification. The formulation and development of quantum theory still has many unwritten chapters, particularly with regard to Pauli's role in them. Pauli's invention of the exclusion principle is still mysterious. His brand of philosophy in his later years differed distinctly from that of the Copenhagen school. Did Pauli change his orientation before or after 1930, when the Copenhagen interpretation was largely complete? His name is linked to those of Niels Bohr and Werner Heisenberg in the Copenhagen school, but what were their respective roles? In the index to Mara Beller's book, *Quantum Dialogue: The Making of a Revolution*, there are about twenty centimeters of references to Bohr, eighteen to Heisenberg, and less than six to Pauli, and although this is hardly an objective measure of their respective contributions, my impression remains that Pauli's role in the development of quantum theory is still poorly understood by historians of physics.[7] His unique contributions up to 1930 were his invention of the exclusion principle, his introduction of spinor mathematics for the electron, and his introduction of the neutrino hypothesis. These seem to be contributions that have little to do with one another, suggesting a need for further inquiry. I support Beller's dialogical analysis, but Pauli's dialogical role needs further clarification, since he was at the nexus of the dialog. Pauli was known as the "conscience of theoretical physics" (Weisskopf's characterization) for good reason, because physicists used him to test their reasoning and intellectual rigor. Moreover, Albert Einstein was the principle opponent of the Copenhagen hegemony, yet Einstein and Pauli enjoyed a collegial relationship, especially during Pauli's years at Princeton. When Pauli received the Nobel

---

[7] Mara Beller, *Quantum Dialogues: The Making of a Revolution* (Chicago: University of Chicago Press, 1999), pp. 355-362.

Prize for Physics in 1945, Einstein even publicly declared Pauli to be his heir apparent in theoretical physics![8]  Pauli was awestruck.  To my knowledge, Pauli is the only member of the Copenhagen school to have published an article on theoretical physics with Einstein--an article dealing with, at first glance, an esoteric topic in unified field theory.[9]  Further discussion of Pauli's and Einstein's philosophies and their congruence need to be explored, as a means of clarifying the continuing controversies over the completeness of quantum theory.

Third, Pauli's story promises to offer a unique opportunity to study the intersection of rationalism and nonrationalism, the intersection of science and mysticism.  Pauli is widely known for his acerbic personality and critical comments to physicists; he is less well known for his wide-ranging intellectual interests and collegial relationships as illustrated by those he displayed and enjoyed during his Princeton years.  He collaborated there with the distinguished historian of art, Erwin Panofsky; he published an essay on the Kepler-Fludd polemic, and Panofsky helped to translate material for it from Latin into German.[10]  Pauli's essay was published together with Jung's essay in which Jung introduced his concept of synchronicity.  Concerning Pauli's Kepler-Fludd essay, historian of science Robert S. Westman remarked in 1984:

> Until now, no one has asked publicly why Pauli wrote such an essay, why he encoded his analysis in Jungian terms, and what his relationship to Jung might have been.[11]

Pauli's essay also revealed his collegial relationship with the Jungian psychologist Marie Louise von Franz.  Thus, Pauli's simultaneous collaborations with an art historian, two Jungians, Einstein, and several quantum theorists point to a man of remarkable personality and of highly diverse interests.  Were Pauli's interests in science and mysticism and the relationships between the rational and the nonrational unrelated and mutually exclusive in his psyche?  I believe they were not.  Pauli intertwined his rational and nonrational thinking.  He only shielded these two sides of his psyche from many of his colleagues and from the public.

Pauli's complex personality might be better understood by assuming that he was both a hard rational scientist and an intuitive mystic, and that he kept these two sides separate within his psyche.  Many examples in the literature support this assumption. Pauli's mathematical-physical genius was recognized by Max Born:

> He was a genius, comparable only with Einstein himself.  Indeed, from the point of view of pure science he was possibly even greater than Einstein, even if as an entirely different type of person he never, in my opinion, attained Einstein's greatness.[12]

Abraham Pais quotes Pauli in speaking about Johannes Rydberg:

> I think one has to admit that Rydberg's speculations were sometimes rather wild, but on the other hand they were always controlled again by his study of the empirical material.[13]

Pais thus implies that Pauli valued empirical data above all. Historian Robert S. Westman, in contrast, declared:

> Put in Jungian terms: The crisis of visualization in quantum mechanics apparently resonated with the conflict in Pauli between the "feminine," intuitive, emotional, picturing part of himself and the "masculine," measuring, quantifying, critical part.  Those who knew Pauli observed that he had a hypercritical streak in him.[14]

Westman concludes by quoting Jung, who summarized the themes that Pauli tried to enunciate in his essay on Kepler:

> We shall hardly be mistaken if we assume that our mandala aspires to the most complete union of opposites that is possible, including that of the masculine trinity and the feminine quaternity on the analogy of the alchemical hermaphrodite.[15]

---

[12] Max Born, *The Born-Einstein Letters* (New York: Walker and Co., 1971),  p. 228.
[13] Abraham Pais, *Inward Bound: Of Matter and Forces in the Physical World*  (New York: Oxford University Press, 1995), p. 173.
[14] Westman, "Nature" (ref. 11), p. 218.
[15] *Ibid.,* p.219.

The confusing array of perspectives on Pauli continues.  Mara Beller, in the first part of her book cited above, considers that Pauli provided Heisenberg with penetrating analyses of the fundamental quantum-theoretical issues that had to be addressed by a new theory,[16] but she later seems to flip in her perspective, writing that:

> He [Pauli] advanced imaginative, metaphorical ideas of wholeness and acausality to a mystical abyss that Bohr felt no inclination to approach.[17]

What is going on here?  Is Pauli Weisskopf's critical physicist, Born's mathematical genius, Pais's hard-core empiricist, Westman's and Jung's intuitive Jungian, or Beller's mystical metaphysician?  If Pauli embodied all of these sides in his personality, is that why his widow and Weisskopf shielded his "mystical" papers from view?  Was there something else that was being hidden?  Adding to the confusion in describing Pauli's philosophical perspective, Jungian-flavored writers such as Arthur Koestler and F. David Peat cite Pauli in support of their mystical perspectives.[18]  I conclude that in Pauli the historian has a unique opportunity to explore the intersection of science and mysticism, the intersection of the rational and the nonrational.

I will not attempt to answer all of the questions posed above in my dissertation.  I hope simply to contribute to a deeper understanding of Pauli's physics, philosophy, and psychology, and I hope to show that Jung and Westman were on the right track.  Pauli's multifaceted philosophy and psychology did indeed influence his physics.  In fact, the three were so intimately intertwined that trying to separate them has proven to be elusive.

**Pauli in Jungian Perspective**

Where can one go to find answers to the questions posed above?  My approach differs substantially from that of other studies in the history of science.  In brief, I will attempt to use methodologies that Pauli himself might have endorsed after his encounter with Jung. Thus, I will use aspects of Jung's psychological concepts to provide a heuristic perspective to why Pauli was receptive to Jung's ideas.  This is not to say that I will adopt a Jungian perspective, either exclusively or completely.  I will use some of Jung's methods, however, in an attempt to appreciate Pauli on his own terms, because Pauli believed in many of them.

---

[16] Beller, *Quantum Dialogues* (ref. 7), p. 79.
[17] *Ibid.,* p. 257.
[18] Arthur Koestler, *The Case of Midwife Toad* (New York: Random House, 1973); *The Roots of Coincidence* (New York: Random House, 1972); F. David Peat, *Synchronicity: The Bridge between Matter and Mind* (New York: Bantam Books, 1987).

Psychology is still a young science. It might be described as a field in which professionals in it tend to adopt either of the extremes of empiricism or metaphysics. Psychologists either accumulate empirical facts about human behavior and make predictions in a logical-positivist fashion, or they generate a teleological metaphysics that serves to attain the same ends. Jungians are on the metaphysical extreme, but the pragmatic results they have achieved in several areas have been impressive. For example, Jung's concepts of the unconscious and collective unconscious have garnered support in electroencephalograph (EEG) or "lie-detector" technology.[19] Clinical application of Jung's psychological methods also has been of proven merit in therapeutic applications. Enthusiasts swear by it.[20] Another use of Jungian concepts has been in the comparative study of religions, as in Joseph Campbell's work.[21] Another example is in poetry as in Robert Bly's poetry of men's issues.[22] Finally, Jungian concepts have been employed in the classification and interpretation of human personalities, as in Isabel Myers and Katherine Briggs's popular typology.

I regard this last area as the one in which Jungian concepts have achieved their most impressive pragmatic results. The mother and daughter team of Katherine Briggs and Isabel Myers has refined and developed Jung's personality typology into an industry. The Myers-Briggs personality typology is used convincingly in many settings, ranging from amateur attempts by married couples to understand their spouses, to professional business psychologists to organize work forces. In my view, historians of physics also should recognize the pragmatic value of this psychological tool, especially in situations where perceptions of a historical figure differ. Two physicists may simply be looking at the world from the perspective of their different personalities. A case in point is in the differences of opinion expressed about the merits of experimental and theoretical physics. I believe that the personality types of experimentalists and theorists differ, so their goals for physics differ.[23]   Pauli knew and embraced Jung's early concepts; Myers and Briggs extended them to provide a psychological dynamic of four dimensions with eight polar opposites, a thesis that would have delighted Pauli.   The mature Pauli loved any four-part dynamic inside a whole entity, in this instance a "quaternian" dynamic inside the whole "mandala" that makes up one's personality.  Myers and Briggs advanced a classification scheme of personality based upon the axes of Extroversion-Introversion, Sensing-Intuition, Thinking-Feeling, and Judging-Perceiving.  I will discuss the Myers-Briggs typology further below.[24]

---

[19] Mary Ann Mattoon, *Jungian Psychology in Perspective* (New York: The Free Press, 1981), pp. 33, 45.
[20] Opponents are equally as vociferous.
[21] Joseph Campbell with Bill Moyers, in Betty Sue Flowers, ed., *The Power of Myth* (New York: Anchor Books, 1991), pp. 34, 49, 60, 271. Many other sections deal with Jungian concepts.
[22] Robert Bly, *A Little Book on the Human Shadow* (San Francisco: Harper & Row, 1988).

        The concepts related to Jung and to his school of psychological philosophy that I have used in my dissertation include those of synchronicity, the individual and collective unconscious, archetypal components of the unconscious, the anima and the animus, the Shadow, dream analysis, the process of individuation, and Myers and Briggs's personality typology.  Not all will appear conspicuously but I will introduce some of them for the sake of clarity.  My intent is not to justify their validity, but to use them to better understand Pauli.  Jungian psychological metaphysics may seem unwarranted to some readers, but there is some pragmatic justification for its use.  Brain research now indicates that brain messages are communicated by chemicals and electric currents. Edward O. Wilson in his book, *Consilience: The Unity of Knowledge*, points out that "Fear of snakes is deep and primordial among the Old World primates, the phylogenetic group to which *Homo Sapiens* belongs."[25]   Future brain science based on empirical studies may reveal a pragmatic explanation for Jung's archetypes as electrochemical brain signals; thus, for example, the snake brain message might be some common electrochemical response of the brain in all Old World primates.  The snake imagery releases a pattern of electrochemicals in the brain, which may be similar within this phylogenetic group.  This pattern of electrochemicals, in other words, can be associated

---

[23] My belief is easily tested, although to my knowledge this test has not been performed.  A Myers-Briggs testing of experimental and theoretical physicists should reveal differences of personality type.  Measured differences would not be surprising.  For example, business organizations make use of Myers-Briggs testing of their employees to better fit employees to jobs.   From my own experiences while working in a corporation, it was general knowledge that over a third of all my corporation's employees were STJ personality types.

[24] Mattoon, *Jungian Psychology* (ref. 19), p. 77.  Many other sources deal with Myers and Briggs's methodology.

[25] Edward O. Wilson, *Consilience: The Unity of Knowledge* (New York: Alfred A. Knopf, 1998), pp. 78-83.

with the imagery of a snake, and the snake "archetype" thus is pragmatically equated to the electrochemical pattern. The brain, after all, is not a rational, electronic digital computer but instead is an electrochemical factory.

By using Jung's terms and methodology, especially the derivative Myers-Briggs personality typology, I can interpret remarks by various historical figures. A commonly held belief is that experimentalists differ in personality from theorists, for example.[26] What makes these interpretations so challenging is the uncertainty in knowing their personality types. There also is a problem of interpretation in Pauli's own commentary. According to Jung, an individual's personality may have several voices, so the researcher must decide what personality component is speaking. Each person has a Shadow personality which to some Jungians is more than just the unconscious contents of the person's psyche. In their intuitive and perceptive style, Jungians attach differing connotations to the concept of the Shadow.[27] Typically, the Shadow has connotations of the unsavory aspects of a person's psyche, hidden from the conscious Ego. I will use the term Shadow to mean the unconscious components of a person's psyche and its unrealized psychic potential, without any pejorative connotation. Pauli's Shadow can be seen in his humor, in his quick unfiltered responses, in his criticism of others, and sometimes in his carefully filtered but ambiguous declarations. Pauli's Shadow appears also in his early physics, where he struggled with his physical intuitions. The problem of interpreting Pauli's commentary then becomes one of interpreting which Pauli voice is speaking, his Shadow, his consciousness, his dormant or his dominant personality sides. These all appear to change dramatically during the period from 1900 to 1930.

To gain deeper understanding of Pauli's complex personality, one needs to be ever mindful of Pauli's Shadow. I recommend that when reading Pauli to look for examples of his biting humor. Humor is often a twist in the meaning of a word when applied simultaneously in two or more contexts, a bridging of concepts between two mutually exclusive areas of thought, and a mark of high creativity.[28] In Pauli's quick humor we see his brilliant creativity unfettered and unfiltered. Unfortunately, as we will see, Pauli's creativity in his early years came from his Shadow when his harsh criticisms, for example, emerged.  I recommend that students of Pauli reflect on what thought patterns preoccupied him when he made his sharp criticisms.   By reading vast amounts of the primary and secondary literature, and by comparing Pauli's interests in physics to what was going on simultaneously in his personal life, I have found that long-term patterns of thought and action emerged that often illustrated his Shadow.  Pauli's actions speak louder than his words.

---

[26]In Chapter 4 I discuss Heisenberg's comments on experimental and theoretical physicists in his university education.
[27]Mattoon, *Jungian Psychology* (ref. 19), pp. 25-28; Bly, *Human Shadow* (ref. 21), pp. 1-3.
[28] Albert Rothenberg, *The Emerging Goddess* (Chicago: University of Chicago Press, 1979), p. 37. Rothenberg discusses creativity extensively in his book; jokes serve as indicators.

Pauli was a prolific writer, publishing many articles, including encyclopedia articles, in physics.  He tried to cover everything, summarizing the known experimental data, the accepted theoretical explanations of it, and the remaining puzzles.  This was a task too daunting even for Pauli.  His deepest thoughts, thoughts that still might have resided in his Shadow, can be found especially in the inevitable ambiguity in the summarizing statements in his encyclopedia articles.  I recommend reading them closely and reflecting on Pauli's potential thought patterns in them.

To complicate our quest to sense Pauli's long-term patterns of thought, his writings were addressed to different audiences, the community of physicists, close colleagues in his personal correspondence, and to Jungians, all with different voices.  In trying to analyze Pauli's long-term behavior, it is possible but very difficult to glean patterns from his Shadow.  I have found secondary sources to be especially valuable here.  Those who knew or studied Pauli and then wrote about their impressions of him gained perspective and recognized patterns in Pauli's behavior.  I am not sure if Pauli himself had the ability to do so.  Secondary sources thus complement the primary sources.  Their authors, however, tend to display their own narrow perspectives, have their own personality typology, and have their own Shadow patterns.

We also should look for Pauli's Shadow during his periods of crisis, when it emerges and patterns can be recognized.  During his times of psychological crisis or emotional stress, the protective layer covering his Shadow becomes thin.  One finds that Pauli was most creative during his several periods of psychological crises.  I recommend focusing on Pauli's difficult times for clues to his creativity.

The student of Pauli must survey many primary and secondary sources, both in English and in German, to gain some understanding of his complex character.  I do not pretend to have a full command of the German nuances that Pauli often used in his communications.  The sources that I have either perused closely or scanned appear in the bibliography.  It is not complete, although I believe that students of Pauli should read all material related even tangentially to him to see the patterns that emerge.  The bibliography is intended to help students of Pauli begin to learn more about this amazing person.

Finally, I must say a few words about my research methodology, one that Pauli might have endorsed, at least after 1932.  I have followed my intuitions.  For better or worse, Pauli's patterns of creativity came to me after reading many sources, causing an intuitive leap in my mind.  I have had many deep and engaging conversations in my mind with Pauli during my dissertation research.  Only after reading him and engaging him could I substantiate in objective historical references the patterns I initially saw intuitively in him.  I hope that I now have combined my intuitive understanding of Pauli and my objective substantiation of him.  For the period 1900 to 1930, I will discuss, approximately in chronological parallelism, Pauli's intertwined physics, philosophy, and personal life.[29]  My dissertation is not a biography of Pauli, but a discussion of his thought patterns as I have gleaned them intuitively.

To form this template for presenting Pauli's life and work with all its complexity has been an excruciating process.  I found it to be especially difficult to determine a precise chronology for the development of his philosophical ideas.  I therefore placed them in the chronological periods where they might have surfaced and were relevant to his physics at the time.  I hope that in this way I have been able to impose some order on the complexity of Pauli's life and work.

---

[29] The significant events in Pauli's life are important to track chronologically. Some confusion and error have crept into the literature, and I have attempted to remove them; see Appendix. Pauli Timeline.

## Chapter 2. The Pauli Family in Prague and Vienna

### Pauli's Father and Grandparents

There may be no better way to first encounter the name Pauli than by marking it by a confusion. Wolfgang Pauli the experimental scientist was not born with the name Wolfgang Pauli, and Wolfgang Pauli the experimental scientist was not Wolfgang Pauli the theoretical physicist. Wolfgang Pauli the experimental scientist, disciple of Ernst Mach, university professor and respected colleague of many prominent Viennese intellectuals, was born Wolfgang Pascheles in Prague. He was the father of Wolfgang Pauli the theoretical physicist. This confusion illustrates a continuing theme in our story. Father and son, the experimental scientist and the theoretical physicist, had different types of personalities, and the father's contaminated the son's. I will argue in later chapters that Pauli the theoretical physicist was confused over who he really was and needed to be.

Wolfgang Joseph Pascheles, father of physicist Wolfgang Pauli, was born in Prague, the capital of Bohemia, on September 11, 1869. To distinguish between father Wolfgang Pauli and son Wolfgang Pauli, Junior, I will call the father "Pauli's father" and the son simply "Pauli." Pauli's father was born into a respectable Jewish family in Prague, the son of Jacob W. Pascheles and Helene Pascheles neé Utitz.[30] Soon after Jacob died and shortly before Pauli's father married, he changed his surname to Pauli, and converted to Roman Catholicism.

Pauli's grandfather Jacob was the son of Wolf Pascheles, the owner of a bookshop in Prague. Jacob inherited the bookshop and operated it so successfully that he and his wife Helene could reside on the Old Town Square of Prague with its ancient clock of the seasons, the statue of John Huss,[31] and the 14th-century Tyn Church containing the tomb of Tycho Brahe.[32] Jacob Pascheles was a respected elder in the Jewish community of

Prague and attended the "Gypsy Synagogue," where on June 13, 1896, he presided over the bar mitzvah of Franz Kafka, whose family also lived on the Old Town Square. The "Gypsy Synagogue" (*Zigeunersynagoge*) was so named because it was located on the Zigeunerstrasse.[33] Because of the cabalistic interests of Franz Kafka, one may ask if they also were of interest to Jacob Pascheles.

In light of his later interests, one may ask if Wolfgang Pauli may have been influenced indirectly by his grandfather in his formative years. A grandfather often provides a word or two of guidance to a receptive grandson or stimulates interest in exploring his heritage. We know that Pauli became intrigued with the Jewish cabalistic tradition, but not at what point in his life. Pauli never met his paternal grandfather, but may have been influenced indirectly by him to study the Jewish Cabala. When his grandmother Helene Pascheles visited the Pauli household in 1916, "they were told, for the first time, that their name was really Pascheles and that their father was Jewish."[34] That event may have triggered the young Pauli's interest in learning more about his Jewish heritage, the city of Prague, and even his own father.[35] Pauli's maternal grandfather also was Jewish and also was born in Prague, but his maternal grandfather did not practice his faith in his later years and Pauli's maternal heritage appears to be less significant to him than his paternal heritage. Pauli thus had three Jewish grandparents, but by all indications he was reared by his parents as a Roman Catholic.[36]

Pauli's maternal grandfather, Friedrich Schütz, was born in Prague in 1845, and became a writer, poet, journalist, and playright, and after moving to Vienna became a member of the editorial board of the *Neue Freie Presse*, the influential liberal Viennese newspaper.[37] He was a non-practicing Jew, presumably politically liberal, and had an artistic temperment. Although he died in 1908, he had some influence on the young Pauli, taking him on walks in Vienna and visits to antiques shops.[38] Some of Pauli's artistic and aesthetic interests could very well have been ignited by this man. He married

Bertha Dillner von Dillnersdorf in 1875, who by contrast exerted a strong influence on the young Pauli, nurturing his artistic side during his formative years.   She would practice the piano for hours with the young Pauli, playing and singing, thus conveying an image of a kind, nurturing grandmother.

**Pauli's father and Ernst Mach**

Pauli's father was a student and disciple of Ernst Mach.  Mach, incisive critic of Newton's concepts of absolute space and absolute time, and a forerunner of the Vienna Circle's school of logical positivism, proffered a philosophy of science in which only empirical evidence directly derived from sense perceptions should be admitted into the body of science, with scientific theories describing this evidence in the most economical way possible.  Mach thus doubted the existence of atoms, since atoms could not be directly sensed or perceived, which allied him with the intensely critical opponents of Ludwig Boltzmann:

> Boltzmann's work in statistical mechanics was strongly attacked by W. Ostwald and the energeticists who did not believe in atoms and wanted to base all of physical science on energy considerations only.  He also suffered from misunderstandings of his ideas on the nature of irreversibility on the part of others who did not fully grasp the statistical nature of his reasoning.  Boltzmann was fully justified against both sets of opponents by the discoveries in atomic physics which began shortly before 1900 and by the fluctuation phenomena, such as Brownian motion, which could be understood only by statistical mechanics.
>
> Ill and depressed, Boltzmann took his own life on September 5, 1906, at Duino, Trieste.[39]

Mach also opposed Albert Einstein's theory of relativity, saying in 1913, "I must … as assuredly disclaim to be a forerunner of the relativists as I withhold from the atomistic belief of the present day…[and he added that relativity seemed] … to be growing more and more dogmatical."[40]

---

[39] *Encyclopedia Britannica,* Vol. 3 (ref. 3), p. 893.
[40] Quoted in Abraham Pais, *Subtle is the Lord: The Science and the Life of Albert Einstein* (New York: Oxford University Press, 1982), p. 283.

Mach, in sum, opposed metaphysics whether it appeared in questions of the reality of atoms or in the foundations of space and time.  He would have been aghast at numerological mysticism, or at least numerological metaphysics, as a guide in exploring the inner workings of the atom.  Yet those very methods were proving useful, for example, in the spectroscopic work of Johann Jakob Balmer (1825-1898) and Johannes Robert Rydberg (1854-1919).  Young Pauli would become intrigued with the works of Balmer and Rydberg, while Mach would have regarded such numerology as a return to superstition and unfounded religion.  Pauli's father, Mach's disciple, thus also would likely have degraded such numerology and advocated a belief in sound logic and empirical facts as a model of physics to his son.

Wolfgang Pascheles: Jewish Heritage and Conversion to Catholicism

Pauli's father transmitted a heritage to his son that could not be easily discarded.  A son ruminates about his origins and heritage and decides what to reject or accept.  Pauli's father discarded his Jewishness.  His son was born into Catholicism, and he would have to discover for himself his family roots in Prague and his Jewishness.  In a letter of 1950 to a colleague of Jung's, Pauli mentioned his interest in Prague and in that almost archetypal figure, the golem: "The Jewish ancestors were in Prague for a long time, certainly a very characteristic city (the 'Golem' by Meyrink had always fascinated me) where, however, I never was." [41]  Prague was home to important cabalistic activity, which attracted Pauli:

> Perhaps the most famous Jewish occultist was the nineteenth-century Rabbi Lowy of Prague, creator of the golem, a supposedly superhuman being---which some historians of science claim as the precursor of the modern robot.
>
> According to the legend, Rabbi Lowy, a master of Hebrew letter permutation, shaped a block of inert matter into the form of a man, which he animated by breathing into its nostrils the sacred Names of God.  Harmless enough at first, the golem functioned as the rabbi's servant, running errands and doing odd jobs around the house.  But further empowered by the master's meditations with the passage of time, the creature began to develop spectacular strength and a will of its own.  Rumor had it that the rabbi was attempting to endow the golem with a soul, but that he was having trouble keeping its animal instinct in check and the creature was going on rampages.  Things got so out of hand that the rabbi had to lock the golem in the attic at night.  Finally, unable to control it, he had to destroy it.  But the legend of the golem continued into the twentieth century, and it still persists.  Rabbi Lowy's house in Prague has been turned into a museum, and, in keeping with his instructions, the attic containing the dismantled parts of the golem has never been unlocked.  Part of the legend's power is attributed to the strange events surrounding Rabbi Lowy's house during the Nazi invasion of Prague.  According to eyewitnesses, the invading soldiers sent in to round up all the city's Jews destroyed every Jewish synagogue and home but Rabbi Lowy's.  Adolf Hitler himself had given the order to leave it standing.  Having been warned by his astrologers that destroying the house and releasing the golem would lead to his defeat, the maddest occultist of them all deferred to the power of the Kabbalah.[42]

As I mentioned earlier, Pauli's grandfather Jacob Pascheles presided in Prague over the bar mitzvah of Franz Kafka, who was seriously interested in the Cabala.  Kafka, however, found his exposure to Judaism in his youth to be boring, traditional, and common without much deeper meaning.  His bar mitzvah was only a required ceremony, a custom that provided little insight into his later existential journeys.  Refugee peasant Jews passing through Prague brought with them, in Kafka's eyes, an authentic and deeply rooted Jewishness with a sense of belonging and community that apparently was lacking in Prague.  Kafka's training in the Cabala came not from Jacob Pascheles but from an eccentric friend:

It was his growing interest in this authentic strain of Judaism that led to Kafka's friendship with Georg, alias Mordechai, Langer, one of the more colorful eccentrics on the Prague scene.  Langer was a medieval Jewish mystic born into the wrong century, as well as into the wrong, thoroughly assimilated family....  Georg, on the other hand, left home in adolescence, spent years in a Hasidic community in Hungary, and finally, rather to the consternation of his respectable

> family, returned to Prague, a bearded Hasid in outlandish garb determined to devote his life to probing the mysteries of the Cabala.... Kafka later became one of his students.[43]

One can infer that Jacob Pascheles was one of the respectable Jews of Prague who would not have educated his grandson in the ways of the Cabala had he lived long enough to do so; he died on November 23, 1897, three years before the birth of his grandson.

The Cabala was later of great interest to Pauli. Through Jung he developed a long friendship with Gershom Sholem, the great authority on the Jewish Cabala.[44] Pauli, in fact, would maintain his interest in the Jewish Cabala longer than he would in Jung. In the final year of his life, Pauli became estranged from Jung, but he would talk to Gnostic scholar Gilles Quispel:

> A few months before he died, Pauli told the Gnostic scholar Gilles Quispel that while he could accept "the God of the Gnostics ... I could never accept the existence of a personal God. No such being could possibly endure the suffering of humanity." According to Quispel, Pauli, in searching for a meaning to his life while confronting his death, came to reassert his Jewish tradition.[45]

More questions than answers surround Pauli's conversation with Quispel, since there is no indication that Pauli practiced Jewish traditions either in his childhood or in his adulthood. His Jewishness was a part of his Shadow; the Cabala with its numerology may have entered Pauli at an early time, but it had to surface into Pauli's consciousness.

Life in Prague during the late nineteenth century when Pauli's father was growing up contained existential *fin-de-siècle* undercurrents of social, political, and intellectual tensions that were to erupt later in the Great War. We see in the Pascheles family a heritage of intellectual community, a middle-class environment, and a commitment to education and to long-standing Jewish traditions. During Pauli's father's childhood, refugee Jewish peasants from the countryside were migrating through Prague. Anti-Semitism was on the rise. Germans, Czechs, and Jews clashed in ethnic riots. Socialist movements gained momentum in response to the European industrial revolution. Kafka wrote of his existential perceptions of Prague in terms of symbolic struggles between light and darkness, good and evil, meaning and nothingness. There would have been many good reasons for Pauli's father to leave Prague, to seek his fortune in nearby Vienna, to leave his family religion, and to embrace the Roman Catholic faith. The Pascheles home at No. 7 on Prague's Old Town Square (*Altstädter Ring*[46]), had been a Paulan convent earlier, which may have been the reason he changed the family name to Pauli.[47]

Pauli's Jewish heritage and family background in Prague influenced his physics in later years, but at present our knowledge is too limited to establish direct connections to it. All we can say at this point is that Pauli later became enamored with his Jewish roots and with the Cabala, and relations to them can be found in his physics. For example, Jewish philosophy is related historically to the Newtonian concepts of absolute space and absolute time,[48] areas that remained fascinating to Pauli. In general, Pauli valued early philosophers more than contemporary ones.[49] We also know that Ernst Mach's philosophical perspectives influenced Pauli's father directly, so he may have tried to impress on his son antiatomic, antirelativistic, antinumerological, and proempirical perspectives. We will see, however, that Pauli's Jewish roots in Prague, including his interest in the Cabala, engaged him more than Mach's philosophy. Pauli became expert in atomic physics, relativity theory, and quantum theory; he also would reflect on the reasons for his father's religious conversion, forcing him as a youth to abandon Judaism and embrace Catholicism.

**Pauli's Father and His Son**

The most intense developmental period in the life of a human being is childhood, especially the earliest years. Language skills and personality develop, conceptual understanding emerges, emotional stability either strengthens or becomes dysfunctional. A sense of personal identity develops. The child's earliest environment is a mixture of people, places, and events. Nurtured by parents and other family members, the child grows in a physical surrounding that provides its unique sense of place. Time unfolds in a series of events that define its unique milieu. Pauli's childhood environment may have been atypical, filled either with beneficial conditions for the development of genius or with detrimental conditions that lead to emotional crises. We have no reason to believe, however, that as a product both of his environment and genetic makeup he was atypical in his human development, that his needs for nurturing and role modeling were different from those of any child becoming a functioning, self-confident adult. He was just a child born into an upper-middle-class family, growing up in *fin-de-siècle* Vienna. Let us look more closely at Pauli's family.

---

[46] Visual circles including rings were important symbols to Pauli.
[47] F. Smutný, "Ernst Mach and Wolfgang Pauli's Ancestors in Prague," *Gesnerus* **46** (1989), 183.
[48] Max Jammer, *Concepts of Space* (New York: Dover, 1993).
[49] Pais, *Genius of Science* (ref. 1), p. 245.

Wolfgang Joseph Pascheles studied medicine at the Charles University in Prague where Ernst Mach was a professor and Mach's son Ludwig also was a student. Pauli's father was attracted to Mach's style of physics. According to Abraham Pais, Pauli's father's contact with Mach was the most important event in his intellectual life.[50] Pais continues:

> Already as a boy he [Pauli's father] had shown scientific interest in physics and chemistry. In his early student years he spent every free moment in the physics institute of the famous physicist Ernst Mach who became his teacher and model, and who remained his fatherly friend until the end of his [Mach's] life. He [Mach] also repeatedly presented brief scientific communications of young Pauli's [Pauli's father's] results to the *Kaiserliche Akademie der Wissenschaften* in Vienna.[51]

Physics thus was an important foundation of his scientific career.[52]

Wolfgang Joseph Pascheles obtained his doctor's degree in medicine from the Charles University on April 24, 1893. He practiced medicine in Vienna starting in 1893, but soon turned to a career in chemistry, a decision that was likely influenced by his mentor Ernst Mach, who moved to Vienna in 1895. Pauli's father started at the University of Vienna with an assistantship in the Medical Faculty in 1898. His decision to change his career to chemistry and to convert from Judaism to Catholicism may have

---

been related to the professional restrictions faced by Jews in *fin-de-siècle* Vienna.[53] His medical practice had involved the wealthy of Vienna,[54] through which he may have met his future wife. He received permission to change his name to Pauli in July 1898, and he converted to Catholicism in March 1899, shortly before he married Camilla Schütz on May 2, 1899. Their son Wolfgang was born on April 25, 1900.

After leaving medicine, Pauli's father built an impressive career as an expert in the physical chemistry of proteins. He became full professor and director of the Institut für medizinische Kolloidchemie at the University of Vienna in 1922, a position, according to Pais, that was created especially for him in the Medical Faculty. He would write hundreds of articles and several books on colloids and proteins over the following decades: "We owe to Pauli [Pauli's father] the first insights into the exact connections between the constitution, the structure and the stability of colloids, as well as their chemical-physical behavior."[55]

Pauli's father exhibited the *fin-de-siècle* attitude of enlightenment and progressivism, following a career in science, marrying into a cultured Viennese family, and becoming a proponent of Ernst Mach's philosophy. By first pursuing a university degree in medicine and then a scientific career in experimental physical chemistry, his career reflected an attitude of scientism in which human progress could be attained through science, trusting observation and measurement, and by negating metaphysics. Pauli's father was a successful university researcher, a respected scientist specializing in the physical chemistry of proteins. In a book of the early 1920s he explained his motivation for pursuing this research: "The biologist attempts to arrive at a general expression for the profound and diverse phenomena of life, and finds that the rich variety of the reactions of the proteins confronts him as one of his greatest difficulties."[56]

In his Preface, Pauli's father pointed out that proteins, as nonvitalistic chemical systems, are especially important in the organization of living matter and vital processes, and are unique in displaying the special properties of living matter. It seems that he believed that biological processes can be reduced to chemistry and physics. Extensive discussions on the mechanisms of biological evolution were then taking place in Vienna. The neoLamarckian scientist Paul Kammerer was stirring up controversy by propounding nonDarwinian mechanisms for evolution.  I will discuss Kammerer's speculations later because of their possible relationship to Pauli's and Jung's concept of synchronicity.[57] Since Kammerer's reasoning was guided by arguments of design in evolution, while Pauli's father's reasoning was guided by an antivitalist philosophy,  Pauli's father likely opposed Kammerer's neoLamarckism.   His son Pauli, however, later would espouse neoLamarckian evolutionary views in biology.[58]

We can gain some understanding of Pauli's father's scientific philosophy and work in biochemistry by examining his book of the early 1920s noted above, which was based on lectures he delivered in 1912-1913.  He is a literate and capable scientist who applies his knowledge of chemistry and physics to biology.[59]  He is concerned with physical explanations of the phenomena of life, treating proteins as systems of inert chemical compounds whose reactions exhibit patterns that can be deciphered by applying the laws of chemistry and, especially, of physics.  He cites such important scientists as Avogadro, James Clerk Maxwell, Rudolf Clausius, Jean Perrin, and Robert Brown (discoverer of the "Brownian motion") whose works he clearly has studied carefully.  His descriptions are verbose and impersonal, referring to himself in the third person, and focusing on experimental facts as if emulating Mach's philosophy:

> Pauli and his coworkers have explained the course of the viscosity curve of hydrochloric acid and albumin by the same ionic conceptions as those employed by Lagueur and Sackur.  The work of Manabe and Matula has finally settled the parallelism between the viscosity curve and the ionization curve as determined by the electroenergetic method (Table 21A, Fig. 14).
>
> We give next a very complete series of careful measurements of the viscosity of horse serum albumin (free from salts) by Pauli and R. Wagner, which include the initial decrease.[60]

Such detailed and complex measurements and the accompanying table, as well as the focus on minutia, convey a sense of Pauli's father's chemical expertise and his concern with precision experimental measurements. He rarely offers general reflections or philosophical speculations that might explain the grandeur of the theory or the big picture; his commitment is to empirical rigor. He must have spent an enormous amount of time in his laboratory. He was after the details, the experimental facts. He writes as if he were a logical positivist, for whom theoretical formulae are merely a shorthand for the description of experimental facts. Pauli's father focussed on factual measurements and data compilation, seeking an antimetaphysical science of biology based upon objective experimental chemistry and physics.[61]

Pauli's father pursued an enormously productive career at the University of Vienna for decades. In 1932, for example, he published two articles in *Die Naturwissenschaften* that reveal him to be a first-rate, hardworking, observant, and literate scientist. He cites numerous leading physicists, showing that he is cognizant of state-of-the-art experimental physics as it pertains to his field of research. Some regarded his scientific work as worthy of a Nobel Prize.[62]

The scientific interests and style of Pauli's father stand in contrast to those of his son at mid-career. Perhaps one way to characterize their scientific differences is to note how the 29-year-old Pauli perceived the field of solid-state physics. Pauli applied the new quantum mechanics to collections of atoms and molecules,[63] some regarding his work as founding the field of solid-state physics. Yet, it held no fascination for him. It was messy and "dirty." It apparently promised to yield no pure or pristine philosophical or physical insights. He thus left this field to others, concentrating instead on the elegant mathematical foundations of the new quantum theory. He would not dirty his hands with solid-state physics. Pauli's father, by contrast, did get his hands dirty. His brand of physical chemistry of the colloids and proteins was by its very nature dirty and messy, and not only in terms of mathematical theory. He got his hands dirty in the laboratory making experimental measurements. He collaborated with similarly committed colleagues. We see in him a disciple of Ernst Mach, an antimetaphysical scientist unconcerned with what could not be directly proven by experiment. No doubt, however, in contrast to Mach, he accepted the reality of atoms as part and parcel of his chemical studies.

---

[61] His text, *Colloid Chemistry of the Proteins* (ref. 27 ), based on lectures delivered in 1912-1913, is a milestone in the history of medicine. There is an English translation in the special books collection of the Bio-Medical Library at the University of Minnesota.
[62] Enz, *No Time to be Brief*, (ref. 3), p. 11.
[63] *Ibid.,* p.157.

Regarding the relationship of Pauli's father's to his son, the few clues in the historical literature suggest a strong difference in their personalities, a strained and a likely distant relationship. One can speculate about causal patterns. For example, during his son's formative years, Pauli's father was busy pursuing his demanding career, presumably leaving much of the parenting to his wife. He did assist in parenting by asking Mach to make recommendations for his son's scientific education, by obtaining the science books Mach recommended, and by taking his son to visit Mach. His son Pauli was a child prodigy. If the example of Mozart suggests a pattern, then fathers of child prodigies become strong authoritarian figures for their children. Pauli's early education in physics thus might very well have been forced upon him. I suspect that Pauli's strained relationship with his father followed a long-term pattern and was caused by a few critical events. Pauli's second wife Franca commented much later on the reasons for Pauli's animosity toward his father, first, because Pauli's father had converted to Catholicism, and second, because his father had remarried.[64]

Pauli's father left his wife Bertha, Pauli's mother, in 1927 to live with another much younger woman, the sculptress Maria Rottler. Soon thereafter, on November 15, 1927, Pauli's mother committed suicide by poisoning. Pauli's father then married Maria Rottler in 1928. Pauli referred to her as the "wicked stepmother."[65] Despite their strained relationship, however, Pauli did not become estranged from his father. In 1931, when the 31-year-old Pauli was suffering from emotional distress, his father recommended that he go to the Jung Clinic in Zurich for therapy. Then, in 1933, when Pauli met Franca Bertram, who would become his second wife, Pauli took his bride-to-be to his father's and stepmother's home in Vienna at Christmastime.[66] They married with his father's blessing on April 4, 1934, in London. Pauli's father lived in Vienna until he was forced to leave in 1938, while his wife remained in Vienna for the duration of the war. Pauli orchestrated his father's relocation to Zurich, but to thus enable them to live close to each other in relative safety, his father had to leave all of his laboratory equipment behind in Vienna. Then, when Pauli left for America in 1940, his father remained in Zurich, where he died in 1955.[67]  Pauli wrote to Carl Jung:

> On November 4, 1955, my father died of a weak heart at a ripe old age.  This leads to a considerable change in the unconscious, and I suspect that in my case it also means a transformation on the shadow.  For the shadow with me was projected onto my father for a long time, and I had to learn gradually to distinguish between the dream figure of the shadow and my real father.  Accordingly, the bond between the light anima and the shadow or Devil ... often used to appear projected onto the "wicked stepmother" (my father's much younger second wife) and my father.  The inner archetypal situation behind the external situation was always very clear with me.[68]

In this same letter, in which he recalls many of his dreams, Pauli also describes pairs of opposites that he sees in science and in Catholicism.[69]  He mentions no grief or sense of loss after learning about the death of his father in this significant letter to his intimate colleague Jung.

Another clue suggesting that Pauli and his father were distant but not estranged appears in a witty remark he made when his first wife left him for a chemist.  After they divorced in 1930, Pauli remarked: "If it had been a bullfighter—with someone like that I could not have competed—but such an average chemist!"[70]  His witticism assumes a disquieting seriousness when we recall that his father was a chemist. Pauli also dropped the "Junior" when signing his name to his publications after becoming a full professor in Zurich in 1928.  The common explanation is that earlier he had to distinguish himself from his father in his scientific publications.[71]  Was there also, however, a degree of professional rivalry between son and father?

Other evidence too suggests that Pauli and his father had a distant but not estranged relationship.  Pauli, in his dream analysis with Jung, frequently associates the archetypal female anima, the feminine personality, with intuition and feeling, and the masculine animus with emotionless reason and rationality. Pauli would report many dreams to Jung in which images of old men--Arnold Sommerfeld, Niels Bohr, and Albert Einstein-- appeared as masculine sages.[72] In many of his dreams, Pauli seems to want approval from these masculine sages. Perhaps, therefore, Pauli's father never gave him the validation and emotional encouragement he sought and needed to become a self-confident person. This is not an unusual thesis. In the latter part of the twentieth century, psychologists, poets, and clergy have focused on the common need of adult men to understand the nature of the father-son emotional dynamic.

Tom Owen-Towle in his recent book, *Brother-Spirit: Men Joining together in the Quest for Intimacy and Ultimacy,* speaks of the need for men to seek intimacy with other males: "By intimacy I refer to warm, close bonds grown through sharing our minds and hearts. By birth we are inescapably men; by choice we become brothers to other men, women, animals, divinity."[73] Owen-Towle goes on to describe the common father-son wound:

> My thesis is that the father-son dynamic and resolution influences greatly all male-male interactions which follow. This is true in biblical lore and our contemporary lives, whether we are speaking theologically or biologically, about our "heavenly" or "earthly" fathers. As with the case of God-Adam, then Cain-Abel, so also with their descendants. The blockage we men experience in our relationships with our fathers is worked out, usually in unsatisfactory and often damaging fashion with our brothers. To put it another way, because we are unresolved in our vertical, authority relationships, our horizontal, peer encounters suffer a consequential price.[74]

Pauli's life is filled with examples of his critical interactions with his peers in physics that might be interpreted as a remnant of his relationship with his father. An image comes to mind here that appears in Owen-Towle's book when he quotes from the writer Nikos

---

[72] Meier, *Atom and Archetype* (ref. 16), pp. 107, 149, 143.
[73] Tom Owen-Towle, *Brother Spirit: Men Joining Together in the Quest for Intimacy and Ultimacy* (San Diego: Bald Eagle Mountain Press, 1991), p. xvi.
[74] *Ibid.,* p. 3.

Kazantzakis's autobiography, *Report to Greco,* where he offers a vivid description of the common father-son dynamic:

> I had never faced my father with a feeling of tenderness. The fear he called forth in me was so great that all the rest---love, respect, intimacy---vanished. His words were severe, his silence even more severe. He seldom spoke, and when he did open his mouth, his words were measured and well weighed; you could never find grounds to contradict him. He was always right, which seemed to make him invulnerable.... An oak he was, with a hard trunk, rough leaves, bitter fruit, and no flowers. He ate up all the strength around him; in his shade every other tree withered. I withered in his shade similarly. I did not want to live beneath his breath.... This is why I was forced to write down all I wished I had done, instead of becoming a great struggler in the realm of action--from fear of my father. He it was who reduced my blood to ink.... I had feared only one man in my life: my father.... He alone remained always as I had seen him in my childhood: a giant. Towering in front of me, he blocked my share of the sun.[75]

This father-son dynamic—the son being distant from the father and seeking his approval—is near universal and not unique to the Paulis. We can speculate about the degree to which the young Pauli suffered from a distant relationship with his father, since events in Pauli's later life are consistent with a deep struggle with his father. In stereotypical fashion, the psyche of a child is a delicate flower opening from darkness into light. The child Pauli sees the ominous, strong, authoritative but distant figure of his father. By contrast, the mother's image is one that stereotypically nurtures and delicately assists a child's psyche. If the Pauli family was typical, then the young Pauli in his formative years was exposed to the strong and distant personality of his father and the close and kind personality of his mother. Associate then these persons with the hard rationality of the father-scientist and the emotional religiosity of the mother, and we have key elements of the polar opposites that Pauli the physicist struggled to reconcile during the balance of his life.

There are several potential influences of Pauli's relationship with his father on his physics. Pauli continued to struggle, trying to do too much too well. He tried consistently to perfect the rational and empirical in his physics; he struggled consistently to release his intuitive theories. His father was the empiricist; Pauli had to discover his intuitive physicist voice on his own. His father orchestrated Pauli's physics education to be guided by Ernst Mach, who was opposed to both the concept of atoms and to the theory of relativity. Pauli's father was a disciple of Mach, so one can assume that Pauli's early logical-positivist leanings originated with his father. Pauli's process of individuation from the personality of his father can be seen in Pauli's struggle in his physics: he struggled to follow his intuition while simultaneously trying to be perfectly rational and to adhere to empirical facts.

---

[75] *Ibid.,* p. 8.

**Pauli's Mother, Grandmother, and Sister**

Bertha Camilla Pauli neé Schütz, Pauli's mother, was the daughter of a well-to-do Viennese family. Her father was Friedrich Schütz, a fallen Jew, playright, writer, and editor; grandson Pauli would bear the middle name Friedrich after him. His grandmother was Bertha Schütz neé Dillner von Dillnersdorf, of noble descent**,** a Catholic, and a singer in the Imperial Opera in Vienna. She was talented across a wide spectrum of artistic endeavors. She imbided in her grandson Pauli an interest in piano, singing, and aesthetics; she would play the piano for hours for him.[76] She was a warm and nurturing presence for the young Pauli, who much later would recall her womanly influence in a dream mixed in imagery with that of his friend and colleague, the Jungian analyst Maria-Louise von Franz.[77] Pauli's image of anima, the ideal archetypal woman, was partly generated by his grandmother. He would write about his dream of piano playing in 1953, which was important to Jungians because of Pauli's dreams of the anima.[78] Grandmother Schütz was an influential person in Pauli's formative childhood.

Pauli's mother Bertha was a Catholic by birth and had an influential and distinguished career, becoming a correspondent for her father's newspaper, the influential *Neue Freie Press,* a feminist and pacifist, all the while maintaining a close and loving relationship with her son. Charles Enz describes her as a strong personality, motivated by beauty and righteousness, a progressive and liberal woman having an extensive knowledge of dramatic literature.[79]  She earned an education when it was very difficult for women to do so, attending a well-known high school (*Gymnasium*) for girls and passing her final examinations (*Matura*) in classics in 1905, at the age of 27, five years after the birth of her son.  She and her husband both left the Catholic church in 1911 for unknown reasons; she was described thereafter as Protestant (*Evangelisch*).[80]  She was a passionate socialist and adamantly opposed to the Great War, as well as a passionate writer on topics united by the common thread of "beauty and righteousness."  With her strong personality, her influence on her son Pauli must have been profound since, as Abraham Pais remarked: "It is known that Pauli had strong ties to his mother."[81]  The scant historical clues suggest that she took an active and nurturing role in the rearing of her son, and later of her daughter.

---

[76] Enz*, No Time to be Brief* (ref. 3), p. 4.
[77] *Ibid.*
[78] H. Atmanspacher, H. Primas, E. Wertenschlag-Birkhäuser, ed., *Der Pauli-Jung Dialog und seine Bedeutung für die moderne Wissenschaft* (Berlin: Springer, 1995), *Anhang B: Klavierstunde*, pp. 317-345. Several articles discuss Pauli's dream of *Die Kalvierstunde*.

Pauli's mother, despite her strength and passion, developed an "existentialist anxiety,"[82] which seems to have contributed to her suicide by poisoning on November 15, 1927, at the age of 47.  As noted above, Pauli's father had left his wife before her suicide to live with a much younger woman whom he later married.[83]  I will assume that Pauli's mother's suicide was precipitated by his father's affair.  We do not know when this affair began, nor do we know the length and degree of Pauli's parents' marital discord.

That both Pauli's father and mother abandoned Catholicism raises images of religious conflict in their household.  Moreover, the young Pauli went on to adopt the socialist and pacifist convictions of his mother,[84] while he also embraced his father's adherence to logical rigor and empirical facts.  Pauli's parents thus seemed to view the world quite differently.  Pauli's youthful veneer of Catholicism, in contrast to his later attraction to his hidden Jewish roots, implies a lifelong process of personal inquiry, confrontation, and validation.  Historians may never know the extent to which the Pauli household was involved in dialogues or heated confrontations over religion and politics.  To the child, reality rests with the parents.  Only an adult can choose a personal religion and his or her politics. Pauli's philosophical interests mixing polar opposites may have originated in the Pauli household.

---

[79] *Ibid.*, pp. 9-10.
[80]  I thank Dr. Stuewer for helping to clarify this translation.
[81] Pais, *Genius of Science* (ref. 1), p. 214.
[82] Enz, *No Time to be Brief* (ref. 3), p. 10.
[83] Charles Enz, "Wolfgang Pauli (1900-1958): A Biographical Introduction," in C. Enz and K. von Meyenn, ed., *Wolfgang Pauli: Writings on Physics and Philosophy* (New York: Springer-Verlag, 1994), p. 18.
[84] Enz,, *No Time to be Brief* (ref. 3), p. 14.

Pauli's sister, Hertha Ernestina Pauli, was born on September 4, 1906, six and a half years after him. According to Charles Enz, her middle name also honored Ernst Mach.[85] Like her mother, she took an interest in writing, in strong women, in art and beauty, in peace and justice. She went on to become a biographer; perhaps her best-known biography was of Bertha von Suttner, the Austrian pacifist and winner of the Nobel Prize for Peace in 1905, in whom she saw qualities similar to those of her mother. She also wrote a biography of Alfred Nobel with the revealing subtitle *Dynamite King-Architect of Peace*,[86] and another of the abolitionist Sojourner Truth,[87] in both of which she reveals her deep commitment to social justice. In general, in contrast to her brother's sterile physics, her biographies convey a deep interest in politics and social causes that are in keeping with her own life. Her biography of Bertha von Suttner was banned by the Nazis, and she had to flee Austria by a perilous route over the Pyrenees into Spain. She left Europe for the United States in 1940 and eventually married another writer and emigrant, Ernst Basch (E.B. Ashton) of Munich. She became an American citizen in 1952 and lived with her husband in a small farmhouse at 102 Woodhull Road, Huntington, Long Island, New York. They had no children. She died on February 9, 1973, at the age of 67 in Southside Hospital, Bay Shore, Long Island, New York.[88]

Tantalizing questions remain to be answered about Hertha Pauli. What was it like to grow up in the shadows of her grandparent's prominent positions in Viennese society, her mother's career, her famous chemist father, and her famous physicist brother? How did she respond to the breakup of her parents' marriage, and what was her reaction to the suicide of her mother? Did she introduce her brother to his first fiancée? When and why did she begin her career as a biographer of people committed to social causes? How did she survive the Nazis? Charles Enz and others provide some information on her,[89] but much still remains to be investigated.

---

[85] *Ibid.,* p.12.
[86] *Ibid.,* p.18.
[87] Hertha Pauli, *Her Name Was Sojourner Truth* (New York: Appleton-Century-Crofts, 1962).
[88] Enz*, No Time to be Brief* (ref. 3), pp. 17-19.
[89] *Ibid.*, pp. 11-19.

Pauli scarcely mentions his sister Hertha in his published correspondence.  In one of his letters to Jung, he associates her birth with the birth of his anima, the unconscious side of his psyche.  The anima also is the ideal archetypal female.  Her birth when he was nearly seven years of age becomes another sign, since he associates the number seven numerologically with the anima.[90]  She was born, however, when he was six and not seven--a Freudian slip in Pauli's memory?  In any case, her birth represented a significant transition for him, since parental attention then had to be shared.  Probably owing to their difference in ages, it seems that as an adult Pauli had a warm but emotionally distant relationship to his sister.  His few published comments about her presence in his life are ones of love and devotion.

**Pauli's Women and Jung's Anima**

In many of Pauli's dreams that he shared with Carl Jung, one can discern a connection to his sister.[91]  He describes his image of her as the ideal symbolic female, the anima.  Pauli had a problematic relationship to women throughout his life, and there is no easy way to describe his thoughts on the human female.  Since to Jung the anima, the archetypal ideal female, is buried in a person's Shadow—a concept that Pauli endorsed—it is worth exploration.  The qualities Pauli sees in the anima also are connected to the way in which he pursued physics.  In Jung's archetypal psychology, the anima is the source and bearer of intuition, visual imagery, aesthetic feelings, kindness and warmth, nurturing, and feminine love.  The anima also has a dark side, one of treachery, mystery, deceit, and cunning.  Jung's psychological metaphysics, of course, was subjective and based on the intuitive patterns he saw in his clinical studies.

The earliest women in Pauli's life were his mother Bertha, his maternal grandmother Bertha, and his sister Hertha, in order of importance in his formative years.  His paternal grandmother Helen Pascheles first appeared in his life in 1916 and first made him aware of his Jewishness.   He had a difficult time emotionally when his mother committed suicide in 1927.  He did not like his new and much younger stepmother.  When he sought psychological counseling at the Jung clinic in his early thirties, Jung reported that his patient was having difficulty in relating to the women he had courted.[92] Pauli's first marriage ended quickly. He had an extramarital affair during his second marriage.[93] In his later years, he had a close working relationship with the Jungian Maria-Louise von Frantz[94] (of note, his mother's nickname also was Maria[95]). He had excursions into pornography, and he had strained relationships with women prior to his emotional problems in his late twenties and early thirties—which led him to seek help in the Jung clinic. During his professsional career, he held female physicists such as Lise Meitner and Chien-Shiung Wu in high regard. To state the obvious, Pauli's relationships to women were complex, emotional, messy, tangled, and perpetually compelling. He found love, nurturance, reassurance, and idylic purity in his mother, maternal grandmother, and sister. He found attraction, nurturing, disappointment, repulsion, joy, animosity, humanity, enigma, and love in his wives and lovers. What is striking is Pauli's later attraction to Jung's concept of the anima as seen, for example, in a letter of February 27, 1953, that he wrote to Jung.[96] In it he used Jung's concept of the anima in his psychological-philosophical musings about several topics--including theoretical physics. Pauli was drawn to a pure form of theoretical physics that was manifested by beauty, symmetry, wholeness, intrigue, and inscrutable mystery, and he associated the anima with its pursuit, with a release of his creativity through intuition in a quest for beauty and the ideal. Pauli thus spoke of his relationship to creative theoretical physics as a metaphor for his relationship to the anima, the archetypal ideal woman.

Pauli's parents left the Catholic church in 1911 but kept their son and daughter associated with it. Several questions come to mind about this presumably emotional event and its consequences. Was Pauli well grounded by then in Catholicism within his peer group and within his cultural environment, and did his parents' action thus precipitate unsettling emotions in him? A young adult goes through a period of questioning and determining his or her truly core values, including religious ones. In the absence of further documentation, his parents' action will remain of unknown significance for his psychological development.  Pauli remained a Catholic until he himself left the church in 1929 for unknown reasons, but ones that presumably were related to his mother's suicide in 1927 and his father's remarriage in 1928.[97]

**Pauli and Mach**

Apart from his immediate family, the person who exerted the most influence on the young Pauli was the physicist-philosopher Ernst Mach.  Pauli's father was close to Mach both personally and professionally, so close that he asked Mach to be Pauli's godfather, and he christened his son Wolfgang Ernst Friedrich Pauli when he was baptized as a Roman Catholic.  Mach exerted his influence further through Pauli's father by recommending educational materials in physics and mathematics for his godson.   Pauli later attributed his own antimetaphysical stance to Mach's influence (without, however, commenting on Jung's metaphysics).[98]

Pauli's father may have been more impressed with Mach than the young Pauli himself.  In  the margins of one of the texts Mach suggested, Pauli noted that to obtain a vivid picture of classical culture, the classical authors should be read in their original language, which was contrary to what Mach had espoused.[99]  Also, in a letter to Jung in 1953, Pauli remarked:

> What Mach wanted [in his positivism], although it could not be carried out, was the total elimination of everything from the interpretation of nature that is "*not* ascertainable *hic et nunc.*"  But then one soon sees that one does not understand anything--neither the fact that one has to assign a psyche to others (only one's own being ascertainable) nor the fact that different people are talking about the same (physical) object (the "windowless monads" of Leibniz).  Thus, in order to meet the requirements of both instinct and reason, one has to introduce some *structural elements of cosmic order,* which "in themselves are not ascertainable."  It seems to me that with you [Jung] this role is mainly taken over by the archetypes.[100]

The elderly Mach, who had suffered a stroke in 1898 that left him paralyzed on his right side and unable to speak clearly, and who had retired from his professorship in Vienna in 1901, probably did not strike the young Pauli as a warm and empathetic figure. Mach equipped his apartment with laboratory apparatus of all kinds, and would demonstrate to the young Pauli the mental illusions and errors that occur when following one's intuition instead of carefully controlled experiments. Pauli seems to have enjoyed Mach's experiments and instruction, but also to have had reservations about Mach's way of thinking about the world. Pauli later recalled his visits to Mach:

> Working on the assumption that his psychology was a universal one, he recommended everyone to use that inferior auxiliary function [intuition] as "economically" as possible (thought economy). His own thought processes closely followed the impressions of his senses, tools, and apparatus.[101]

Mach was enormously popular in Vienna for his progressive philosophical ideas on physics; for his support of socialism and economic justice; for his opposition to militarism and racism. According to Walter Moore:

> The success of Mach in Vienna may be found in the persistent search of the romantic German mind for a monistic or unitary picture of man and nature…. The phenomenalism of Mach was well adapted to the superficial upper-middle-class society of Vienna.[102]

To the young and impressionable Pauli, Mach appears to have been an authoritarian, judgmental person in contrast to Pauli's mother and grandmother Schütz. Here was an old man, disabled, a living symbol of successful science but living in seclusion, yet someone whom Pauli's father greatly admired and who encouraged his son, in not too subtle a fashion, to follow as a mentor whether or not he found it personally attractive to do so.

**_Fin-de-Siècle_ Vienna**

The Vienna of Pauli's childhood was the home of Ernst Mach, Sigmund Freud, Carl Jung for a time, Gustav and Alma Mahler, the young Arthur Koestler, the Vienna Circle of philosophers, the Lamarckian biologist Paul Kammarer, musicians, opera stars, artists, leading physicists, and many other intellectuals. European high culture, in a word, flourished in Vienna during Pauli's childhood and adolescence between 1900 and 1918, and through his father and mother, he enjoyed access to many of these leading intellectuals. But these also were years of intense social pressures and sweeping political changes. The Great War swept across Europe, ethnic minorities passed through and some settled in Vienna, the Hapsburg Monarchy came to an end, anti-Semitism increased, and uncontrollable inflation was on the horizon. These were tumultuous times in Vienna, as Arthur Koestler has described:

> Vienna before the First World War is as remote as the vanished continent of Atlantis. It was a glittering world of opera, theatre and concerts, of picnics on the Danube, summer nights in the vineyards of Grinzing, and love affairs light as fluff. It was also a world of corruption and decadence, creaking in all its multinational joints, waiting to fall to pieces.[103]

The population of Vienna rose from about 440,000 in 1840 to over 2,000,000 in 1910 and reached its peak at almost 2,500,000 in the early years of the Great War. After the Great War, Vienna no longer was the imperial capital of a monarchy of 52,000,000 people, but a city in a country of vastly smaller population. Changing political borders, economic crises, the German Anschluss, and persecution of the Jews would reduce the population after the Great War and continue to affect Vienna, where Pauli's father continued to live.

Pauli spent his youth and adolescence in economically and emotionally stressful times. Outward appearances of a comfortable upbringing in a stimulating and caring environment need to be tempered by reflections of the times. His parents turned to pacifist and socialist politics. He himself petitioned to be excused from military service owing to a "weak heart," an obvious ruse.[104] As an adult he would refer to Vienna then as a "_geistige Einöde_," a spiritual desert. He explained to the physicist Valentine Telegdi

when asked when and why he left Vienna: "Me already in 1918. I always had a good intuition."[105]

In Pauli's youth Vienna was a center of physics. Ernst Mach and Ludwig Boltzmann taught and Paul Ehrenfest and Erwin Schrödinger studied at the University of Vienna. Einstein from nearby Prague met with Mach in Vienna in 1911 after lecturing at the university, quite likely with Pauli's father in attendance.[106] Theoretical physicist Fritz Hasenöhrl, Schrödinger's teacher, taught at the University of Vienna and was killed in 1915 in the Great War. The doyen of Viennese physicists, experimental physicist Franz Exner, director of the Second Physical Institute, was likely in close contact with Pauli's father.[107] Pauli's father thus no doubt was familier with many developments in physics at the time and passed his knowledge on to his son, supplementing Pauli's education in relativity and quantum physics. Guided by his father's efforts to stimulate Pauli's imagination in physics, Pauli would reject Mach's views on relativity, atomism, and positivism.

Vienna was an active center of physics, but classical Machian "economical" physics[108] was still predominant. The grounding assumption of classical physics, that all would become known, was being shattered by new developments in relativity and quantum physics. In addition, physics was no longer the innocent discipline of abstraction. With the Great War, physicists would be directly involved in the war effort across Europe as all factions encountered the destructive potential of science.

**Pauli and Myers-Briggs Personality Typology**

Pauli became recognized as a genius in physics, which is sufficient reason to examine his formative years for patterns of uniqueness. He also experienced an emotional or psychotic breakdown in his early thirties, which provides an additional reason to attempt a psychological diagnosis of him. Because psychological analysis was still a young discipline when Pauli entered the Jung Clinic in Zurich, the methodology of his treatment was still in flux. Indeed, Pauli's work with Jung helped to refine that methodology, which Pauli subsequently endorsed.  I will attempt to understand the genius and personality of Pauli by using the tools of personality typing as pioneered by Jung.

---

[105] *Ibid.,* p. 15.
[106] Ronald Clark, *Einstein: The Life and Times* (New York: World Publishing, 1971), pp. 159-160.
[107] Moore, *Schrödinger* (ref. 73), p. 35.
[108] *Ibid.,* p. 42. Moore stated: "Mach believed that the aim of science was the ordering of the elements in the most economical way."

Jung believed that human personalities could be dissected into metaphysical categories, and that such analysis could be used to determine behavior.  Jung's metaphysical categories, however, were at an early stage of development, so I will use the closely related personality typology of his disciples Isabel Myers and Katherine Briggs, which is now more familiar and accessible.  Moreover, Myers and Briggs's personality typology has components arranged into a quaternity, something that Pauli would have loved.  In their typology, four dimensions are required to understand the unity of personality.  These four metaphysical dimensions of personality concern the ways humans process information, the filters through which humans look at the world.  I offer here a brief description of Myers and Briggs's typology that will prove useful in my analysis of Pauli's personality.  More refined descriptions are widely available.[109]

Myers and Briggs's first category of personality typing concerns the relative "place" in which information is processed, as in "external" *versus* "internal." The Extrovert prefers externally processed information, usually in contact with other people; the Introvert prefers internally processed information, usually in solitude.  Each person has a placement indicator for his or her personality along a continuous scale, between the extremes of completely Extrovert (E) and completely Introvert (I).  If one's personality tests indicate placement closer to the Extrovert (Introvert) extreme, that person would be labelled an Extrovert (Introvert).  A person also may have a balanced personality, the indicator then lying in the middle of the scale:

Extrovert (E)___________________________(I) Introvert

Myers and Briggs's second category deals with how a person values the information he or she receives.  The Sensing personality values information received from the objective senses; the Intuitive personality values information received subjectively from the

"mind's eye." Each person's personality can be placed somewhere along this continuum between the two extremes of Sensing (S) and Intuitive (N):

Sensing (S)______________________________(N) Intuitive

Myers and Briggs's third category deals with the methods a person uses to process information. The Thinking personality uses only rational, nonemotive methods of logic. The Feeling personality processes information emotionally, empathetically. A continuum exists here as well between the two extremes:

Thinking (T)______________________________(F) Feeling

Myers and Briggs's fourth and last category deals with where the processed information is "stored" and the firmness with which it is categorized. The Judgmental personality needs to categorize information rigidly, to place the processed information into firm boxes within the "mind's eye." The Perceptive personality continually tries to make new connections among the categories of information by placing, removing, and rearranging "stored" information within the topical categories of the "mind's eye." Each person's personality also can be "typed" along this continuum:

Judgmental (J)______________________________(P) Perceptive

In Myers and Briggs's testing and typing, each person has a unique personality that can be placed into one of sixteen approximate categories. Some personalities test near the middle of the continuums. Some also may have an oddity in their categorization; Pauli was one example, as I will soon describe. Myers and Briggs's typology is an inexact metaphysical scheme, yet the behavior indicators that emerge have proven to be remarkably successful.

Myers-Briggs typologists have refined their testing procedures so that people can be categorized based upon the ways in which they behave in the world. Here are some examples. An outgoing, experimental scientist (E) who values sensory information (S), processes that data with extreme rational logic (T), and categorizes the evidence so processed into rigid classifications (J), likely would be classified as an ESTJ personality. A reclusive (I), intuitive (N) person, with strong emotional responses to situations (F), who is creatively engaged in the arts and easily changes his or her mind (P), might be labelled an INFP personality.

I will describe Pauli's father as an ESTJ personality because of the way he behaved, while Pauli's mother appears to have been an INFP personality because of her artistic career and writing, her sense of beauty and justice, and her pattern of changing her mind. My categorizations may be incorrect, but Pauli's parents seem to exhibit these patterns. In addition, the metaphysical "personality distance" between ESTJ and INFP is the largest one within the typology system, and so selecting his parents as extremes serves to emphasize my perception of Pauli's situation.

Pauli himself is more difficult to type. During his early career he is outgoing, values experimental data over intuition, thinks about everything with hidden emotion, and is very opinionated and judgmental. Some of these traits continue throughout his life, but beginning in his twenties he exhibits other traits. He becomes more of a loner in his thoughts, begins to place high value on his intuition, has emotional releases implying pent up psychic pressures, and changes his mind about his earlier conclusions. Thus the heart of my thesis is that Pauli was naturally an INFP personality but was coerced to adopt an ESTJ personality during his childhood and adolescence. This helps me to see a pattern in Pauli's behavior, particularly during his career in the 1920s and at the time of his emotional crisis around 1929. In his youth, Pauli's consciousness was an ESTJ personality; his INFP side was contained within his unconscious Shadow. His INFP side was emerging during his twenties, but was affected traumatically by the events that led to his emotional crisis in his late twenties. After his time in the Jung Clinic starting in 1932, he felt validated and began to explore the natural side of his personality, but he never fully integrated the two extremes. He went through most of his life with an abnormal "two-valued" personality type, displaying his ESTJ personality to most of his professional colleagues, while revealing his INFP personality only to his closest friends. Pauli's INFP personality emerges especially with colleagues who shared his Jungian perspective.

The most influential people in the life of a child are the child's parents.  In Pauli's case, my supposition that his parents, both strong-willed, were of opposite personality types and were both involved actively in his parenting and education, indicates a stressful situation for him.  His mother and grandmother nurtured and encouraged his INFP traits with their music and artistry.  His father, and Mach, were instilling ESTJ traits at the same time as they were exposing him to the world of physics.  When the parents are involved actively in the raising of their child, their emotional responses and philosophical positions become instilled in their child's developing psyche.  The child's *tabula rasa*, however, receives this information in ways that do not necessarily reflect the parents' intentions or priorities.  A minor parental squabble observed by the child may initiate a traumatic impression.  A profound insight may come to the child from an unintended parental comment.  The infant does not follow a cognitive development that is rationally organized, but instead one that is emotionally laden.  The child's cognitive progression is developmental, with some rational or other capabilities emerging only later in adolescence.  Thus, psychic concepts that are unambiguous to the adult may be loaded with all sorts of emotional connotations to the child.  Jean Piaget, for example, has studied the developmental stages followed by the child in learning complex concepts.  The child learns the concept of time, in particular, not in a straightforward, linear route, but instead by mixing it up with the child's emerging powers of spatial ordering and coordination.[110]  A temporal sequence of events has far different meaning for the child than for the adult.  The child's interpretation of observed events, and thus the concept of time, are beyond rational logic and instead follow the developmental, emotional pattern.  Thus one can assert that many childhood impressions of seemingly inert concepts, like those associated with the world of physics, may be laden with deep nonrational content.[111]  Pauli's physics concepts were emotionally laden, one reason perhaps being his exposure to them in early childhood.

Pauli was a prodigy in the rational subjects of mathematics and physics.  His childhood emotional development is another matter.  Even if Pauli was normal in the early stages of his emotional development, he no doubt encountered the trying times so familiar to all parents of adolescents. If, as Carl Jung might maintain, his psychic energy was finite and limited, and if his conscious psychic energy was being consumed by his pursuit of the deeply rational, then his feelings and emotions were being forced into his unconscious. His developmental cognitive patterns were such that his consciousness was advanced, but the emotional content of his unconscious psyche was being arrested: He had developmental asynchrony characteristic of an adolescent. He might have associated the inert concepts of physics with emotional connotations. We know that he later would bring to his physics a deeply serious, almost religious perspective to his pursuit of the meaning and reality he sought in physics. He would seek not just the impersonal, inert structure of the physical world; he also would seek deep teleological messages within physics and connected to philosophy, psychology, and other disciplines. He was not playing games. As I already cited, Heisenberg in 1963 would recall when reflecting on his friendship with Pauli:

> You know physicists really do very serious things; they think about the structure of the world. After all, that's what we do. So then why is it that so many physicists are in disagreement with that way of thinking, also with this side of a man who took these things very seriously? It was not for Pauli a kind of funny game. It was certainly not meant as opium; it was the contrary of opium for Pauli.[112]

---

[110] Jean Piaget, *The Child's Conception of Time* (New York: Ballantine Books, 1971).
[111] This is my conclusion from reading Piaget's book.

Pauli underwent therapy in the Jung Clinic partly in an effort to deal with his buried emotions. Pauli enthusiast F. David Peat has pointed out that:

> Jung found his patient to be: ...a university man, a very one-sided intellectual. His unconscious had become troubled and activated; so it projected itself onto other men who appeared to be his enemies, and he felt terribly lonely, because everyone seemed to be against him.
>
> He had lived in a very one-sided intellectual way, and naturally had certain desires and needs also. But he had no chance with women at all, because he had

> no differentiation of feeling whatsoever.  So he made a fool of himself with women at once and of course they had no patience with him. [113]

Pauli learned his physics at a young developmental age when the purely physical (if there is such a thing) became mixed with his emotional cravings.  He could have grown up unaware of the mixed messages coming from his father and mother.

Robert Bly, the Minnesota poet who has made a lifelong venture out of Jungian psychological dynamics, has applied the concept of the unconscious to the human condition.  He speaks of the Human Shadow.  He refers to the Shadow, that part of our unconscious psyche, as that psychic "long bag we drag around behind us…. We spend our life until we're twenty deciding what parts of ourself to put into the bag, and we spend the rest of our lives trying to get them out again."[114]  If we apply Bly's concept of the Human Shadow to young Pauli, then we can discern in his childhood causes for his emotional and cognitive struggles as an adult.  His Shadow bag may have been filled beyond his ability to empty it.  Many issues may have entered his Shadow during his childhood and adolescence.  The birth of his sister and his resultant jealousy was a likely one.  The hidden Jewishness of Pauli's family, disguised as Catholic until his parents left the Catholic church when he was eleven, was another.  Pauli's natural INFP personality being coerced to adopt an ESTJ mantle was another.  Pauli being forced to adopt Mach's philosophy when he was beginning to sense contradictions in it, for example, in the close connection between mathematics and physics,[115] was another.  The ominous *fin-de-siècle* cultural milieu in Vienna prior to the Great War, with its accompanying eroticism, psychic energy, and chaos, was still another.  The tumultuous political undercurrents in Vienna prior to the Great War, and the effects on the young Pauli of the Great War were still more issues that he may have stuffed into his Shadow.

Support for my contention that Pauli was troubled by his Shadow comes from Pauli's own words.  He wrote of his dream of July 20, 1954, to Carl Jung in a self-analysis in which his dream reminded Pauli of how his wife approached life situations differently than he did:

---

[113] Quoted in F. David Peat, *Synchronicity* (ref. 63),  p.17.
[114] Robert Bly, *A Little Book on the Human Shadow* (San Francisco: Harper & Row, 1988), pp. 17-18.
[115] Moore, *Schrödinger* (ref. 73),  p. 45.

In this connection I should like to point out that in my case, as far as I am in a position to judge, the evaluation of the functions [Jung's personality functions, akin to the Myers-Briggs personality types] in the general function schema has shifted somewhat in the course of my life. It seems to me that in earlier years the thinking function was the most differentiated one, and feeling was correspondingly the inferior function. These days, I regard intuition as my most differentiated function, and accordingly it seems to be going better with the feeling side, and *the inferior function is extraverted sensation* (The weak link with reality) [Pauli's italics; Jung added the phrase in parentheses].[116]

In the next chapter, I will explore Pauli's first exposures to the hidden symbolism of mathematics and physics. His intuition sees a deeper level of reality as he recognizes specific, repeated symbols. Much later, these symbols will become validated for him by Jung as mandalas and quaternian processes.

---

## Chapter 3. Pauli's Adolescence and Gymnasium Education, 1913-1918

### Adolescence

Adolescence, that awkward period of expansion of the child's world, culminates in separation from parents with the goal of reaching independent adulthood. Some children do better than others at this process. In the case of Pauli, there are few indications of external traumatic events during his adolescence, but his internal responses to events were crucial to his sense of identity and self-confidence. Some of the events and factors that may have occupied his attention were the Great War, his ambiguous religious identity, his coerced personality type, and his individuation process in separating from his parents and adopting his own world perspective. To explain Pauli's later struggles, I will assume he was affected by these events and factors. To explain his later artistry in physics, I propose that his attempts to come to peace with them became intertwined with his physics.

The Great War directly affected Pauli's country, family, and friends. Abraham Pais noted that, "It was a man-willed slaughter the likes of which had never been witnessed before, the number of killed, wounded, and missing exceeding thirty-seven million…."[117] The emotions, politics, and moral principles of scientists were challenged as never before. Science, especially chemistry, was placed in the service of war. In Germany and Austria, physicists were divided over their country's justification for the war, with an influential majority supporting their political leaders. Only a handful opposed the war, notably Einstein and, as it turns out, the young Pauli.[118]

The war affected Pauli's life in several ways. Fritz Hasenöhrl, professor of theoretical physics at the University of Vienna, was killed by a grenade on the Italian front in 1915, which influenced Pauli's decision to move from Vienna to Munich to continue his studies under Arnold Sommerfeld.[119] Of Pauli's graduating class of twenty-seven boys from the Döbling Gymnasium in 1918, all but Pauli entered military service;

Pauli was exempted with the dubious excuse of "faint-heartededness."[120]  Pauli's mother, the firebrand feminist, pacifist, and socialist, opposed the war and her son took up her cause:

> After the eruption of World War I, a passionate interest for politics awoke in him [Pauli] which certainly was also nourished by his socialistically oriented and literarily active mother.  The longer lasted the war the keener became his opposition against it and, generally, against the whole "establishment."[121]

In reflecting on how the adolescent Pauli internally processed his emotions associated with the war, one suspects that his Shadow may have inclined him to avoid it and to enter physics.

Pauli learned about his Jewish heritage at age sixteen from his Jewish grandmother, just when many adolescents question and form their religious identity.  Raised as a Catholic by parents who left the church when he was eleven, Pauli now had to contend with hidden religious complexity.  Which spiritual direction should he pursue, the scientism of his father, the Jewishness of his ancestors, the Catholicism of his environment?  Whether or not he was aware of them at the time, his spiritual questions later would permeate his thoughts in physics, when he would invest physics with animistic qualities as, for example, as Heisenberg reported, when Pauli associated the electron's "two-valuedness" with the devil.

I speculated in the last chapter about Pauli's coerced personality type.  In his pioneering work on personality type, Carl Jung studied cases where children adapted their natural personality types to those of their parents, to gain the parental responses they desired.  In most cases, the child's natural personality eventually predominated, but in abnormal cases the result was harmful:

> Under abnormal conditions … when the mother's own attitude [personality type] is extreme, a similar attitude can be forced on the children too, thus violating their individual disposition, which might have opted for another type if no abnormal external influences had intervened.  As a rule, whenever such a falsification of type takes place as a result of parental influence, the individual becomes neurotic

---

[120] *Ibid.*, p. 15.
[121] *Ibid.*

later, and can be cured only by developing the attitude consonant with his nature.[122]

We do not yet know the exact nature of Pauli's later treatment at the Jung Clinic, but I suspect it involved helping Pauli free his dormant intuitive INFP side of his personality. Pauli's father, with his devotion to Ernst Mach and experimental science, was pushing his adolescent son in a direction at odds with his natural affinities, which were to his mother, to the arts and humanities. Pauli's dormant INFP personality would later hinder him from following his intuition in creative physics.

Pauli's individuation process of separating from his parents and becoming his own man in his own "mind's eye" was not unusual. It is completely normal for an adolescent or young adult to individuate from his parents. In Pauli we see this in his politics: as a teenager, his politics imitates his mother's; as an adult, his politics prompted withdrawal. In another area, the teenager Pauli was exposed to the experimental physics of his father and the positivism of Mach; the adult Pauli was a theoretical physicist who explored Platonism. Some time between adolescence and adulthood, Pauli individuated. Pauli's individuation process is the subject of my present speculations, because his physics appears to be two-sided, its ESTJ side reserved and conservative, its INFP side wildly intuitive and heated.

**Döbling Gymnasium Education**

Pauli's graduating class from the Döbling Gymnasium became known as the "class of geniuses," since it included two future Nobel Prize winners in science, two famous actors, an important conductor, three university professors, two directors of medical schools, a politician, and several industrialists. Something special was occuring in the Döbling Gymnasium. It was noted for its humanistic studies, but Pauli's specialty was mathematics and physics, which surfaced as early as his fifth grade.[123] By the time he graduated at age eighteen, he was recognized as a genius in mathematics and physics, with his humanistic education still being important to him.

---

[122] Carl Jung, *Psychological Types* (Princeton: Princeton University Press, 1990), p. 332.
[123] Enz, *No Time to be Brief* (ref. 2), p. 27. I interpret this to mean the fifth year at the Gymnasium, which makes Pauli's age 14 or 15.

Pauli's education did not take place entirely at school; an adolescent acquires knowledge from many sources, sometimes with unintended consequences. I assume that Pauli's mother and maternal grandmother continued to lead him in appreciating the arts, theater, opera, and music. The young Pauli was exposed to leading intellectuals and the highest cultural and artistic venues that Vienna had to offer, ones in which his mother and grandmother were intimately involved. The stimulating environment in Vienna offered the impressionable adolescent Pauli many potential learning experiences to nurture his intuition, feelings, and perceptions. The appreciation of aesthetics and emotions, the general areas that appeal to the Myers-Briggs INFP components of one's personality, undoubtedly continued to grow in Pauli during his adolescence. We do not know if he was conscious of these thought patterns, or if instead he was placing them into his Shadow, but he later would write to his German literature teacher about his appreciation of the humanities and aesthetics:

> That the mention "with distinction" is written on it [my diploma] somewhat surprises me even today. For, my talents were specifically in mathematics and physics while otherwise I was a mediocre pupil. You will be surprised when I now say: I am glad to have attended the humanistic high school and [to] have learned Latin as well as German … but later I became interested in scientific texts from the 17[th] century, as well as in the Greek philosophers (*Naturphilosophen*)…. I also have the best remembrance of your lessons in German literature.[124]

Pauli thus did not recall that he did well in his nonmathematical and nonscience courses, but he learned from them.

At the Döbling Gymnasium, Pauli took six years of mathematics and two years of physics and chemistry from Rudolph Kottenbach, earning a grade of "sehr gut" in all of his classes.[125] Kottenbach's philosophy of physics and mathematics education has been described as follows:

> In a detailed article for the 1906/07 Gymnasium [humanistic high school] yearbook, Kottenbach advocated a "thorough reform of the natural science instruction of our middle school students [for ages 10 to 16 years for grades 5 to 10]," and an "increase in the number of physics lessons." Using a compilation of stereotypical problems in the 1913/14 yearbook, he illustrated his teaching concept. In this context Kottenbach showed the value of practical use of mathematical tools, which until then had been mainly neglected in mathematical instruction: "Proofs in arithmetic, proofs in geometry comprised the main context of mathematical instruction, even--*horribile dictu*--in the examinations of the final matriculation test. That the addiction to proofs is not suitable to the spirit of youth, that it takes the joy out of instruction of youth, nobody will doubt.... [The] lively, fiery spirit of youth is frightened by proofs à la Euclid as they would be by the monotony of slave-like services imposed on mathematics by the needs of life, technology, navigation, astronomy, and geodesy." [126]

Kottenbach's influence may have resurfaced later in Pauli's retort to John von Neumann about his attempts to find a proof of the completeness of quantum mechanics: "Well, if a proof was important in physics, you would be a great physicist."[127]

Several physicists also influenced Pauli during his formative years at the Döbling Gymnasium. I already discussed Ernst Mach. Mach and Pauli's father guided Pauli's studies in physics, and arranged tutors for him since the Döbling Gymnasium concentrated on humanistic studies. Mach himself was too elderly and frail to serve as Pauli's tutor. Charles Enz has concluded instead that Pauli was tutored in relativity theory by the theoretical physicist Hans Adolf Bauer of the University of Vienna, who also seems to have recommended to Pauli that he read Erwin Schödinger's and Hans

---

[126] *Ibid.*, p. 51. My translation of the following passage: "In einem detaillierten Beitrag zu dem Jahresbericht 1906/07 des Gymnasium setzte sich Kottenbach für eine 'gründliche Reform des realistischen Unterrichtes an unseren Mittelschulen' und eine 'Erhöhung der Zahl der Physikstunden' ein. Anhand einer Zusammenstellung von 'mathematischen Leitaufgaben' in dem Jahresbericht 1913/14 illustrierte er sein Lehrkonzept. An deiser Stelle weist Kottenbach auf die Notwendigkeit der praktischen Handhabung des mathematischen Werkzeuges hin, die in dem bisherigen Unterricht allzusehr vernachlässigt wurde: 'Beweise in Arithmetik, Beweise in Geometrie bildeten den wesentlichen inhalt des mathematischen Unterrichts, ja sogar—horribile dictu—Prüfungsstoff bei Maturitätsprüfungen. Dass diese Beweissucht dem Geiste der Jugend nicht angemessen ist, dass sie der Jugend den Unterricht verleidet, wird niemand in Zweifel ziehen.... Den lebhaften, feurigen Sinn der Jugend schrekken Euklid'sche Beweisse wie das öde Einerlei der Sklavendienste, die der Mathematik von der Not des Lebens, von Technik, Schiffahrt, Astronomie und Geodäsie auferlegt werden.' "
[127] Thomas S. Kuhn interview of Werner Heisenberg, Session Two, February 7, 1963, Archive for History of Quantum Physics**,** Niels Bohr Library, the University of Minnesota, and other repositories**,** p. 2.

Thirring's works on general relativity. Pauli later cited all three physicists in his 1921 encyclopedia article on relativity.[128] Pauli was lucky to be studying relativity theory in Vienna: Bauer, Thirring, and Schödinger were doing research in it.[129] Pauli's preparation in relativity was sufficient for him to publish on it by 1918, before he left Vienna for Munich and Sommerfeld.

Mach also recommended Wilhelm Wirtinger to Pauli's father as Pauli's tutor in mathematics.[130] Considered the greatest mathematician in Austria, Wirtinger may well have opened Pauli's eyes to the beauty of mathematics, and to question Mach's positivism. Walter Moore, in describing Schrödinger's renunciation of Mach's philosophy, could also have been describing Pauli's:

> There are a number of fairly obvious defects in presentational phenomenalism [positivism]. For instance, it fails to explain the close relationship between mathematical reasoning and theoretical physics; mathematical operations and symbols do not denote empirical sensations, and yet one cannot do science without them. Also, experiments are planned interactions of the scientist with the environment; how can they be explained as mere collections of sensations? Mach fails to explain the enormous predictive power of physical theories; how can it be that [Paul A.M.] Dirac predicts a positive electron and [Carl David] Anderson finds it in a cloud chamber?[131]

Pauli would become fascinated with the deep meaning of mathematical symbols. His "mind's eye" was opened through his explorations of complex numbers, Maxwell's equations, and special-relativity transformations.

### Symmetry and Kernels

There is a common thread in Pauli's physics. After Pauli's death, Ralph de Laer Kronig and Victor Weisskopf noted that Pauli focused on invariants and symmetry in physics:

> The tendency towards invariant formulations of physical laws, initiated by
> Einstein, has become the style of theoretical physics of our days, upheld and

developed by Pauli during all his life by example, stimulation, and criticism. For
Pauli, the invariants in physics were the symbols of ultimate truth which must be
attained by penetrating through the accidental details of things. The search for
symmetry and general validity transcended the limits of physics in Pauli's work; it
penetrated his thinking and striving throughout all phases of his life, in all fields
of philosophy and psychology. Some of his writings dating from his last years
reflect his thoughts in this direction.[132]

Exactly when Pauli's interest in invariants and symmetry began to permeate his physics is
an open question. Heisenberg claimed that Pauli displayed this interest early in his life. I
conjecture that it arose during his formative years between 1913 to 1918, and then
became part of his personal philosophy.

Before I discuss Pauli's interest in invariants and symmetry, I must introduce some
necessary concepts. Pauli later adopted Jung's belief in mandalas in part as a visual
representation of the concepts of invariance and symmetry. Instead of referring to the
visual images that the early Pauli encountered in his physics as mandala images, I will
introduce the term "kernels" for them. To my knowledge, Pauli never used the term
"kernels" in this context, but he did use the term "mandala" after being introduced to
Jung's psychological philosophy. I propose that Pauli saw kernels between 1913 and
1930 rather than mandalas, and that these kernels then served as the basis for his adopting
Jung's concept of mandalas later. Only after being treated in the Jung Clinic did he feel
comfortable with the concept of mandalas to denote nonrational components in physics,
and with intuitive and aesthetic reasoning. I use the term "kernel" to remove the vitalistic
connotations of Jung's mandala concept; I mean by it the images of geometrically
circular entities that appear in mathematics and physics, and when grouped together form
a "conserved system" connected visually in the "mind's eye." Additionally, kernels often
group together internally four polar entities in a two-dimensional symmetry, with a
continuum of values between the polar extremes. Alternately, kernels may have more
than two dimensions, and more than four polar entities. A kernel, then, is devoid of
spiritual or vitalistic components and is a physical or mathematical grouping visualized as a symbol in the mind.  Kernels often produce a sense of awe, beauty, and mystery when a student first encounters them in mathematics or physics.  A kernel is a symbol of the entire group of intertwined entities that make up a physical system larger than the sum of its parts.  For example, the Bohr-Sommerfeld atom, or the Copernican solar system, might be viewed as kernels.  I will attempt to show that Pauli may have seen kernels in his studies of physics between 1913 and 1918, and that these kernels prepared him to later adopt Jung's mandala symbolism as part of his personal philosophy.

Examples of mandalas that Pauli saw in his later years, as they appear in his letters to Jung, may help to elucidate his earlier noticing of kernels.  There are several examples of Pauli's use of mandalas, which he alternately refers to as "quaternios."  In a letter of 1950, he draws four physical concepts on two axes in a four-part mandala, as shown below.[133]  This image has a holistic quality and an internal "symmetry-breaking" feature, as he showed later when he modified it in a letter of 1952, in which the "poles" of "energy" and "space-time" are broken up into three-dimensional momentum plus energy and three-dimensional space plus time.[134]

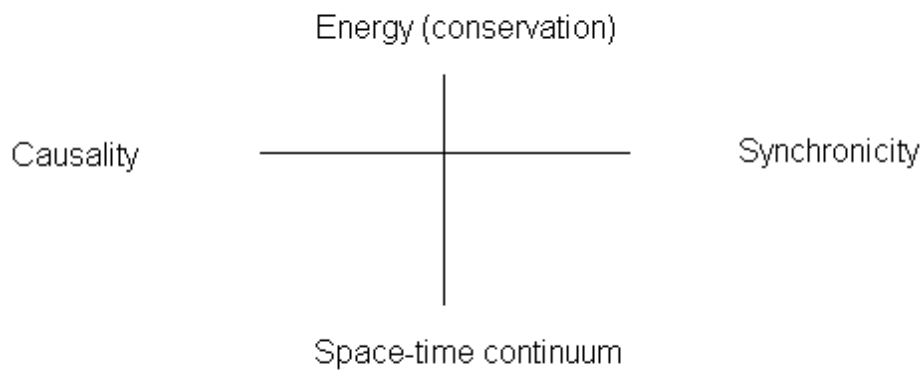

Figure 1. Pauli's "physics-concepts" mandala, from Pauli's letter to Jung of November 24, 1950.

---

[133] C.A. Meier, *Atom and Archetype: The Pauli/Jung Letters 1932-1958* (Princeton: Princeton University Press, 2001), p. 57. My redrawing.
[134] *Ibid.,* pp. 57, 80.  See Chapter 7 for this image.

In a letter of 1953 to Jung, Pauli discussed his relationship to physics and psychology:

> I can attempt to represent my relationship to physics and psychology through the quaternio [below] in which the people stand for mental attitudes and you [Jung], of course, represent your analytic psychology.[135]

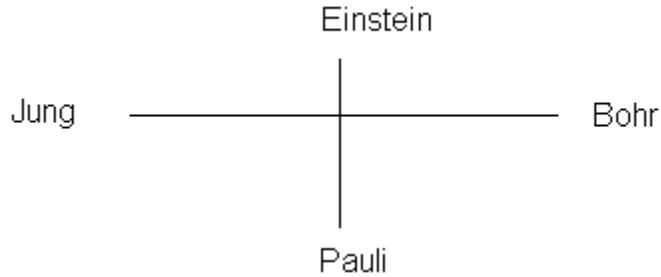

Figure 2. Pauli's "personality-type" mandala, from Pauli's letter to Jung of May 27, 1953.

Then, in a letter to Jung of 1956, Pauli discussed a dream in which opposite psychological symbols are arranged in quaternian fashion, with a "path" through the psychological space of the "mandala-of-countries" symbolism.[136]

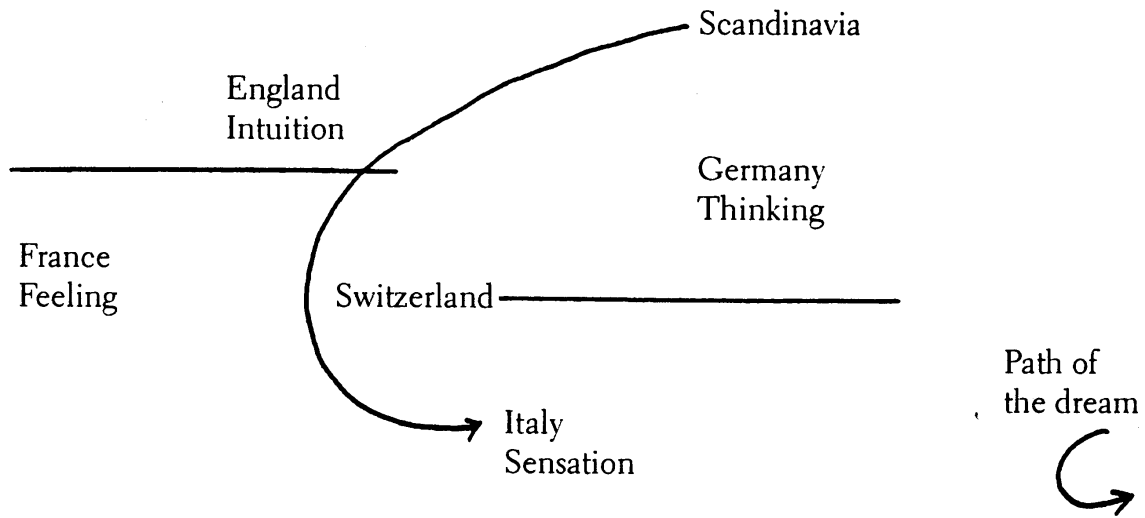

Figure 3. Pauli's "country-type" mandala, from Pauli's letter to Jung of October 23, 1956.

The later Pauli thus embedded all kinds of non-physical qualities into these mandalas, but their visual representation is key and must be remembered in what follows.

Thus, I now will identify similar examples as kernels in mathematical physics that Pauli may have encountered between 1913 and 1918, as follows: (1) the complex numbers $z = x + i\,y$ and $z = |z|\,e^{\,i\,\theta}$; (2) Maxwell's equations in electrodynamics; and (3) the Lorentz group in special relativity. In later chapters covering later periods in Pauli's life, I propose that Pauli saw kernels in other areas: (1) the mathematical quaternians used by Sommerfeld and Felix Klein to describe the motion of a top; (2) the angular momentum of the electron and its four quantum numbers; (3) the momentum and position variables in quantum systems, that is, the Heisenberg uncertainty relationship; (4) 2 x 2 Pauli spin matrices and perhaps the 4 x 4 Dirac matrix used to model the electron; and (5) the energy-general relativity, and charge-electromagnetism relationships that form a group in the theory of beta decay, that is to say, the conserved quantities that Pauli used to justify his neutrino hypothesis.

*Complex and Imaginary Numbers*

What may have stimulated Pauli intellectually and philosophically in his early mathematical education? One of the striking impressions many beginning students of mathematics receive concerns the close and surprizing relationship between trigonometry, imaginary numbers, and natural logarithms, which typically produce a feeling of awe and wonder, not unlike the Pythagorean religiosity that ancient Greeks encountered when probing the mysteries of geometry and number systems. Imaginary numbers were introduced in the Renaissance by Geronimo Cardano [Jerome Cardan] (1501-1576) in his influential work of 1545, *Ars magna*, which stimulated research on algebraic systems.[137]

Imaginary numbers and complex numbers bear names that conjure up images and connotations of mystery that reflect the amazement a student feels when first exposed to them.  They are abstract concepts, and elicited awe from their inception.  Gottfried Leibnitz (1646-1716), for example, remarked that imaginary numbers are a kind of amphibian, halfway between existence and nonexistence, resembling in this respect the Holy Ghost in Christian theology.[138]  Similar to the later Pauli, the theologian-mathematician Leibnitz was interested in Platonic-like elemental cognitive symbols as building blocks for a universal language.  Pauli had predecessors.

---

[137] Carl B. Boyer, *A History of Mathematics* (New York: John Wiley & Son, 1968), p. 310.

In introductory algebra courses, imaginary numbers are interpreted typically as related to rotations.  E.T. Bell noted in a popular essay:

> As an operation, multiplication by $i$ x $i$ has the same effect as multiplication by -1; multiplication by $i$ has the same effect as a rotation by a right angle, and these interpretations ... are consistent.

Bell goes on to note:

> Although the interpretation by means of rotations proves nothing, it may suggest that there is no occasion for anyone to muddle himself into a state of mystic wonderment over nothing about the grossly misnamed "imaginaries."[139]

I suggest that students who feel the mystery in $i$ are likely of the INFP personality type.  Pauli likely felt that same mystery when he first encountered this fascinating mathematical symbol.

The connection of imaginary numbers to the base of the natural logarithms $e$ and to $\pi$ has a long history, and also points to interest among mathematicians in the intellectual attraction that some call mysticism.  The relationship  $e^{i\pi} + 1 = 0$  was known to Leonhard Euler--to whom we largely owe the symbolism for the three important mathematical symbols $e, i,$ and $\pi$.[140]  In fact, the symbol $e$ was chosen to honor Euler.[141]  The mathematical connection that relates these symbols with their disparate histories and applications is striking to any student when first encountering them.  The chord that is struck in a student, whether it resonates as a mere example of a mathematical triviality or as a Platonic truth about the mathematical basis of reality, depends upon the personality of the student and the student's philosophical orientation. Pauli's naural personality type, INFP, was one in which deep connections and meanings were primary, in contrast to the utilitarian aspects of mathematical equations valued by some other personality types.

When forming a complex number as a real number plus an imaginary component, mathematicians discovered that a diagram could be constructed to illustrate it. A circle drawn in a two-dimensional plane with axes of real numbers and orthogonal imaginary numbers could be used to describe a complex number, as a vector from the origin to a point on the circle, as shown below.

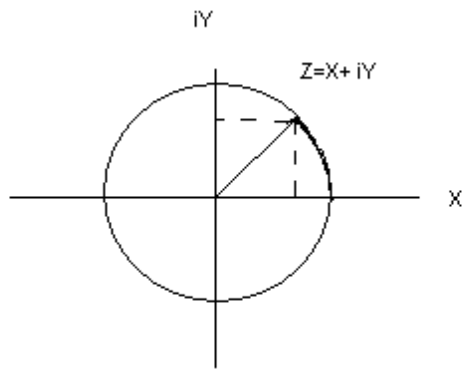

Figure 4. Author's drawing of the "kernel" symbol of a complex number.

Trigonometric functions combine to form an expression for $e$ and the complex number $Z = X + iY = | Z | e^{i\theta}$, where $\theta$ is the angle formed between the vector $Z$ and the $X$ axis. The circular visualization of the complex-number plane involves four mathematical symbols, $e$, $i$, 1, and $\pi$, and may have been the first mathematical "kernel" or mandala that Pauli recognized during his early education. The symbols $e$, 1, and $\pi$ are of similar real-number character, but $i$ introduces the need for an almost mystical perspective to appreciate it fully. And how does one comprehend the deep relationship between these seemingly disparate symbols and mathematical traditions? Are they a product of the human mind, or does the mathematical mind uncover an aspect of deep reality? The symbol for the imaginary number $i$, the square root of the real number -1, is related to $e$, the base of the natural logarithms, and to $\pi$. This striking visual connection between $e$, $i$, and trigonometric functions that include $\pi$ and represent rotations is mysterious. The circle has the image of a mandala, the mathematics is that of a four-part constituency, and the beauty is expressible by no means other than as a visually complete symbollism, visible only to the "mind's eye."  Pauli was exposed to this mandala-like image or kernel and to its mathematics early in his education.

Pauli's first exposure to imaginary numbers likely occurred during his early adolescence, as they are part of algebra, and he had learned calculus by the age of fourteen.[142]   Pauli had the benefit of a wonderful instructor of mathematics and physics at the Döbling Gymnasium, Rudolph Kottenbach.[143]   Imaginary and complex numbers also were topics of active mathematical research, something his mathematics tutor Wilhelm Wirtinger would have been inclined to communicate to him.  Not long before Pauli's birth, mathematicians succeeded in proving that $\pi$ was transcendental, and thus that it is impossible to "square the circle" using geometric methods.  In 1882, C.L.F. Lindemann in Munich proved that $\pi$ was not algebraic, putting an end to discussions of whether $\pi$ could be a root of a second-degree equation, a test that determined whether $\pi$ was algebraic.[144]   Thus, in Pauli's youth $\pi$ and imaginary numbers were still topics of special interest to both teachers and students, particularly in regard to their aesthetic features and even their mystical interrelationships.

That Pauli had a lasting impression of the imaginary number $i$ can be seen in comments that he made in his philosophical reflections after 1930.  He shared his comments with Jung and with his close friend, the Jungian psychologist Maria-Louise von Franz.  Von Franz noted Pauli's interest in the imaginary number $i$ in an exchange with Pauli after he had shared one of his dreams with her in which a woman possessed a ring with mathematical symbolic importance, the "ring $i$" of the imaginary number. Von Franz commented:

> All these questions he [the master?] does not pose.   So we have to ask them: is the "Ring $i$" a trap to catch the master or is the "Ring $i$" a vessel of understanding?  In quantum thoery it specifies a formula which includes the irrational in a symbol of totality, in a holistic "cosmogramm."  But the formula has a catch.   If one squares  $i = \sqrt{-1}$, although a negative, one obtains a rationally understandable negative number -1.   So one can make the irrational disappear through a slight of hand.   This formula does not correspond to reality because the irrational that we call the collective unconscious or the objective psyche can never be rational.   It remains always creatively spontaneous, not predictable, not manipulatable.  Each holistic formula is in that sense also a trap, because it brings about the illusion that one has understood the whole.[145]

The imaginary number $i$ was for Pauli a symbol of the door opened by the nonrational into new understanding, providing new perspectives that could not be proven using conventional rational methods.   There is a point beyond which one cannot go rationally, but it is nonetheless powerful to go beyond into the nonrational, to see inner connections in the abstract realm beyond the rational, similar to the door that opens up into the unconscious psyche of the mind.

Pauli was impressed with the visualizability of the relationships among $e$, $i$, and $\pi$, and he would write about his impressions in his later years in his philosophical speculations.  He likely first encountered this visualization, with its sense of wonder, in his teens when he was learning mathematics.  In an unpublished article of June 1948, he wrote of "modern examples of 'background physics',"[146] by which he meant visually symbolic representations of physical concepts, symbols of a psychologically objective nature and independent of the person viewing them.   These were the cognitive archetypes of Jung, and Pauli was exploring their appearance and role in physics.   One of the examples he gave in 1948 was that of the visualizable symbol generated by $e$, $i$, and indirectly $\pi$, in his self-analysis of one of his dreams.   He likely first encountered that same mathematical symbolism in his adolescence, and now in 1948, he was interpreting his awe-inspiring adolescent experience.   His mystical awe arose from the image of a circle, which he referred to as a mandala that was produced by the mathematical relationship of the symbols *e, i,* and $\pi$. Pauli had dreamed of an egg-shaped image that first split into two eggs, then into a third, and the third into two again to generate four eggs. In his dream the four eggs then changed into mathematical symbols of trigonometric functions. Pauli here was interpreting the numerological relationship of a kernel to a quaternity, where the number four indicated a process of splitting from one into four components:

> I [Pauli] say, "The whole thing gives $e^{i\delta}$, and that is the circle." The formula vanishes and a circle appears…. One becomes two, two becomes three, and out of the third comes the One as the fourth. The last mentioned typically comes about for me through mathematics. The formula
>
> $(\cos \delta/2 + i \sin \delta/2) \Big/ (\cos \delta/2 - i \sin \delta/2) = e^{i\delta}$
>
> is mathematically correct, and in the representation of complex numbers through distances $e^{i\delta}$ is a number that always lies on the "unit circle" (the circle with radius 1) ….
>
> The imaginary unit $i = \sqrt{-1}$ is a typical symbol since it is not continued under the ordinary numbers; the introduction of this symbol gives many mathematical theorems a simple and distinct form. In this dream it has the irrational function of uniting the pairs of opposites and thus producing wholeness.
>
> Without going into mathematical detail, I should nevertheless like to stress here that I cannot acknowledge an antithesis between a mathematical and a symbolic description of nature, since for me the mathematical representation is a symbolic description par excellence.[147]

---

[145] Marie-Louise von Frantz, "Reflexionen zum <<Ring $i$>>," in H. Atmanspacher, H. Primas, and E. Wertenschlag-Birkhäuser, ed., *Der Pauli-Jung Dialog und seine Bedeutung für die moderne Wissenschaft* (Berlin: Springer, 1995),  pp. 331-332.  My translation of the following passage: "All Diese Fragen stellt er nicht.  So müssen wir sie stellen: ist der Ring $i$ eine Falle, um den Meister zu fangen oder ist der Ring $i$ ein Gefäss des Verstehens?  Er stellt in der Quantentheorie eine Formel dar, welche das Irrationale in einem Ganzheitssymbol einbegreift, ein holistisches Kosmogramm.  Die Formel hat aber einen Haken.  Wenn man $i = \sqrt{-1}$ ins Quadrat erhebt, erhält man eine zwar negative, aber rational verstehbare Zahl (-1).  So kann man das Irrationale durch einen 'tour de passa-passe' zum Verschwinden bringen.  Die Formel entspricht in diesem Punkt nicht der Wirklichkeit, denn das Irrationale, das, was wir das kollektive Unbewüsste oder die objektive Psyche nennen, kann nie rational werden.  Es bleibt immer kreativ spontan, nicht voraussagbar und nicht manipulierbar.  Jede hölistische Formel ist in diesem Sinn auch eine Falle, weil sie die Illusion erweckt, man habe das Ganze verstanden."
[146] Meier, *Atom and Archetype* (ref. 17), p. 187.

Pauli was giving here an example of his orientation to Platonic ideals as an adult, of a deep reality being described by mathematics. Note the appearance of the four-part symmetry in the mathematical symbolism; the four parts equating to a whole or unit expression; and the appearance of the factor ½ . The factor ½ forms an important clue to Pauli's perspective on electron spin and spinors.[148]   He had learned mathematics early, was gifted in using it as a symbolic system for the communication and expression of

---

physical meaning, and he saw in mathematics a supreme system for the symbolic expression of cognitive patterns. Others might use sketches to assist visualization; Pauli saw the interconnected meanings of mathematical symbols in his "mind's eye." Mathematics and physics were his media to express his philosophical and artistic interests. Regardless of his affinity for the aesthetics of mathematics, similar to the aesthetics of fine music, the young Pauli struggled against the message from his father and from Mach to view mathematics more as a tool to be used than as something to be enjoyed in its own right. Both Pauli's father and Mach may have appreciated the aesthetics of mathematics, but their message to the young and impressionable Pauli was to use mathematics to evaluate facts and measurements, that mathematics serves as a shorthand to describe the external world of physics, rather than to be valued for the pure joy that mathematics gives as a glimpse of deep reality.

*Maxwell's Equations*

A second example of a kernel and its associated intuitive visualization that Pauli may have experienced during his Döbling Gymnasium days concerns the mathematics of electromagnetism. Maxwell's four equations of electromagnetism constituted an almost complete mathematical description of it, almost complete because as Einstein recognized later the lack of symmetry in the equations between a moving magnetic field and a moving electric field was problematic. By introducing it Einstein was led in part to his theory of special relativity. Pauli's attraction to relativity theory could have developed on the basis of firm knowledge of Maxwell's equations and electrodynamics.

The connection that Maxwell's equations provide between electicity and magnetism to explain optical phenomena points to a deep reality in the physical world. Maxwell had seen that the equality of the empirical value of the speed of light and the theoretical value derived from his equations indicated that light consists of electromagnetic waves. Theory thus was more than a mere shorthand to summarize empirical knowledge; theory revealed new and unexpected connections between phenomena. A three-dimensional sketch of an electromagnetic wave is an example of a visual kernel that might have impressed the young Pauli. Electric and magnetic fields vary in sinusoidal fashion, and in one's "mind's eye" a circular kernel forms, traveling forward while the electric vector **E** and magnetic vector **B** oscillate transversally, forming a pulsating, circular whole as shown in the figure below taken from a standard physics textbook.[149]

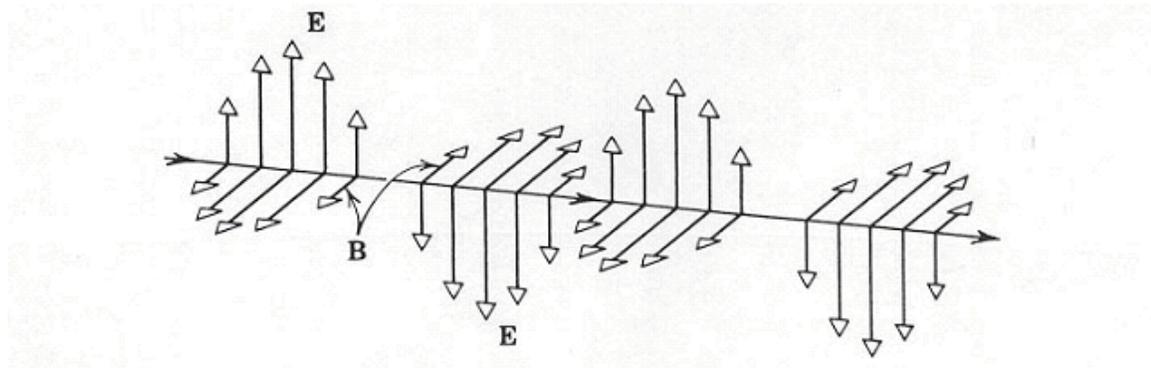

Figure 5. Electromagnetic "kernel" from a popular physics textbook showing a plane-polarized wave with orthogonal **E** and **B** vectors moving to the right at the speed of light.

Pauli's encounter with Maxwell's theory of electromagnetism, in addition to acquainting him with the above kernel, also may have been his first encounter with the need for another whole, the quantum of electric charge. Remember that during his Gynmnasium education Mach was resisting the concept of the atom, and hence that of the electron, one of its building blocks. In the 1940s Pauli would lecture his own physics students:

> It is … by no means true that field physics has triumphed over corpuscular physics, as can be seen from the fact that electricity is atomistic in nature…. There is no explanation for the fact that only integral multiples of a certain charge occur. The existence of an elementary charge has, until now, in no way been made plausible. It is an open problem in theoretical physics. The electron itself is a stranger in the Maxwell-Lorentz theory as well as in present-day quantum theory.[150]

---

[149] David Halliday and Robert Resnick, *Physics for Students of Science and Engineering,* Part II (New York: John Wiley & Sons, 1962), p. 1055.
[150] Wolfgang Pauli, in C. Enz, ed., *Pauli Lectures on Physics.* Vol. 1. *Electrodynamics* (Mineola, N.Y.: Dover, 1973), p. 2.

*Special relativity*

In his encyclopedia article of 1921 on the theory of relativity, Pauli illustrated how the Lorentz transformation could be visualized as a rotation in an abstract mathematical space, as shown in the figure below.[151] The space-time distance between two four-dimensional points, *x, y, z, ict* and *x', y', z', ict'* is invariant, giving Pauli in his "mind's eye" another opportunity to visualize a kernel, with the transformation between coordinate systems perceived visually as a rotation. Pauli adapted the above figure from Hermann Minkowski's famous lecture of September 21, 1908,[152] which Pauli would have encountered during his Gymnasium education, since he was completely familiar with relativity theory before he began his studies with Arnold Sommerfeld in Munich in 1918.[153] The above image of 1921 reminds one of a kernel, in which the space-time distance has four components and is invariant under a rotational transformation. Some components within the circle change, but not the underlying measure or radius of the circle. The fourth component, moreover, involves the imaginary number *i*, an extension in mathematical complexity from the other three components.  If Pauli had mystical stirrings, they certainly could have occurred here.

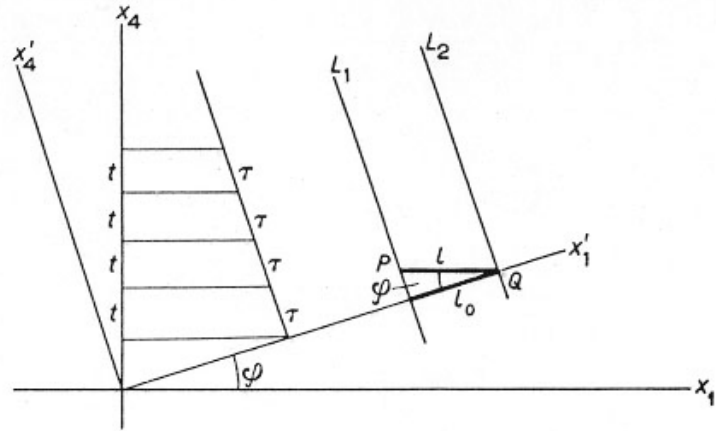

Figure 6. Pauli's illustration of a Lorentz transformation.

Pauli was learning about the mystical aesthetics of mathematical physics that point to deeper levels of reality.  Because of his complex adolescent development, he had no comfortable personal philosophy into which to place these thoughts.  Was he Catholic, Jewish, atheist, positivist, or Platonist?  He probably fit into none of these categories at the time.  Whether or not these kernels, imaginary numbers, wholes greater than the sum of their parts, awareness of underlying deep reality, and the like were entering his consciousness or instead were going into his Shadow, it is worthwhile to summarize these philosophical challenges for the young Pauli.

## Pauli and Jung's Mandalas

In Carl Jung's psychological philosophy, everything is metaphor and symbol: Time and logic are abstractions; dreams and memories are as valid as conscious thoughts; their meanings and relationships are where one finds them.  The adult Pauli would have found prescient meaning in his Jewish grandparent's home address in Prague: Number 7 on the Altstädter Ring.  The Old Town "Ring" and the number 7 would live in Pauli's Shadow. As Pauli sought his religious and philosophical grounding as an adolescent, he would find his intuitions to be in conflict with his reason.  The appearance of kernels in his studies of mathematics and physics would be transformed later into mandalas.  Dynamical systems of four parts, such as complex numbers, electromagnetic fields, and Minkowski space-time, would indicate a "quaternian" way of perceiving the world.  Mathematical group transformations would be visualized as rotations of a mandala.  Even his own personality would be transformed from being outwardly focused to inwardly consumed, from believing that rationality and the senses were supreme to equally valuing feelings and intuitions.

The topics of a philosophical nature that I believe first appeared in Pauli's adolescence need further identification.  I see them surfacing in his teens as he became aware of kernels, quaternities, rotations, and deep reality.  The kernels of Pauli's adolescence later would become the mandalas of Jung.  One might think of kernels as being inert physical wholes larger than the sum of their parts, implying a unity of mathematics and physics, the "all is number" of Pythagoras. Kernels can be inferred most easily through visualization, with their indicator the circle expressing unity. Pauli's kernels came in four-part divisions. Another indicator of the presence of kernels was visualization of rotation of these four components.

The four mathematical symbols *e, i, π,* and 1 comprise the components of the kernel that formed the unit circle in the complex-number plane. The electric and magnetic fields of the four Maxwell equations comprised the kernel that could be visualized as a traveling electromagnetic wave. The four space-time dimensions comprised the kernel found in Minkowski's view of special relativity, and could be visualized in the Lorentz transformations as rotations. Whenever Pauli encountered rings, circles, wholes dissectible into four parts, and spheres, he became intuitively drawn to them.

By the time Pauli met Jung, Pauli was ready for Jung's concept of mandalas. Jung's mandalas extended the idea of kernels for Pauli by including animate conscious components. Jung uses the Sanskrit word *mandala* as a symbolic or "magic circle"[154] that captures in the "mind's eye" the psychic center of a transformational process**.** This idea defies precise definition since the person seeing the mandala image in his or her "mind's eye" is supposed to experience a mystic feeling when viewing the symbol, a feeling going beyond a reductionist attempt at verbal definition. Jung's mandalas also generally have four components. Thus, the mandala is the "One" whole that is larger than the sum of its four parts.

The four components of most kernels that Pauli saw in his physics later developed into his numerological attraction to the number 4. In his youth, however, I think Pauli was struggling with positivist messages that undermined his trust in his intuition. The number 3, likely derived from his Catholic upbringing and attachment to the Trinity, was another numerological sign to which the young Pauli was sensitive. The number 3 represented an opposition to the number 4; either three or four components existed in a physical system, but not both. The three components of Cartesian space were completely rational and perceptible by the senses. The fourth component in Minkowski space-time, *ict*, involved a nonrational extension that was not perceptible by the senses; it extended the three to form a unity of four components. Thus, to the youthful Pauli, numerology of

---

the number 4 clashed with his positivist upbringing and rationality. Four components were not subject to proof; instead they had to be accepted on the basis of intuition, symmetry, and aesthetics. He was not ready to accept the numerology of the number 4 until after he met Jung.

Pauli's mature philosophy embraced faith in a deep reality and was a product of his finding symmetry in kernels and mandalas. As an adult he referred to this philosophy as *Hintergrundsphysik* (background physics) reflecting a combination of quantum-mechanical phenomena and psychological phenomena.[155] He never refined his philosophy into a cogent system. There is a convenient but rarely used term to describe faith in a deep reality, one that seems to fit Pauli's mature philosophical ruminations: Christopher Norris's term "alethic reality." By "alethic reality," Norris means a reality that is objective and truth-based but transcends verification.[156] Norris's term "alethic" seems to be absent in dictionaries, but I will use it as he does.[157] Alethic reality is deep down, but unattainable by rational means.

The meaning of "alethic reality" is close to that of Pauli's later *Hintergrundsphysik*. Arthur Koestler, a great admirer of Pauli and a "scientific mystic" in his own right, has described what he calls "reality of the third order," which corresponds to "alethic reality":

> The "hours by the window" [Koestler was imprisoned during the Spanish Civil War] … had filled me with a direct certainty that a higher order of reality existed, and that it alone invested existence with meaning. I came to call it later on "the reality of the third order." The narrow world of sensory experience constituted the first order; this perceptual world was enveloped by the conceptual world which contained phenomena not directly perceivable, such as gravitation, electro-magnetic fields, and curved space. The second order of reality filled in the gaps and made sense of the sensory world. In the same manner, the third order of reality enveloped, interpenetrated and gave meaning to the second. It contained "occult" phenomena which could not be apprehended or explained either on the sensory or on the conceptual level, and yet occasionally invaded them like spiritual meteors piercing the primitive's vaulted sky.[158]

---

[155] Meier, *Atom and Archetype* (ref. 17), p. 179.

[156] Christopher Norris, *Quantum Theory and the Flight from Realism* (New York: Routledge, 2000), p. 4.

[157] The *Oxford English Dictionary* defines only the word "alethiology" as "The doctrine of truth, that part of logic which treats of truth." It cites a passage of 1827 from the Scottish philosopher Sir William Hamilton (1788-1856): "The first part [of logic] treats of the nature of truth and error, and of the highest level of their discrimination, Alethiology." I thank Dr. Stuewer for pointing this out to me.

Or, as Koestler wrote later:

> Just as one could not feel the pull of a magnet with one's skin, so one could not hope to grasp in cognate terms the nature of ultimate reality. It was a text written in invisible ink; and though one could not read it, the knowledge that it existed was sufficient to alter the texture of one's existence.[159]

Pauli's mature comments on the nature of quantum reality always supported the Copenhagen interpretation of quantum mechanics, yet paradoxically it appears that he also believed in a reality beyond the empirical. His was alethic reality, a reality beyond quantum mechanics and human understanding. The physical kernels that Pauli first saw as an adolescent affirmed his belief in an alethic reality, but no rational means existed to prove it. Pauli's use of mandala visualizations appeared in his use of symmetry methods and group theory. For the adult Pauli, his alethic reality was a blending of the "mind stuff," Jung's psychological archetypes, with the inert physics behind the quantum. Early in his mathematical education, Pauli discovered the power of mathematical symbolism, how that symbolism points to a Platonic alethic structure of reality, how that reality was often superior to pictures or words, and how certain mathematical symbols have special significance. Pauli saw in the mathematics of complex variables, electrodynamics, and relativity stage settings for alethic reality.

The adolescent Pauli was brilliant in mathematics and physics, but he lacked the self-confidence that is necessary to excel in them. It required his mentor Arnold Sommerfeld in Munich to awaken Pauli's intuition, enabling him to value numerology as a tool in theoretical physics. In Sommerfeld Pauli also found a father figure that helped him to individuate from his parents, to become his own man in physics and in the world. Pauli was, however, a poor learner. He was confident that his talents would allow him to rationally extract the core of physics; he was arrogant with others who were not like-minded; and he completely lacked confidence in his intuitions that still lacked the rigor of logic. Sommerfeld seems to have filled Pauli's Shadow with the value of intuition, mysticism, and creativity, but Pauli's consciousness was still too strong for their fruitfulness to emerge.

---

[158] John Beloff, "Koestler's Philosophy of Mind," in Harold Harriss, ed., *Astride the Two Cultures: Arthur Koestler at 70* (New York: Random House, 1976), p. 82.
[159] Arthur Koestler quoted in Harriss, *Astride the Two Cultures* (ref. 42), p. ix.

## Chapter 4. Pauli's University Education in Munich, 1918-1921

**Pauli in Munich**

Pauli's personal life during his years at the University of Munich seems to have been free of crises: The war had ended; he was away from his parents and enjoying his independence; he was gaining recognition for his brilliance in mathematics and physics; he thrived under his mentor, Arnold Sommerfeld; and he found a lifelong friend in Werner Heisenberg. If one digs deeper, however, there are several issues that he may have struggled with emotionally as a late adolescent that led to his later psychological crises with their impact on his physics. Germany lost the Great War, which had devastating consequences for the political, economic, and social life in Germany and Austria, which no doubt affected Pauli's family. There also was great unrest in postwar Munich when Pauli was there;[160] Pauli no doubt lost friends and classmates in the war; anti-Semitism was on the rise, which probably upset Pauli since he had recently become aware of his Jewish heritage; Pauli's parents separated later, so his home environment may have been stressful much earlier; Pauli's parents, especially his father, seem to have been pushing Pauli to succeed, which may have continued to affect him in Munich; and Pauli in his late adolescence was likely struggling to develop his social skills and sexuality. During his university years in Munich, Pauli would begin his avid nightlife, going out on the town until the late hours, then working intensely for hours on physics, sleeping late in the morning, and missing classes. Considering Pauli's sensitivity as a young man, he may have had good personal and psychological reasons for leaving Vienna to bury his thoughts in physics in Munich. Pauli's Shadow likely was continuing to enlarge in Vienna and did not begin to empty until he met Arnold Sommerfeld in Munich.

**Sommerfeld and the Bohr-Sommerfeld Atom**

---

[160] Charles Enz, *No Time to be Brief: A Scientific Biography of Wolfgang Pauli* (New York: Oxford University Press, 2002), pp. 49-52.

Arnold Sommerfeld was a teacher *par excellence*. He stimulated many of his students to become renowned physicists, some winning Nobel Prizes, and to carry on his tradition in teaching. In his seminars, he would engage his students in current problems in theoretical physics, and ask some to help in revising his famous bible of spectroscopy, *Atombau und Spectrallinien*, after the publication of its first edition in 1919. He introduced his students to other prominent physicists, helping them gain self-confidence. He was a caring man; many of his students loved him dearly. Pauli directed his scathing criticism and biting sarcasm to all physicists save one: Sommerfeld. Not even Einstein escaped Pauli's biting tongue, as when he declared after one of Einstein's lectures: "You know, what Mr. Einstein said is not so stupid."[161]

Sommerfeld once remarked that he had little to teach the young Pauli, but in my view Sommerfeld was the person most instrumental in developing Pauli's skills as a theoretical physicist. Sommerfeld conveyed to Pauli, through his erudition and by example, new ways of looking at physics, new ways of balancing intuitive aesthetic beauty and empirical facts. Sommerfeld opened Pauli's "mind's eye." Some of the areas in which Sommerfeld assisted his student, which I will discuss below, were to expose Pauli to the power and limitations of the Bohr-Sommerfeld atomic model; to Johannes Kepler's style of creative scientific reasoning; to the mathematical methods of quaternians and matrices; to the mathematical limitations of continuous functions in relativity theory and the need for a new quantum physics; and, most importantly, to impress upon Pauli the need to follow his intuitions.

Pauli decided to study theoretical physics under Sommerfeld at the University of Munich not without some trepidation. Getting a job was of concern to both students and parents during those difficult postwar years. There also was conflict between experimental and theoretical physicists in Munich, as Heisenberg reported:

> My father knew about some troubles within the University between experimental physics and theoretical physics. Here in Munich, there was really a rather difficult situation between Willy Wien, who was an experimental physicist, and Sommerfeld. Now, this difference between the two men was, perhaps, a political difference. I would say Willy Wien was very much on the right side of politics, and Sommerfeld on the left side, if these terms left and right mean anything. Besides that Willy Wien considered experimental physics as the center of physics, and in some ways he disliked theoretical physics. He still, of course, had been a theoretical physicist himself … you know Wien's law. But certainly one would say that he disliked every physics which was not as clear as classical physics, so quantum conditions and that sort of thing, that he considered as a kind of weak talk which meant almost nothing. And I do remember that Willy Wien gave, when he was rector of the University, a speech about atomic physics and never mentioned the name of Sommerfeld. Everyone felt that that is a thing that one can't do because Sommerfeld after all was a very famous and certainly a very good physicist. So there was lots of trouble between the two, and my father was worried because he saw this trouble and saw that men like Wien just disliked theoretical physics as a kind of subject. So he felt, "Should my son go into a line which is still so much under suspicion among the physicists at the universities."…

He [Wien] was against the non-classical nature. It was this thing which people criticized as Sommerfeld's mysticism. I mean, you know he [Sommerfeld] was enthusiatic for having integral numbers and that sort of thing. He knew, of course, quite well that quantum thoery was not consistent. I mean, as he did classical physics he was a very good physicist and has written a number of excellent papers which are completely clear. On the other hand, he had an enormous instinct, or intuition for how physics really is. Therefore, he didn't mind contradictions when he knew, "Well, finally this must be so." So I was always greatly impressed by this ability of Sommerfeld's to see quite early what are the important problems and how they will finally be solved….

So when Bohr's paper[s] had appeared in 1918 and 1920, and so on, he knew that Bohr was right about the system of the elements, in spite of the fact that nothing came out of the mathematics. I mean, it was all intuition from Bohr, but Sommerfeld at once could see, "Well, that is the right way to go." And Wien disliked this tremendously because he felt, "Well, only people who have that kind of intuition can take part in the game. And the people who haven't, well, they just can't help it, that's it." I would say that was one of the reasons … that you could not use the standard methods of doing things, you could not use your classical way of calculating things. It was such a funny situation; you always had to work in a kind of fog of uncertain knowledge, and so on. And Sommerfeld liked it; he felt "Now I see how things are connected, and that's enough." Well, the two personalities, Wien and Sommefeld, were very different in this way, and therefore, it was quite natural that they wouldn't understand each other.[162]

In Sommerfeld and Wien, we see the INFP theoretician and the ESTJ experimentalist personalities in conflict.[163] Pauli, the forced ESTJ personality, was being attracted to Sommerfeld's INFP physics and was beginning to see his own INFP Shadow. To Pauli, Bohr and Sommerfeld represented the epitome of physics: Their names appear repeatedly in Pauli's later writings, accompanied by expressions of reverence and awe. Pauli would always greet Sommerfeld as "Herr Professor!" and Pauli would ask to see Bohr one last time on his deathbed.[164]

Einstein called Bohr's theory of the atom of 1913 the "highest musicality of human thought," and Sommerfeld extended it in 1916 by introducing the so-called Sommerfeld quantization conditions. The Bohr-Sommerfeld atom, so named because of Sommerfeld's theoretical extensions of Bohr's musicality, was much more of a visual than physical model. One of its striking features was its visually aesthetic appeal. It was a visually beautiful model filled with classical electron orbits, as seen in the figure below.[165] The young Pauli criticized it sharply for that very reason: There was no firm theoretical basis on which to calculate these classical orbits.

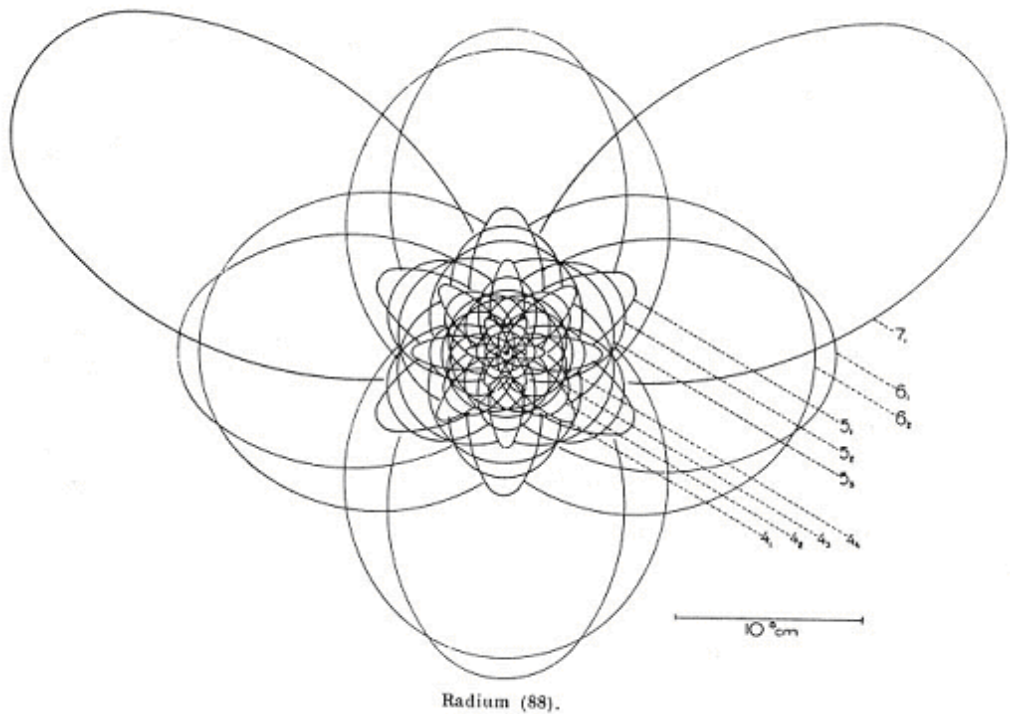

Radium (88).

Figure 7. A Bohr-Sommerfeld atomic "kernel" for the radium atom as depicted in a popular book of the period.

In Pauli's later years, however, he was not such a harsh critic of such models: He then saw mandalas in them and the power of their explanatory metaphysics.  Thus, for example, in a  paper of 1954 on Johannes Rydberg and the periodic system of the elements, he included the following drawing by Rydberg:[166]

[166] *Ibid.*, p. 77.

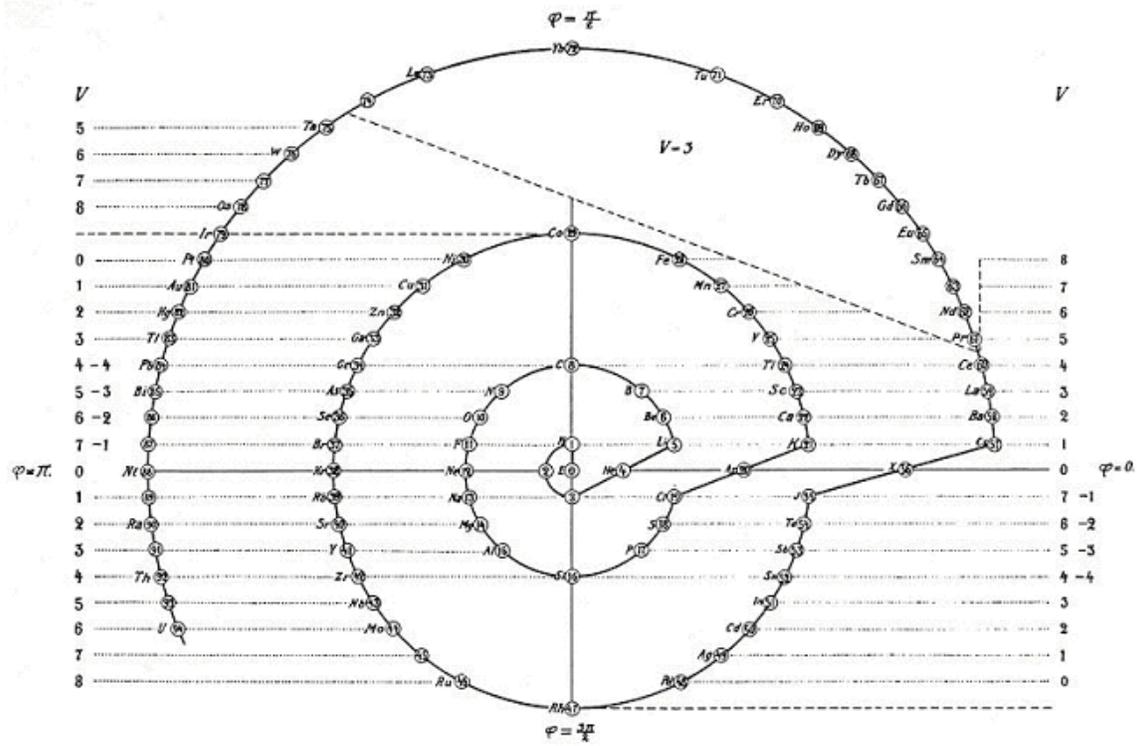

Figure 8. Rydberg's "mandala."

The Bohr-Sommerfeld model of the atom is reminiscent of Johannes Kepler's elliptical planetary orbits: Both were refinements of a mandala-like circular model. Kepler's model was a refinement of Copernicus's model, which had replaced Ptolemy's owing in part to its greater aesthetic appeal.[167] A circle was more pleasing aesthetically to the "mind's eye" than Ptolemy's complex epicycles, deferents, and equant points. However, one's "mind's eye" had to be receptive to such a mandala-like circular figure. Pauli's father and Willy Wien were unlikely to have been, and Pauli's ESTJ side of his personality similarly resisted. Sommerfeld, by contrast, appreciated such visually intuitive models, and likely revealed the value of such imagery to Pauli's INFP Shadow, by revealing the value of combining such visual intuitions with hard theoretical and mathematical logic. Sommerfeld's ability to combine intuition with logical reasoning enamored him to his student Pauli.

In the early 1920s, as it became apparent that the Bohr-Sommerfeld model of the atom was inadequate, Pauli advocated the abandonment of such intuitive and visualizable models, and their replacement by mathematical models based upon solid physical principles.  Pauli thus is often seen as one who did not value visual knowledge, but instead valued most highly knowledge that was based upon concrete philosophical and physical principles.  Instead of intuitive, visual knowledge supported only by one's personal sense of aesthetics, the basis of physical theory should be rational knowledge based upon empirical facts.  On closer inspection of Pauli's personal philosophy, however, that conclusion is supportable only during his early years.  He changed later. He  continued to place a high value on his rational side, but he supplemented it with intuitive, visual knowledge that he often found difficult to describe in words.  He struggled to appreciate intuitive knowledge.

---

[167] Pauli discussed Kepler's aesthetics in 1952. See Meier and von Meyenn, *Wolfgang Pauli* (ref. 6), p. 225.

That Pauli valued visual knowledge internally while denying it publicly, can be seen by considering his treatment at the Jung Clinic.  Jung opens his text on psychology and alchemy, the volume in which he includes comments on Pauli's treatment, by explaining why he added many illustrations to his text:

> The wealth of illustrations ... is justified by the fact that symbolical images belong to the very essence of the alchemist's mentality.  What the written word could express only imperfectly, or not at all, the alchemist compressed into his images; and strange as these are, they often speak a more intelligible language than is found in his clumsy philosophical concepts.  Between such images and those spontaneously produced by patients undergoing psychological treatment there is, for the expert, a striking similarity both in form and in content.[168]

Pauli supported Jung's views, because Jung's analysis of Pauli's dreams was the basis for Pauli's treatment and improvement, which required his endorsement as a patient.

During Pauli's treatment, as documented in Jung's text, Pauli had a dream of a man with a pointed beard, on which Jung commented.  The dream imagery stimulated associations for Pauli that helped to root out some of the psychological issues that were troubling him.  As Jung explained above, such imagery was far more powerful than could be described in words, and in Pauli's dreams it produced difficult emotional  associations. Thus, by rooting them out from Pauli's unconscious, the source of his dreams, Pauli's

---

[168] Carl Jung, *Psychology and Alchemy* (London: Routledge & Kegan Paul, 1953), p. vii.

consciousness had a chance to address these issues.  Jung analyzed one of Pauli's dreams as follows:

> The "man with the pointed beard" is our time-honoured Mephisto whom Faust "employed" and who was not permitted to triumph over him in the end, despite the fact that Faust had dared to descend into the dark chaos of the historic psyche and to steep himself in the everchanging, seamy side of life that rose up out of that bubbling cauldron.
>
> From subsequent questions it was discovered that the dreamer himself had recognized the figure of Mephistopheles in the "man with the pointed beard" .... [The] man with the pointed beard represents the intellect, which is introduced by the dream as a real *familiaris*, an obliging if somewhat dangerous spirit.  The intellect is thus degraded from the supreme position it once occupied and is put in the second rank, and at the same time branded as a daemonic.  It had always been daemonic--it was only that the dreamer had not noticed before how possessed he was by the intellect--the tacitly recognized supreme power.  Now he has a chance to view this function, which has so far been the uncontested dominant in his psychic life, at somewhat closer quarters.   Well might he exclaim with Faust: "So that was the poodle's kernel!"  Mephistopheles is the diabolical aspect of every psychic function that has broken loose from the hierarchy of the total psyche and now enjoys independence and absolute power....  But this aspect can only be perceived when the function becomes a separate entity and is objectivated or personified, as in this dream.[169]

To appreciate Jung's analysis, we need only note Pauli's fascination with Mephistopheles. In the famous play in Bohr's institute in Copenhagen, Pauli was cast in the role of Mephistopheles.  "Das ist der Pudels Kern" remained one of Pauli's favorite expressions, as Heisenberg recalled in his 1963 recollections of his friend.

In 1948 Pauli commented on visual knowledge and the superiority of emotional attachments to visual symbols as follows:

> Not only alchemy but also the heliocentric idea furnishes instructive examples of the problem as to how the process of knowing is connected with the religious experience of transmutation undergone by him who acquires knowledge (*Wandlungserlebnis des Erkennenden*); it transcends natural science and can be apprehended only through symbols, which both express the emotional, feeling aspect of the experience and stand in vital relationship to the sum total of contemporary knowledge and the actual process of cognition. Precisely because in our times the possibility of such symbolism has become an alien idea, it may be considered especially interesting to examine another age to which the concepts of what is now called classical scientific mechanics were foreign, but which permits us to prove the existence of symbols that had simultaneously a religious and scientific function.[170]

---
[169] *Ibid.*, pp. 66-67.

Sommerfeld, to extend Bohr's model of the atom mathematically, used Lagrangian and Hamiltonian methods in the "Kepler problem." He introduced the so-called Sommerfeld quantization conditions that singled out allowed electronic orbits as determined by integer multiples of Planck's constant $h$. Sommerfeld treated the electron's motion relativistically and, to simplify the mathematics, formed the "fine-structure constant" that governs the fine-structure splitting of the hydrogen atom's spectral lines. Just as in Kepler's refinement of Copernicus's planetary orbits, Sommerfeld worked intuitively toward a deeper understanding, expecting that his theoretical model would be vindicated physically and empirically. Enter Pauli. The limitations of the Bohr-Sommerfeld model now were apparent, and its physical details required exploration. He would have to learn Sommerfeld's techniques to gain deeper insights. Sommerfeld would introduce Pauli to Kepler, and Pauli's silent INFP side would listen.

Both Pauli and Sommerfeld were attracted to Kepler's Platonism, but Pauli had to struggle to accept it, and then to exploit it. Behind the integers and numerical coincidences were keys to the mysteries of Nature. All that was required was to use mathematics to create a model that emulated alethic reality. This mathematical model might be far removed from the familiar verities of the senses; it might deal only with a level of alethic reality. As long as the alethic mathematical model could be checked by observation and measurement, these could be taken as supreme tests of the model's validity. The model if it was aesthetically beautiful, if it had connections to the mind of God, and if it was consistent with empirical results, then it would mimick the underlying alethic reality. Pauli resisted this approach. Kepler, however, had taken his Platonic solids and harmony of the world as proof that God was a geometer, and revealed Himself to the human mind through mathematics. Sommerfeld expressed his veiled admiration of Kepler in the introduction to his celebrated *Atombau und Spectrallinien* of 1919, which Pauli would later quote when he discussed Sommerfeld's contributions to quantum theory:

> What we are nowadays hearing of the language of spectra is a true music of the spheres within the atom, chords of integral relationships, an order and harmony that becomes ever more perfect in spite of the manifold variety. The theory of spectral lines will bear the name of *Bohr* for all time. But yet another name will be permanently associated with it, that of *Planck*. All integral laws of spectral lines and of atomic theory spring originally from the quantum theory. It is the mysterious *organon* on which Nature plays her music of the spectra, and according to the rhythm of which she regulates the structure of the atoms and nuclei.[171]

Sommerfeld's student Pauli was beginning to become receptive to Kepler and Platonism, and although most of his philosophical writings appeared during his "philosophical period" long after he had met Carl Jung, Pauli's philosophical interests according to Heisenberg were present much earlier and emerged privately. His interest in Kepler and Platonism thus was likely solidified during his years with Sommerfeld, since Sommerfeld was similarly inclined and would show Pauli their relevance to theoretical physics.

In 1955 Pauli wrote about the numerological meaning of the number 4, essentially calling for a reconsideration of Platonism:

> For [Pythagoras] ... wherever number is, there also is soul, the expression of the unity which is God. Whole-number relationships, as they occur in the proportions of the frequencies of the simple musical intervals, are harmony, that is to say they are what brings unity into contrasts. As part of mathematics number also belongs to an abstract supersensuous eternal world which can be apprehended not by the senses but only in contemplation by the intellect. Thus for the Pythagoreans mathematics and contemplative meditation (the original meaning of "theoria") are very closely connected; for them mathematical knowledge and wisdom (sophia) are inseparable. Special significance was attached to the *tetraktys*, fourfoldedness, and there is a traditional oath of the Pythagoreans: "by him who has committed to our soul the *tetraktys*, original source and the root of eternal Nature"….[172]

We have here the fundamental explanation of Pauli's attraction to the number 4 and the four components of his physical kernels. Pauli was attracted to Kepler in part because he was a bridge in the history of ideas between "Trinitarian" and "quaternian" thinking. Pauli's analysis of the Kepler-Fludd polemic in the 1940s is connected to this issue. For Pauli, the Kepler-Fludd polemic mirrored the debates over the philosophical interpretation of quantum mechanics:

Modern quantum physics has come closer to the quaternary point of view, which was so violently opposed to the natural science that was germinating in the 17th century, to the extent that it takes into greater consideration the role of the observer in physics than is the case in classical physics.[173]

**Sommerfeld's Fine-Structure Constant and Numerology**

Pauli's numerological Platonism focused on the numbers 2, 4, and 137. His focus on the number 2 may have stemmed from his Catholic upbringing where "two-valuedness" was an indicator of doubt, and of the Devil. His focus on the number 4 began to emerge in his physical kernels as an indicator of symmetry, but he would not accept its significance until after his discovery of the exclusion principle. He first learned about the number 137 from Sommerfeld, and he would try to decipher its significance until the day he died.

Sommefeld introduced the dimensionless fine-structure constant $\alpha$, numerically equal to about 1/137, in his extension of Bohr's theory of the hydrogen atom where it governed the splitting of its spectral lines. [174] It is comprised most importantly of the four fundamental physical "quanta" of the charge of the electron $e$, the speed of light $c$, Planck's constant $h$, and the Pythagorean symbol of mysticism $\pi$. During the course of his life, the fine-structure constant appealed to Pauli as a symbolic integer number, a theoretical coincidence, a kernel with quaternity components, and as a symbol of mystical foreboding. It had captured his attention already by the time he published his encyclopedia article on relativity theory in 1921. The visualizable kernel formed by the quaternity of $e, c, h,$ and $\pi,$ however, eluded Pauli. He never indicated that he had gleaned an image of it, nor did he decipher the mystery of the interconnectedness that he associated with the number 137.

In 1956 Pauli commented on his thoughts on the fine-structure constant, as given in his *Theory of Relativity* of 1921:

> The reader of the original text ... will see that I was already at that time very doubtful regarding the possibility of explaining the atomism of matter, and particularly of electric charge, with the help of classical concepts of continuous fields alone. In this connection it should be remembered that the atomicity of electric charge had already found its expression in the specific numerical value of the fine structure constant, a theoretical understanding of which is still missing today. Particularly, I felt rather strongly the fundamental character of the duality (or, as one says since 1927, the complementarity) between the measured field and the test body used as a measuring instrument.[175]

We see here the beginning of Pauli's fascination with the number 137, which he owed to his teacher Sommerfeld. Sommerfeld looked for integer numbers to pop out of complex physical equations, which would then indicate causal mechanisms and deep relationships, as did, for example, Balmer's formula for spectral lines of hydrogen. In Sommerfeld's extension of the Bohr atom, the electron moves relativistically about the nucleus, which requires the introduction of a second quantum number that accounts for the splitting of the spectral lines. The amount of this splitting is governed by the fine-structure constant

$$\alpha = 2\pi e^2/hc$$

whose inverse is numerically equal to approximately 137.

That this combination of natural constants reduced to this numerical result seemed to be a striking coincidence not only to Sommerfeld and Pauli, and but to others as well, notably Arthur S. Eddington, who proposed a detailed theory in an attempt to unify elementary-particle physics and cosmology using the fine-structure constant as one of its cornerstones.[176] Other theorists were unimpressed with Eddington's theory. The fine-structure constant became central to contentious arguments about the methodology that should guide theoretical physics.[177] Guido Beck, Hans Bethe, and Walter Riezler published a spoof of Eddington's use of the fine-structure constant in 1930.[178] Pauli too was almost certainly an object of their spoof, and just at that time he was thinking tentatively about his bold neutrino concept. The effect of their spoof on Pauli is unknown, but I suspect it may have contributed to his later hesitancy to publicly disclose his interest in Jung's mysticism.

Fascination and controversy over the number 137 continued. Richard Feymann commented on it in his book, *QED: The Strange Theory of Light and Matter*, of 1980.[179] To Feymann, the relationship of the electron-photon coupling constant to the fine-structure constant was only fortuitous, although he was convinced that a theoretical derivation of the coupling constant was necessary for a complete understanding of quantum electrodynamics. The near coincidence of $\alpha$ with 1/137 was not significant to him. To Pauli, however, there were two more important aspects of that coincidence: the relationship of the fine-structure constant to what I might call "coincidence in equations," and the relationship of the number 137 to cabalistic numerology. The former likely was what was striking to the theoretician Pauli, the latter to the mystic Pauli. Regarding the former, Feynmann implied that a theory that explains the numerical values of fundamental constants is especially convincing. As Valentine Telegdi pointed out to

---

Weisskopf, "This number [137] connects quantum theory (*h*), relativity (*c*), and electricity (*e*)." [180]

In Sommerfeld's relativistic treatment of the orbiting electron in the one-electron atom, its energy is given as:

$$E = -\mu\pi^2 Z^4 e^4 / 2h^2 n^2 \left\{ 1 + Z^2\alpha^2 / n\left(1/n_o - 3/4n\right) \right\},$$

where $\alpha$ is the fine-structure constant, $Z$ the number of electrons, $\mu$ the reduced mass, $n$ the principal quantum number, and $n_o$ the azimuthal quantum number. This equation is exactly the same as the one that P.A.M. Dirac later derived on the basis of a very different relativistic theory of the electron, where the quantum number $n_o$ is replaced by the total angular momentum number $j + \frac{1}{2}$. As Robert Eisberg pointed out:

> The results of a complete relativistic treatment by the Dirac theory can be expressed by the single equation [above]. These results are in exact agreement with the predictions of Sommerfeld's theory. Since the Sommerfeld theory was based on the Bohr theory, it is only a rough approximation to physical reality. In contrast, the Dirac theory represents an extremely refined expression of our understanding of physical reality. From this point of view the agreement between [these] equations is one of the most amazing coincidences to be found in the study of physics.[181]

This coincidence also could not have been lost on Pauli. In fact, he likely saw other "coincidences in equations." He was the first physicist to prove that Heisenberg's matrix mechanics and Schrödinger's wave mechanics are formally equivalent.[182] Pauli also knew that Einstein's formulation of special relativity, based upon his analysis of Maxwell's equations, and Minkowski's formulation based upon a rotational four-dimensional geometry, are equivalent. If physics and mathematics could yield such coincidences in equations based upon widely different trains of logic, an alethic reality had to be behind them, the differing mathematical symbols only redundant.

Pauli saw the numerological significance of the number 137 not only in theoretical physics, but also in Jewish mysticism. He appreciated that the number 137 is the 33rd prime number, which was well known to G. F. Bernhard Riemann (1826-1866) who had investigated prime numbers in his mathematical research. Pauli's mentor Sommerfeld carried on the traditions of Riemann and Felix Klein, which along with pure mathematics were tied to mysticism.[183] The cabalistic numerological connection to number 137 likely occurred to Pauli after his encounters with Jung. Victor Weisskopf noted Pauli's fascination with Jewish mysticism, in which the number 137 has special significance. Pauli had referred Weisskopf to Pauli's friend Gershom Sholem, the foremost authority on the Jewish Cabala. Weisskopf recalled one of his conversations with Sholem:

> When I mentioned this number −137–to Sholem his eyes popped out, and he asked me again if I had really said 137. He told me that in Hebrew each letter of the alphabet has a numerical equivalent and that the Cabala assigned a deep symbolic significance to the sums of such numbers in a given word. The number corresponding to the word *cabala* happens to be 137. Could there be a connection between Jewish mysticism and theoretical physics?[184]

Pauli continued to be fascinated with the number 137 and the fine-structure constant, which he held throughout his life to be a mystery requiring explanation. Pauli shared with Einstein the view that quantum mechanics is incomplete, which Pauli associated with the absence of an explanation of the fine-structure constant. He often concluded his more profound papers by referring directly or indirectly to the fine-structure constant. Thus, the final paragraphs of his Nobel Lecture of 1946 we read:

> At the end of this lecture I may express my critical opinion, that a correct theory should neither lead to infinite zero-point energies nor to infinite zero charges, that it should not use mathematical tricks to subtract infinities or singularities, nor should it invent a "hypothetical world" which is only a mathematical fiction before it is able to formulate the correct interpretation of the actual world of physics.
>
> From the point of view of logic, my report on "Exclusion principle and quantum mechanics" has no conclusion. I believe that it will only be possible to write a conclusion if a theory will be established which will determine the value of the fine structure constant and will thus explain the atomistic structure of electricity, which is such an essential quality of all atomic sources of electric fields actually occurring in nature.[185]

Pauli's fascination with the number 137 thus involved his belief in the numerological significance of integers as a sign of deeper meaning, which had several dimensions: He was attracted to it because it embodied the convergence of different physical theories; he saw the absence of an explanation of it as indicating the incompleteness of quantum mechanics; and its meaning was associated with Jewish cabalistic mysticism.  The number 137 appealed to Pauli's INFP side throughout his life; he was unable to escape his compulsion to it.  Charles Enz, Pauli's last graduate student, visited Pauli on his deathbed and reported that when Pauli learned that his hospital room was number 137, he saw it as a portent of his impending death.[186]

Enz has concluded that the young Pauli would have rebelled against Sommerfeld's belief in the numerological significance of integer numbers emerging from complicated physical equations and indicating an element of mysticism in the physical world.[187] Later, however, Pauli would invest the inverse of the fine-structure constant, the number 137, and the numbers 4, 2 , 1, 8, and others with numerological significance.  We can understand why Pauli might portray different attitudes toward numerology at different times of his life by recalling Pauli's dual personality type, ESTJ and INFP.  To Pauli's mentor Sommerfeld, by contrast, integers were signs of deeper physical meaning but had no mystical connotations.  As his biographer Ulrich Benz has pointed out, Sommerfeld said that:

> "We have here once again a 'preformed harmony' between physics and mathematics bordering on the miraculous. The efficiency of mathematics seems to just fit the needs of physics. In order to not exaggerate the miracle, we want to point out...."  And then he [Sommerfeld] ruthlessly enumerated a few mistaken discrepancies which brought down fanciful readers from the clouds of aesthetics into the world of decimal points. The "preformed harmony," which Sommerfeld often emphasized, belongs to the fundamental principles of the philosophy of Gottfried Wilhelm Leibnitz, who described with it a divine, ordered cosmic world plan…. Of Leibnitz' perspective, Sommerfeld said he felt an "intellectual satisfaction, which brings us … a perception of the beauty and simplicity of natural laws, a downright aesthetical musical joy that we feel each day more distinctly when we listen to the whole number harmony of the foundational phenomena of physics." He stated modern physics had a certain tendency toward Pythagorean number mysticism. He named as recognized number mystics Johann Jacob Balmer, Johannes Rydberg and Walter Ritz, of whom he maintained: "They based their research, consciously or unconsciously, on the demand that the relations of wave numbers in the spectrum be harmonic, as aesthetically compatible as possible with the facts, and the results justified their point of view." Sommerfeld began to subscribe to this attitude in the year 1920. Many a superficial as well as serious critic dismissed this playing with numbers as "atommystical." But Sommerfeld restricted his perspective with caution: "When I talk about number mysticism of old and recent times, I hope not to be suspected of being a proponent of mysticism in the usual sense, as it emerges in the astrological and spiritualistic tendencies of our times. Nothing lies further from me. I speak only of nature's laws and the way to investigate them through insight into their foundations." Because of Sommerfeld's attractions to harmonic number relations, his student Helmut Hönl called him a "Kepler reincarnate." More penetrating but less respectful, the young Pauli mocked his teacher's passion by paraphrasing a popular advertising slogan: "If you want whole numbers, go to Sommerfeld!" [188]

---

[188] Ulrich Benz, *Arnold Sommerfeld* (Stuttgart: Wissenschaftliche Verlagsgesellschaft MBH, 1975), pp. 122-123. My translation is of the following passage:

,,Wir haben hier wieder einmal eine ans Wunderbare grenzende 'prästabilierte Harmonie' zwischen Physik und Mathematik. Die Leistungsfähigkeit der Mathematik scheint geradezu den Bedürfnissen der Physik angepaßt zu sein. Um aber das Wunder nicht zu übertreiben, wollen wir sogleich bemerken:…"

Und dann zählte er schonungslos einege Diskrepanzen auf, die schwärmerische Leser wieder aus den Wolken der Ästhetik in die Welt der Kommastellen herunterholten.

Sommerfeld might follow Kepler in his atomic mysticism, but he was never the Jungian mystic that Pauli would become. The young Pauli might tease Sommerfeld about his numerological interests, but the adult Pauli was deadly serious about numerology.

**Klein and Sommerfeld: Rotations and Quaternians**

In his years with Sommerfeld, Pauli developed his physical intuition, his interest in numerology, his attraction to alethic reality, and his skill in mathematics, quite likely in matrix mathematics, which is especially useful in representations of symmetry. Felix Klein (1849-1925), Sommerfeld's own mentor in Göttingen from 1893 to 1897 and the leading German mathematician of his day, encouraged the work of his students Sophus Lie and Emmy Noether who introduced symmetry methods into physics.[189] The young Pauli was aware of Noether's work, referring to it in his *Theory of Relativity* of 1921.

---

Die „prästabilierte Harmonie", die Sommerfeld oft betonte, gehört zu den Grundbegriffen der Philosophie von Gottfried Wilhelm Leibniz, der damit eine im göttlichen Weltplan angelegte Ordnung beschrieb. Wie Sommerfeld sagte, empfand er eine

„intellektuelle Befriedigung, die uns ... aus der Schönheit und Einfachheit der Naturgesetze entgegenströmt, von der geradezu ästhetisch-musikalischen Freude, die wir empfinden, wenn wir, mit jedem Tage deutlicher, die ganzzahligen Harmonien der physikalischen Grunderscheinungen erlauschen."

Er konstatierte einen gewissen Hang der modernen Physik zur pythagoräischen Zahlenmystik. Als ausgesprochene Zahlenmystiker nannte er Johann Jacob Balmer, Johannes Rydberg und Walter Ritz, von denen er behauptete:

„Sie legten ihren Forschungen bewußt oder unbewußt die Forderung zu Grunde, daß die Zusammenhänge der Wellenzahlen in den Spektren so harmonisch, so ästhetisch einfach sein müßten, als irgend mit den Tatsachen verträglich; und der Erfolg rechtfertigte ihren Standpunkt."

Und dieser Haltung hat sich Sommerfeld in den Jahren im 1920 angeschlossen.
Als „Atommystik" lehnte sowohl mancher oberflächliche Spötter als auch ernsthafte Kritiker die Spielereien mit Zahlenverhältnissen ab. Aber Sommerfeld schränkte seinen Standpunkt vorsichtig ein:

„Wenn ich hier von der Zahlenmystik ältester und neuester Zeit gesprochen habe, so hoffe ich nicht in den Verdacht zu kommen, daß ich der Mystik in gewöhnlichem Sinne, wie sie in den astrologischen und spiritistischen Anwandlungen unserer Zeit hervortritt, das Wort reden wollte. Nichts liegt mir ferner. Ich sprach nur von den Naturgesetzen und dem Wege, sie zu ergründen, nicht von menschlichen Dingen."

Wegen seiner Neigung zu harmonischen Zahlenverhältnissen wurde Sommerfeld von seiner Schüler Helmut Hönl ein „Kepler redivivus" genannt. Eindringlicher, doch weniger respektvoll, persiflierte der junge Pauli die Liebhaberei seines Lehrers frei nach einem Werbeslogan:
„Sind's ganze Zahlen, geh zu Sommerfeld!"

[189] Hans Kastrup, "The Contributions of Emmy Noether, Felix Klein and Sophus Lie to the Modern Concept of Symmetries in Physical Systems," in Manuel Doncel, Armin Hermann, Louis Michel, and Abraham Pais, ed., *Symmetries in Physics (1600-1980)* (Barcelona: Universitat Autònoma Barcelona, 1987), p. 115.

Klein and Sommerfeld published their seminal work on the theory of the spinning top in four parts between 1897 and 1910,[190] employing the mathematics of quaternians, group theory, and matrices. They treated the rotations of a top and other physical bodies using mathematically visualizable hypergeometry. Pauli cut his mathematical teeth by studying Klein and Sommerfeld's theory, with its abstract kernel visualizations of rotations, and continued to use them in his physics.[191] At the same time, Pauli also experienced resistance to these mathematical methods owing to the psychological messages of his father and his godfather Mach, who opposed abstract mathematics and embraced instead measurement and empiricism. Pauli in 1957 reflected on this tension:

> The connected formulation of conceptual systems consisting of mathematical equations and rules whereby they may be linked with data of experience is called a physical theory; within the limits of its sphere of applicability one can then describe it as a "model of reality." As I have explained elsewhere, I regard it as idle to speculate on what came first, the idea or the experiment. I hope that no one still maintains that theories are deduced by strict logical conclusions from laboratory books, a view which was still quite fashionable in my student days. Theories come into being through an understanding inspired by empirical material, an *understanding* which we may best regard, following *Plato*, as a coming into congruence [*zur Deckung kommen*] of internal images with external objects and their behaviour. The possibility of understanding again demonstrates the presence of typical regulatory arrangements, to which man's inner as well as outer world is subject.[192]

In his *Theory of Relativity*, Pauli especially acknowledged his debt to Klein.[193] We can assume he also appreciated Sommerfeld's role, as Sommerfeld wrote the Preface to it. Klein and Sommerfeld worked with quaternian algebra, building on the efforts of William Rowan Hamilton (1805-1865) and Bernhard Riemann. They all recognized the power of quaternian algebra for mathematics and physics, and some of them saw further connections.  Thus, Hamilton saw mystical connections that likely were not lost on either Sommerfeld or the young Pauli's Shadow.  Pauli used quaternian algebra when he introduced spin matrices in 1927, but he might have picked up its mystical connections to Plato's Tetractys much earlier by the way it was visualized.  Hamilton defined quaternians as:

> The quotient of two vectors, or the operator which changes one vector into another....
>
> The SYMBOL OF OPERATION $q( )q^{-1}$ , where $q$ may be called (as before) the operator quaternion, while the symbol (suppose $r$) of the operand quaternian is conceived to occupy the place marked by the parentheses ... can be regarded as a conical rotation of the axis of the operand round the axis of the operator, through *double* the angle thereof....[194]

---

[190] Felix Klein and Arnold Sommerfeld, *Über die Theorie des Kreisels* (New York: Johnson Reprint Corporation, 1965).
[191] Pauli had to have reviewed Klein and Sommerfeld's work when he conducted his comprehensive summary of relativity theory, if not before.
[192] Enz and von Meyenn, *Wolfgang Pauli* (ref. 6), p. 129.
[193] Pauli, *Theory of Relativity* (ref. 16), p. vi.

Klein and Sommerfeld extended the quaternian algebra of the real and imaginary numbers into an algebra that included vectors, matrices, and generalized transformations. Here was a mathematics suitable for describing rotations.  Historian of mathematics Simon Altman has commented on Hamilton's achievements:

> We must stress that Hamilton's everlasting monument ... is his construction of objects that, except for commutativity, obey the same algebra as that of the real and complex numbers: and Hamilton was aware of this–although he could not foresee that quaternions were to receive in 1878, at the hands of [Georg Ferdinand] Frobenius, the supreme accolade of being proved to be the only possible objects with this property.[195]

Klein's and Sommerfeld's treatise, *Über die Theorie des Kriesels*[196]*,* is perhaps the most underappreciated source for the mathematical development of matrix mechanics.  We find in it the mathematics of quaternians and matrices, symmetry analysis, noncommutative matrix-multiplication methods, spin matrices, and a mathematical alethic reality used to describe bodies in motion, namely, the rotation of a spinning top. Sommerfeld no doubt put his and Klein's work into the hands of Pauli who, for example, treated a spinning hydrogen nucleus long before electron spin was considered.[197]  He mastered matrices, vectors, tensors, quaternians, and, in general, Hamiltonian dynamics long before Heisenberg's discovery of matrix mechanics.  Klein and Sommerfeld had used matrix mathematics, infinitesimal rotations, imaginary and complex numbers, and mathematical visualizations to express the deeply embedded rationality behind classical mechanics.  These visualizations often were in an abstract alethic space, and Pauli had to become comfortable with them, such as the one that Herbert Goldstein noted in 1950, when he discussed the equations of motion for a rotating rigid body: "Hence the jabberwockian sounding statement: the polhode rolls without slipping on the herpolhode lying in the invariable plane."[198]  To illustrate the challenge, Goldstein included the diagram below.[199]  He relied on Klein and Sommerfeld's earlier work, where similar illustrations of the physical parameters derived from abstract mathematics appear, such as the one below for the movement of *Pohlbahnen* of a rotating rigid body.[200]

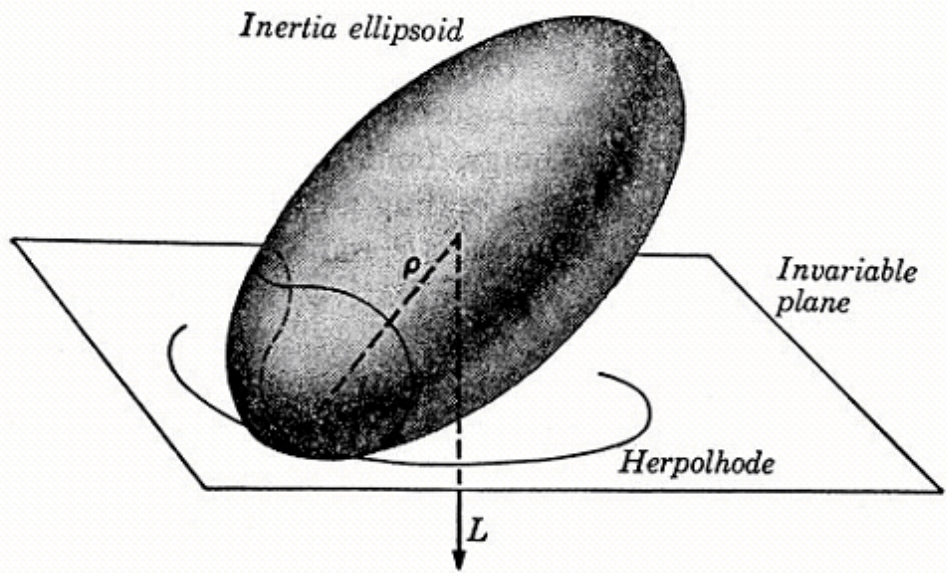

Figure 9.  Herbert Goldstein's diagram of jabberwockian terms in describing the motion of the inertia ellipsoid relative to the invariable plane.

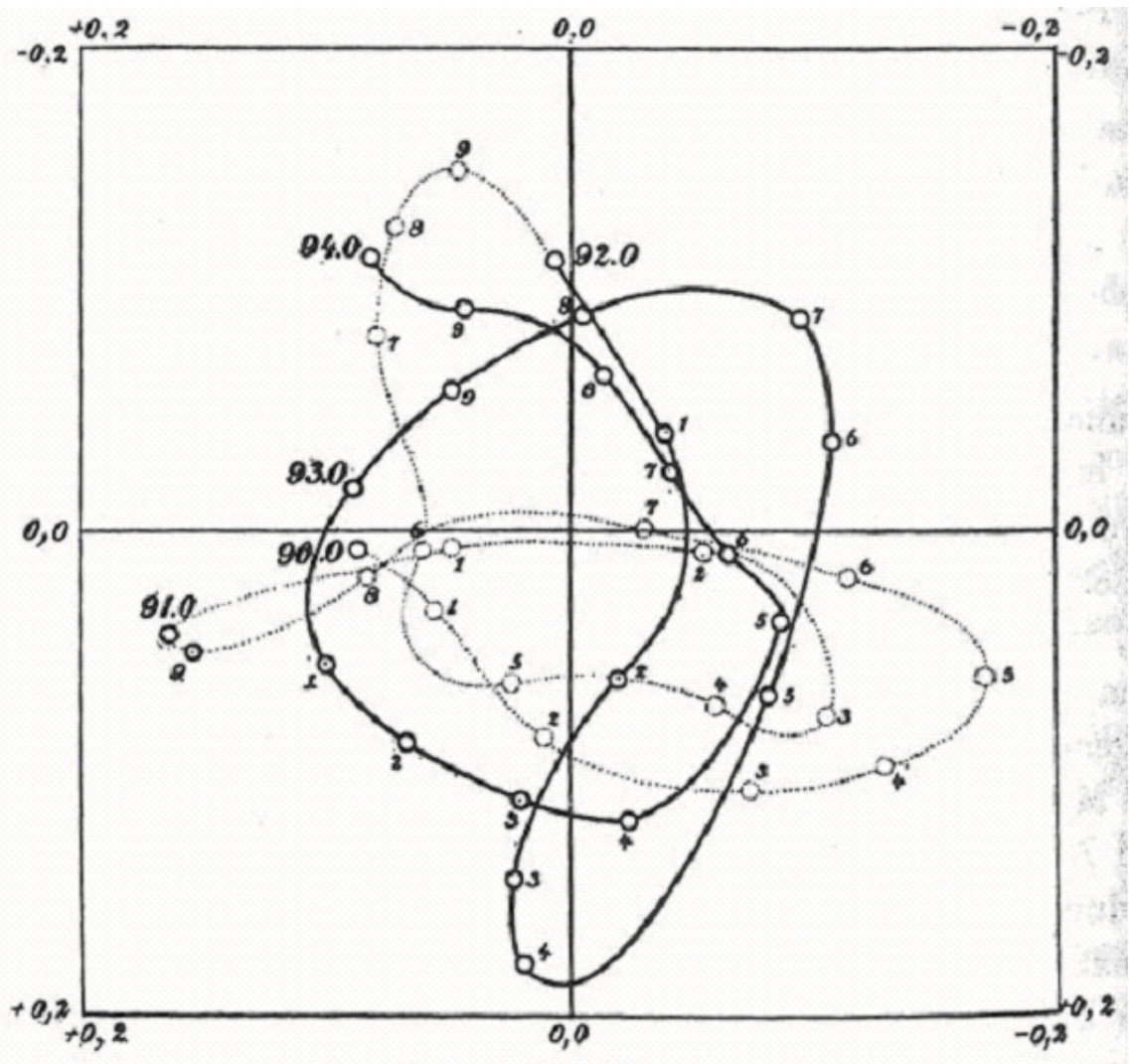

Figure 10. Klein and Sommerfeld's complex diagram for the physical parameters of motion of a top.

Such mathematical visualizations are in an alethic reality, in the sense that one cannot evaluate or measure their components in their abstract space, since the mathematics involves not only real but also imaginary and complex numbers. Their mystical connotations could not be ignored: Imaginary numbers that rotate a real number about an axis move beyond naïve visualizations to abstract ones, hidden from observation and measurement. Visualizations of rotations by quaternians extended the abstractions further and simultaneously illustrated the value of group theory. A measureable, real quantity might emerge from the mathematical derivation, but that derivation proceeded in an abstract space. Mach's direct connections to sensory information no longer held: The end result of a mathematical calculation might be measurable or observable, but not its intermediary steps. Deep reality was hidden and inaccessible at the foundational mathematical level in Klein and Sommerfeld's analysis of rotating bodies. Their readers were transported into an analytic world of difficult but visualizable mathematics that expressed an alethic level of reality.

The mathematics of quaternians can be visualized in the "mind's eye" as a type of abstract rotation similar to that produced by the imaginary number $i$. Quaternians involve a higher degree of abstractness and of rotations, however. Thus, a quaternian Q is defined as Q = $i$A + $j$ B + $k$ C + D. As Klein and Sommerfeld declared:

> By way of the original definition of the word quaternion whereby we base our concept of rotational transformation, *a quaternion means nothing else but the operation of a rotational transformation.* It is unequivocally determined by the amount of the rotation ($T$), by the axis of rotation ($a,b,c$) and the *half* angle ($\omega/2$) of the amount of rotation.[201]

The symbols $i, j$, and $k$ are extensions of the imaginary numbers and are interrelated by:

$$ii = \text{-1}, \quad jj = \text{-1} \ , \quad kk = \text{-1},$$
$$ij = \ k, \quad jk = \ i \ , \quad ki = \ j,$$
$$ji = \text{-}k, \quad kj = \text{-}i \ , \quad ik = \text{-}j$$

The multiplication of two quaternians is noncommutative, as would be expected in a system that models rotations. In the "mind's eye" of a mathematician when learning quaternian algebra, the four elements $i, j, k,$ and 1 that generate the visualizable rotations, and their holistic unity, produce a sense of awe regarding the inherent power of that kernel of symbols. Not surprisingly, Klein, Sommerfeld, and Pauli were struck by the power of quaternian mathematics.[202]

Klein and Sommerfled not only used matrix and group methods to analyze the difficult classical problem of a rotating top; they also connected the alethic-like imaginary numbers and the mystical, almost Pythagorean Hamiltonian quaternians to relativity theory. In the fourth part of their text published in 1910, they treat the Lorentz group and Einstein's theory of special relativity using quaternian mathematics, as Pauli knew, and they used group-theoretical methods that would later reappear in quantum mechanics.[203] As Minkowski did in 1908, so Klein and Sommerfeld showed in 1910 how the Lorentz transformations could be viewed as rotations in four dimensions, that is, how quaternian mathematics could be used to represent a transformation between two coordinate systems.[204]

---

[201] *Ibid.,* p. 58. My translation is of the following passage:
"Als ursprüngliche Definition des Wortes Quaternion legen wir unsern Begriff der Drehstreckung zu Grunde: *Eine Quaternion bedeutet nichts anders als die Operation der Drehstreckung.* Sie ist eindeutig bestimmt durch die Grösse der Streckung (*T*), durch die Axe der Drehung (*a,b,c*) und die Grösse des *halben* Drehungswinkels (*ω/2*)."
[202] *Ibid.,* p. 60. Klein and Sommerfeld made special note that quaternian multiplication is noncommutative.

To Hamilton, quaternians held mystical connotations, which may have fascinated Pauli's Shadow during his years with Sommerfeld owing to their mathematical power in manipulating components of his physical kernels. Quaternians form a system of interrelated mathematical entities. Just as the imaginary number $i$ extends the concept of a real number to include operations that are not allowable for the real numbers, so quaternians extend the algebra of imaginary and real numbers to allow visualization of rotations from one vector system to another.[205] The mathematics of rotations employed in the Lorentz transformations is the mathematics of matrices, as noted by Goldstein:

> Clearly a spatial rotation between two systems at rest relative to each other is included as a subclass of the Lorentz transformation…. [Any] general Lorentz transformation is a product of a space rotation and a pure Lorentz transformation.[206]

Klein and Sommerfled's use of complex numbers, of matrices to describe abstract rotations in a mathematical space, of quaternians, and of symmetry principles, prepared the theoretical physicist Pauli to use these powerful mathematical methods in exploring the physical world. They prepared the ground for a kind of mathematical and Pythagorean abstraction that soon found its way into matrix mechanics. To Pauli, quaternians offered a mathematical tool to handle rotations of abstract mathematical entities, and to visualize them. To him, intuitively visualizable rotations were a clear sign of a buried physical kernel.

**Pauli's Encyclopedia Article on Relativity**

Sommerfeld taught Pauli, inspired him to view numerology as a clue to hidden physics, and asked him to write an encyclopedia article on relativity theory in his place, thus presenting his student with an unparalleled opportunity to make his mark in theoretical physics. Sommerfeld knew that Pauli was up to the challenge, since he already had assisted Sommerfeld in revising his *Atombau und Spectrallinien*.[207] Sommerfeld recognized, in fact, that his student's command of relativity theory might exceed his own.[208] Pauli had written his first article on the general theory of relativity at the age of eighteen in Vienna, before leaving for Munich. Pauli's works throughout his life were noteworthy for their mathematical and physical sophistication, for the breadth and depth of theoretical and experimental knowledge they displayed, and for their critical nature. Pauli was adept at finding shortcomings in the work of others, but in his early years he had neither the intuition nor the courage to exercise his own creativity fully. I believe that his creativity was then restricted by his inner ESTJ messages. He was too critical of his own aesthetic intuitions because of his dichotomous ESTJ-INFP personality.

In his *Theory of Relativity* of 1921, Pauli displayed his brilliance in mathematical physics and his extraordinary ability to penetrate to the heart of relativity theory. Einstein, Hermann Weyl, and Max Born were deeply impressed with his encyclopedia article. The elements in Pauli's treatment of relativity theory that shine through are rotations, quaternians, kernels, and alethic reality. Pauli may have resisted the idea of an alethic reality because it conflicted with Mach's positivism, but, probably under Sommerfeld's guidance, he became impressed with Einstein's willingness to "go beyond naïve visualizations" to explore the limitations of sensory perceptions in redefining the concept of time. By the time that Pauli encountered relativity theory, Einstein had extended his theory to include accelerated motion, and hence, gravitation. Pauli concluded his *Theory of Relativity* by noting that the space-time continuum of relativity

theory could not explain quanta. In 1956 when it was translated into English and republished, he recalled:

> There is a point of view according to which relativity theory is the end-point of "classical physics"' which means physics in the style of Newton-Faraday-Maxwell, governed by the "deterministic" form of causality in space and time, while afterwards the new quantum-mechanical style of the laws of Nature came into play.  This point of view seems to me to only partly be true, and does not sufficiently do justice to the great influence of Einstein, the creator of the theory of relativity, on the general way of thinking of the physicists of today.  By its epistemological analysis of the consequences of the finiteness of the velocity of light (and with it, of all signal-velocities), the theory of special relativity was *the first step away from naive visualizations*.  The concept of the state of motion of the "luminiferous aether", as the hypothetical medium was called earlier, had to be given up, *not only* because it turned out to be unobservable, but because it became superfluous as an element of a mathematical formalism, the group-theoretical properties of which would only be disturbed by it.
>
> By the widening of the transformation group in general relativity the idea of distinguished inertial coordinate systems could also be eliminated by Einstein as inconsistent with the *group-theoretical properties of the theory*.  Without this general critical attitude, which abandoned naive visualizations in favour of a conceptual analysis of the correspondence between observational data and the mathematical quantities in the theoretical formalism, *the establishment of the modern form of quantum theory would not have been possible.*  In the "complementary" quantum theory, the epistemological analysis of the finiteness of the quantum of action led to further steps away from naive visualizations.  In this case it was both the classical field concept, and the concept of orbits of particles (electrons) in space and time, which had to be given up in favour of rational generalizations.  Again, these concepts were rejected, not only because the orbits are unobservable, but also because they became superfluous and would disturb the symmetry inherent in the *general transformation group underlying the mathematical formalism of the theory.*[209]

Pauli thus suggests that relativity theory and its connection to group theory stimulated the development of quantum mechanics. Heisenberg implied the same when he connected the Lorentz group to quantum mechanics.[210] Pauli used tensors and matrix mathematics already in his *Theory of Relativity* of 1921. In his recollections above, he implies that group theory as used in relativity theory, together with Einstein's move away from naïve visualizations, were necessary for the development of quantum mechanics.

In his *Theory of Relativity*, Pauli explained how the Lorentz transformations could be visualized as rotations in an abstract space, citing an article by Klein of 1910 where Klein had shown that quaternians could be used to treat Lorentz transformations,[211] thus indicating that Pauli was familiar with visualization of rotations in abstract spaces composed of alethic dimensions. Pauli also cites Sommerfeld's presentation of 1910 of Minkowski's four-dimensional representation of the Lorentz group.[212] Here were the quaternian components of a physical kernel with three spatial dimensions and one time dimension, the latter being the nonrational *ict*. Pauli also was familiar with Emmy Noether's and Erich Bessel-Hagen's work on invariants in physics.[213] Pauli, in sum, used mathematical methods that went beyond observables and pointed to deep space-time connections. He was comfortable with empirical observations but was more skilled in exploiting mathematical visualizations.

In the concluding chapters of his *Theory of Relativity,* Pauli addressed problems in extending general relativity theory to deal with matter, critiquing the attempts of Weyl and others to explain the quantum of electricity, the charge of the electron. He concluded that continuum mathematics was unable to account for the quantum of electricity, using an argument laced with Machian positivism:

> Finally, a conceptual doubt should be mentioned. The continuum theories make direct use of the ordinary concept of electrical field strength, even for the fields in the interior of the electron.  This field strength is however defined as the force acting on a test particle, and since there are no test particles smaller than an electron or a hydrogen nucleus, the field strength at a given point in the interior of such a particle would seem to be unobservable, by definition, and thus be fictitious and without physical meaning.[214]

Pauli called attention to Weyl's and Einstein's unified field theories, and argued that such continuum theories were of no use in the quantum realm because the field quantities were unobservable, nor were their definitions traceable to observables.  Here Mach's voice surfaced instead of Sommerfeld's.  Historian John Hendry has termed Pauli's philosophical stance at this time "operationalist," citing in support Pauli's encyclopedia article on relativity and a letter that Pauli wrote to Eddington soon after its publication,[215] noting in particular Pauli's point that a field in the interior of an electron cannot be defined operationally.  Pauli's operationalism actually is a curious mixture of Machian positivism and introspective *Anschaulichkeit* (intuitive visualization).  In 1956, as noted above, Pauli commented that Einstein's key contribution was to go beyond naïve visualizations, which implies that in 1956 Pauli was committed to the value of visualization, to *Anschaulichkeit*.  Thus, after 1921 his interest in alethic reality surfaced as his commitment to positivism waned, and after his treatment in the Jung Clinic, he abandoned his operationalist commitments in favor of alethic reality and was content to work only with metaphors.

Pauli's *Theory of Relativity* is marked by Pauli's penetrating command of the scientific literature, but not by creative solutions to open scientific questions.  It reveals, however, Pauli's burgeoning interest in the fundamental constants of Nature and his aesthetic attraction to quaternities, rotations, kernels, and the enigmatic quanta.

**Sommerfeld and Pauli**

Sommerfeld's teaching and style of physics set an example for Pauli in new ways of looking at the world, at odds with those of his scientist father and his physicist-philosopher godfather Ernst Mach. I see Sommerfeld's influence on Pauli's impressionable "mind's eye" as validating Pauli's deepening physical and mathematical intuition, his *Anschaulichkeit*. Sommerfeld exposed Pauli to Kepler's Platonism and numerology in his extension of the Bohr model of the hydrogen atom. He showed that a qualified numerology could be a valid indicator of a deeper physical reality, as in his quantization conditions. Sommerfeld endorsed Pauli's use of Hamilton's quaternians to describe rotations of physical bodies, and acquainted him with the mystical aesthetic underpinnings of quaternian mathematical systems, including their philosophical underpinnings which go back to Plato's Tetractys. Sommerfeld would have been comfortable in pointing out to Pauli the kernels that can be grasped fully only in the "mind's eye."

---

[214] Pauli, *Theory of Relativity* (ref. 16), p. 206.
[215] John Hendry, *The Creation of Quantum Mechanics and the Bohr-Pauli Dialogue* (Boston: D. Reidel Publishing Co., 1984), p. 45.  See Hendry's footnote 45 and his continued discussion of Pauli's philosophical perspective.

Sommerfeld also encouraged Pauli to participate in current physical research and asked Pauli to assist him in revising his *Atombau und Spectrallinien*, and provided the opportunity to become engaged with other students such as Heisenberg. Sommerfeld exposed Pauli fully to the old quantum theory and to his use of Hamilton-Jacobi methods and relativity theory in extending the Bohr model of the hydrogen atom. He displayed to Pauli the fruitfulness of conservation principles and symmetry considerations in atomic physics, especially when treating angular momentum where spin matrices might be employed. He set an example for Pauli in assuming the reality of deep mystical connections in physics, as in his fine-structure constant and its numerological associations. Sommerfeld, in sum, acquainted Pauli with a Keplerian Platonism; numerology as a guide to achieve deeper physical meanings and connections; mathematics as the language of physics; experiment as ruling over theory; visualizable models as valid tools; kernels of physical systems that involve four-part quaternities; and rotations as a sign of a buried kernel. Whether or not Sommerfeld intended to do so, he conveyed to Pauli many of the philosophical perspectives he adopted later. In particular, mysticism as practiced by earlier scientists was not only permitted, it was a valid tool of inquiry.

Pauli's work on relativity theory served to awaken his interest in Platonism. He employed the abstract mathematics of tensors and matrices in four-dimensional space to describe this empirically supported physical theory and to visualize a deep mathematical basis for it.  He used abstract mathematics in an alethic reality.   He saw the importance of visualizing rotations of a whole system of entities, such as rotations of an alethic kernel, to derive empirically verifiable values for many of these entities, such as the projections of a vector onto another coordinate axis in space.  Further, he saw the need for only four-dimensionality in Minkowski space-time, a hint at the numerological significance of the number 4.  Pauli thus was seeing alethic reality, kernel wholes, quaternity numerology, and Keplerian mathematical aesthetics in his approach to relativity, in contrast to his father's and Mach's positivist and operationalist messages.  Pauli became acquainted with alethic reality through the mathematical treatment of Lorentz transformations using rotations of vectors in hypergeometric physical space, the four-dimensional space of Minkowski space-time.  He also was exposed to the mathematical concepts of matrices, symmetry invariants, and group representations.  He might well have begun taking physical entities not at face value as naïvely visualizable, but as kernel wholes that have deeper alethic dimensions and can be visualized abstractly--any mathematical or physical entity can be expressed and analyzed as kernel wholes with internal subdivisions.  He begins to place pure mathematics and visual images on an equal footing with empiricism--and only when the visualizations prove superfluous does he give them up.[216]

In his *Theory of Relativity*, Pauli emphasizes theory over experiment.  He is thoroughly familiar with both, but he emphasizes the deeply aesthetic features of theoretical physics.  This reflects his education and orientation toward the arts, his interest in history and philosophy.  He was attracted to aesthetic beauty, a hallmark of an INFP personality.  Hendry's characterization of Pauli as an operationalist thus requires qualification.  In my view, Pauli pursued theoretical physics primarily seeking aesthetics, beauty, and consistency, and secondarily seeking experimental confirmation of its predictions.  In contrast to Mach who started from observables and sought an economical description of them, Pauli, like Kepler, sought mathematical beauty first and experimental evidence second.  However, again like Kepler, he knew that experiment imposes constraints on theory.  Kepler discarded his theory of the orbit of Mars when faced with Tycho Brahe's observations; Pauli discarded any theory that conflicted with the constancy of the speed of light. The fundamental constructs of physics had to be observable in principle, but not directly as naïve visualizations. Mathematical variables could be abstract, alethic quantities connected ultimately instead of directly to observables. The constructs of Pauli's physics, like the naïve visualizations of classical physics, were subject to reinterpretation, leading to deeper meaning. Minkowski's four dimensions of space-time, for example, led to visualization of Einstein's theory as a kernel of four dimensions. The aether was eliminated, as Pauli wrote in 1956, "*not only* because it turned out to be unobservable, but because it became superfluous…."[217] To Pauli, unobservables were permitted in the mathematical formalism, but only if they were required to describe the physics. Alethic entities were permitted. For example, Schrödinger's wave function $\Psi$ was unobservable, but essential to the theory. The spin matrices that Pauli introduced later were unobservable, but they were necessary for the mathematical description of an electron's internal alethic dynamics. Reality was alethic, as for Schrödinger's wave function and Pauli's spin matrices, but one could never trust that they were the final solution or the deepest level of mathematical or physical symbolism. As we will see, Dirac's extension of Pauli's spin matrices from 2 x 2 to 4 x 4 matrices is a prime example of "going deeper" into alethic reality.

---

[216] Pauli, *Theory of Relativity* (ref. 16), p. 25.

Sommerfeld was Pauli's mentor at a critical stage of Pauli's life and scientific career. Sommerfeld became a role model for Pauli on how to approach life's questions. Pauli had been exposed to conflicting messages from his parents. His father, the extrovert ESTJ personality, looked to the external world for answers to life's questions. His mother, the introvert INFP personality, looked to internal spirituality and aesthetics for answers to them. Pauli's father and mother perhaps would clash in their home owing to their different personality types. Sommerfeld, by contrast, presented to the young Pauli a healthy and balanced personality, perhaps INFP or INTP, but either one with a strong E (Extroverted) component. He conveyed to his students that he was comfortable with both externally derived information and internally perceived aesthetics. He appreciated experiment as the ultimate confirmation of theory, like a modern-day Kepler, and he simultaneously appreciated the beauty of hidden mathematical, even Platonic, truths. Sommerfeld received, recognized, and encouraged Pauli, which must have been a validating experience for him as he separated from his parental home. In Sommerfeld Pauli could observe a successful physicist who was fully engaged in both the external physical and internal psychological worlds, with neither perspective superior to the other but complementary to each other. Pauli no longer had to decide which perspective was superior; in Sommerfeld's court, external facts confirmed internal intuitions and both were valuable. Pauli could commit himself consciously to the importance of both empirical and logical aspects of physics, but subconsciously he was moved by and attracted to the aesthetics and intuitions that Sommerfeld displayed to him. Pauli's early and forced ESTJ personality type could be slowly released in favor of his natural and more comfortable INFP personality type. He did not make that transition easily during his individuation process from his parents, but instead struggled through it, as is typical for a man as he becomes an independent adult. Pauli's Shadow was resistant to Sommerfeld's role modeling. Pauli's INFP personality was awakening under Sommerfeld's influence, but his ESTJ personality was still dominant. The critical juncture in Pauli's transition occurred during his work on the exclusion principle, when he finally converted to Sommerfeld's style, but his individuation process was extended many years beyond that owing to disturbing events that he would experience in his personal life.

---

[217] *Ibid.,* p. vi. My italics.

Sommerfeld's empathetic mentoring as a father figure opened new possibilities for Pauli. He had to make his own mark, to develop his self-confidence. He tested his mettle, as it turned out, by selecting perhaps the most difficult problem in physics at the time, searching for an explanation of the anomolous Zeeman effect. The culmination of his quest, as I will discuss in the next chapter, was his introduction of a fourth quantum number for the electron, a classically nondescribable two-valuedness. For Pauli's ESTJ personality, "classically nondescribable" generated a major conflict in him that shook him to his core. He had to allow his numerological intuitions to emerge, which Sommerfeld would have encouraged, to construct the mathematics necessary to decipher the anomolous Zeeman effect and to formulate his exclusion principle. He had to suspend his operationalism. His INFP personality was surfacing in triumphant manner, but at the price of sacrificing his ESTJ values.  For Pauli's external ESTJ voice, that was still a swindle.[218]

---

# Chaper 5.   The *Pauli Verbot*, 1921-1925

**Pauli in Transition: Göttingen, Copenhagen, Hamburg**

Pauli's explanation of the anomalous Zeeman effect led to his formulation of the exclusion principle in 1925, one of the most far-reaching discoveries in physics.  This was an unsettling period in Pauli's life; it was marked by transitions in his psychological orientation and in his philosophy of physics.  His paper on the exclusion principle, which eventually won him the Nobel Prize in Physics for 1945, actually was more of a description than an explanation of the underlying physics: It thus was a "swindle" in his "mind's eye."  His personal and academic life also was in transition.   He received his Ph.D. degree from the University of Munich in 1921, spent postdoctoral periods in Göttingen in 1921, Hamburg in 1922, Copenhagen in 1922, and returned to Hamburg in 1923 to accept a permanent academic position, embarking on his career with all the uncertainty that usually accompanies academic life.  He experienced all of these pressures in the short span of four years.  Meanwhile, at home in Vienna, his parents were likely experiencing increasing marital discord.  The Great War was over, but the ensuing economic hardships and unsettling political and social conditions continued.  Anti-Semitism was on the rise.  Germany's Shadow was emerging.

The period in a person's life from adolescence to young adulthood is one of great transition.  In his twenties, like any other young man, Pauli had to separate from his parents, becoming his own man and thinking for himself in matters of religion, philosophy, sexuality, morality, and career decisions.  For Pauli, these intellectual and emotional transitions occurred between 1921 and 1925.  He did not make them very gracefully.  In Carl Jung's metaphysics, the psyche and its events often are mirrored by events in the external world.  In Pauli's case, the period of transition in his internal psychic life, career in physics, and the events in the external world mirrored his frenetic mind.  His parents soon would separate, and he soon would court his future wife.  He would indulge in increasingly "anomalous" social recreation.  He would make a transition from student in Munich to professor in Hamburg.  The classical models of the old quantum theory would be replaced with the abstract, noncommonsensical, and controversial models of the new quantum mechanics.  Physics was in a period of profound transition, as was Pauli himself.

Historians have identified three pivotal events in the development of the new quantum mechanics in 1925: Pauli's enunciation of the exclusion principle, Werner Heisenberg's creation of matrix mechanics, and Samuel Goudsmit and George Uhlenbeck's discovery of electron spin.  Pauli was intimately involved with all three.  No one, however, has yet identified the pivotal events in the development of Pauli's psyche at this time.  While he himself may have identified some of these events later when he was in treatment in the Jung Clinic, historians are still unaware of what he was experiencing in his own "mind's eye."  Pauli not only discovered the exclusion principle, he also placed his stamp on Heisenberg's creation of matrix mechanics and on Goudsmit and Uhlenbeck's discovery of electron spin.  His knowledge that he did not receive public recognition for these latter two achievements may have contributed to his later unsettled emotional state.  To support these claims, I will recount Pauli's contributions between 1921 and 1925, the physical and philosophical meanings of his exclusion principle, and his subsequent influence on physics.  This was a period when Pauli became emotionally unglued; the drama associated with his unsettled state of mind would appear later.

Pauli's mental state during his period of transition between 1921 and 1925 was tied to his process of individuation from his parents and to his increasing risk-taking in physics as his personality shifted from an ESTJ to an INFP type.  Instead of following the lead of other physicists, he now began to explore subjects of interest to himself.  He struck out on his academic career.  He explored alternative forms of pleasure and entertainment in the nightlife of Munich, Göttingen, Hamburg, and Copenhagen as he changed his residence.  He also was preoccupied with the normal pursuits of a man in his twenties: women and sexual adventures.  Until a young man has made these transitions successfully in his own judgment, he remains unsure of himself.  He is on a psychological quest.  He is testing his religion, his philosophy, his sexual performance, his way of viewing the world and his place in it--all against the internal messages springing from his Shadow.  The young man who makes these adjustments successfully assumes the mantle of a self-confident adult male; the young man who does not finds that his Shadow erupts uncontrollably, sometimes manifesting itself in criticism of others, inappropriate behavior, and in inappropriate attempts at humor. I believe that Pauli did not make a successful transition to adulthood and self-confidence either in his personal life or in his physics, at least until after he met Carl Jung in 1932. I see Pauli's lack of self confidence continuing even thereafter and never completely disappearing. Pauli's early twenties marked a period when he began to test himself, to remove from his Shadow the issues that made him fearful.

By the time Pauli received his Ph.D. degree at the University of Munich in 1921, his reputation among physicists had grown greatly owing to his brilliant encyclopedia article on the *Theory of Relativity* in 1921. His life and his physics, however, had continued to be determined by others. In Göttingen, his first postdoctoral destination, he worked under Max Born who described him as "childlike," and called him the "little Pauli," when in fact he was not at all physically little.[219] Under Born, Pauli experienced a mathematician's way of doing physics not to his liking. He worked on perturbation theory, as in celestial mechanics, and applied it to Keplerian electronic orbits, probably finding this work exhausting and tedious. He already had shown in his dissertation that the methods of the old quantum theory were inadequate to solve this problem--a new physics was required to do so. His negative attitude toward Born's style of theoretical physics solidified: In 1925 he refused, after no reflection, to work with Born in developing Heisenberg's matrix mechanics further. Earlier, Pauli had referred to Born's style as "the Göttingen effusion of erudition" [*Göttinger Gelehrsamkeitsschwall*], and he now refused Born sarcastically, saying: "Yes, I know you are fond of tedious and complicated formalism. You are only going to spoil Heisenberg's physical ideas by your futile mathematics." [220]

To Pauli, theoretical physics was not doing lengthy calculations of correction factors, or finding mathematical proofs: It required creative insight. Pauli's response to Born thus seems inconsistent with his earlier philosophical attitude in his *Theory of Relativity* where he appears as an "operationalist," exhaustively reviewing and criticizing all of the experimental and theoretical works on the subject. I think Pauli was now seeing the need, awakened under Born, to value creative intuition à la Sommerfeld.  For now, it remained the creative intuition of others; he himself was still too confined by his Shadow to risk disclosing his own creative intuition.

During his stay with Born, Niels Bohr visited Göttingen in June 1922.  If ever there was an intuitive physicist, confident in discussing imagery streaming out of his psyche, it was Bohr.  Bohr's flights of creativity were enormously fruitful.  In December 1922, he would receive the Nobel Prize in Physics for his atomic model of 1913, the "highest musicality of human thought," as Einstein put it.  No wonder that Bohr and Sommerfeld worked so well together: both were intellectual disciples of Kepler.  During his visit to Göttingen, Bohr invited the young Pauli: "Come to Copenhagen next year and work with me."  Bohr's invitation must have struck Pauli to the core of his being: He was being courted by the universally recognized leader in atomic physics.  I maintain that Pauli at this time was still deeply insecure despite his outward brashness: His biting criticism of others is a sign of inner insecurity and lack of self-confidence.  Pauli's fame thus was due, not to his intuition and creativity, but to his encyclopedic knowledge and penetrating criticism; he could keep Bohr on track.

Pauli accepted Bohr's invitation, but not without disclosing his inner insecurity through a revealing boast.  As Pauli later recalled, he responded to Bohr's invitation with the words:

> "I hardly think that the scientific demands which you will make of me will cause me any difficulty, but the learning of a foreign tongue like Danish far exceeds my abilities." … I went to Copenhagen in the fall of 1922, where both of my contentions were shown to be wrong.[221]

In Copenhagen, Pauli found recognition as a rigorously logical operationalist.  His ESTJ side was being reinforced; his INFP side was still dormant.  Creativity in the Copenhagen milieu was flowing from Bohr, not Pauli.   By his example and mentoring, Bohr would hone Pauli's intuition and creativity, but Pauli's ESTJ reservations still were too great to solve  the problem of the anomalous Zeeeman effect.  He recalled his anguish:

> A colleague who met me strolling rather aimlessly in the beautiful streets of Copenhagen said to me in a friendly manner, "you look very unhappy";

---

[221] *Ibid*., p. 89.

whereupon I answered fiercely, "How can one look happy when he is thinking about the anomalous Zeeman effect?"[222]

Pauli would have to set this intractable problem aside, and leave Copenhagen before an alternative to his ESTJ personality emerged; he went to Hamburg where there were diversions to mitigate his anguish owing to his unsatisfactory intellectual performance.

Pauli began his academic career in Hamburg in the fall of 1923 and remained there until 1928, when he moved to Zurich. In Hamburg he met several colleagues who would become lifelong friends: Wilhelm Lenz, Otto Stern, Walter Baade, and Erich Hecke. In Hamburg, Pauli, now on his own, was able to work as an equal at the university, without being subtly led by a father figure like Sommerfeld, Born, or Bohr. He could do whatever he wanted to do. The large port city of Hamburg offered the earthiness of all port cities. Until going there, Pauli did not drink alcoholic beverages.[223] Once there, he probably was attracted synchronistically to the Red Light district, the Reeperbahn in St. Pauli.[224] In Hamburg, the "Pauli effect" also would become evident and noticeable to others: Stern would not allow Pauli into his laboratory for fear that something would go awry simply by his presence. Pauli would revel in his new notoriety.

In these frenetic transition years, Pauli struggled principally with the anomalous Zeeman effect, perhaps the most intractable problem in physics at the time. He also began writing another encyclopedia article on the old quantum theory, which became known as the Old Testament. Even before he finished the Old Testament, however, developments that he would treat in his New Testament of 1932 began to unfold, Pauli himself contributing greatly to them. We still do not know much about Pauli's inner personal and philosophical ruminations during this period. If his psychological development, his physics, and his philosophy were intertwined, then we might speculate about events in his personal life that might have caused him stress. I suspect that during the Christmas holidays of 1924, while Pauli struggled with his exclusion principle, he encountered not only his parents, but also his Shadow.

Pauli's visit home to Vienna for the Christmas holidays of 1924 marked a major transition for him. He had decided to publish his exclusion principle, which eventually would garner a Nobel Prize for him.  His visit home marked another transiton: the beginning of an emotional downturn that would not be reversed until 1932 when he sought help in the Jung Clinic in Zurich.  Except for its magnitude and duration, a stressful transition such as the one Pauli experienced is not unusual in young adults when they separate from their parents, and assume a personality type of their own.  In Pauli's case, he was a twenty-four-year-old male, who was famous for his *Knabenphysik*, but was still tied emotionally to his parents.  I picture Pauli's parents assaulting him along the following lines:

*Pauli's father*: " Well, how's your new career of carousing in the bars and nightclubs of Hamburg going?  Times are hard in Vienna.  Here, we don't have time to play.  Seen any good pornography or drunk any good anisette lately?  Have you found any good Catholic girls to marry?  That will help your career.  Be sure not to tell your girlfriends about your Jewish ancestry! When are we going to be grandparents?  You know your godfather Mach was a freethinker.  He wouldn't fall for that religious garbage!  Oh, your mother is still trying to convert people to her religion and politics.  Have you forgotten what Mach told you about how to do physics? Experiment is the key to avoiding the trap of metaphysics!  Beware of your intuition!  Did you complete that encyclopedia article yet?  Have I told you I'm tiring of your mother and that I'm smitten by a young sculptress?"

*Pauli's mother*: "Have you forgotten what I told you about honoring God?  Don't trust any Catholics.  Are you watching out for the Devil's attempts to trick you?  Trust your inner messages from God.  He will guide you as he did the ancients.  Read Goethe's *Faust* again.  It will do you good.  Beauty, harmony, truth, and social justice will prevail!  What have you been doing for entertainment?  Music, art, theater?  Met any good Protestant girls yet?  Beware of those nightclubs!  Have I told you I have been depressed lately?"

Pauli's life was dramatic and frenetic from the time he left Munich in 1921 until the time he published his paper on the exclusion principle in Hamburg in early 1925.  He moved several times, began his academic career, published several lengthy articles, and made his mark as a theoretical physicis--just when the old quantum theory was replaced by the new quantum mechanics.  Pauli, however, continued to be personally disappointed during this period. He discovered his exclusion principle before Heisenberg created quantum mechanics and before Goudsmit and Uhlenbeck discovered electron spin. His exclusion principle thus was a product of the old quantum theory, and was not fully appreciated by physicists until the new theory had replaced the old. Overshadowed by the success of others, his exclusion principle was a breakthrough in atomic physics, but it took twenty years for him to receive recognition for it by a Nobel Prize. Neither he nor anyone else has yet provided a full theoretical justification for it. He knew that he had failed to decipher the fundamental significance of the anomalous Zeeman effect. As he said, "My nonsense is conjugate to the nonsense which has been customary so far."[225]

Despite his discovery of the *Pauli Verbot*, he remained disturbed and detached from physics, probably owing to personal psychological issues like his concern for his relationship to his parents. As I mentioned earlier, he seems to have been a teetotaler before moving to Hamburg,[226] but that changed there. Historian John Heilbron has described his life in Hamburg:

> "Stern, Pauli, Wentzel, Minkowski are bachelors (quite some boys) and meet each other outside the Institute only in restaurants, cafés, motion picture theatres, and the like." Thus a visitor to the University of Hamburg described what we now call the lifestyle of the physicists there. The city offered special opportunities for dissipation. Josephine Baker, the American cabaret star, was not allowed to perform in Munich; Hamburg welcomed her. Pauli took advantage of his opportunities. He became a dévoté of the cabaret, in Berlin, where he liked a low place called the Catacombs, as well as in Hamburg; he dipped into a culture of alcohol, pornography, and drugs; he married a dancer [when he was in Zurich in 1929]. Pauli balanced these indulgences and avoided further bruises to his psyche by throwing himself at a severely classical problem in physics, the conduction of heat in solids. His disposition improved. "I am always in favor of the varatio delectans." "It has been very good for me to withdraw from atomic physics for a while," he wrote Bohr [on February 11, 1924], and from "problems with which I vexed myself in vain, and which were much too difficult for me."

For almost a year [1924] he had no taste for atomic physics.[227]

Pauli's misgivings about his career and distaste for physics continued, as he wrote in a letter to Ralph de Laer Kronig in May 1925:

Physics is once again at a dead end at this time. For me, at any rate, it is much too difficult and I wish I were a film comedian or something like that and had never heard anything of physics.[228]

Pauli's composure improved by summer of 1925, but his letters suggest that he experienced wild swings in his mood. Excessive drinking can produce that behavior. In July 1925, as he wrote to H.A. Kramers in Copenhagen, he received the news of Heisenberg's creation of matrix mechanics "with jubilation":

I therefore wish him heart-felt success in his endeavour. Thus, I now feel less lonely than about a half a year ago when I found myself (spiritually and spatially) fairly alone between the Scylla of the number-mystical Munich School and the Charybdis of the reactionary Copenhagen coup [*Putsch*] propagated with zealotic [zealous] excesses by [Bohr, John Clarke Slater, and] yourself![229]

Pauli would continue his downward emotional slide. Much later he described this period in his life and in his physics as "a brief period of spiritual and human confusion, caused by a provisional restriction to *Anschaulichkeit*."[230] Pauli's period of emotional and spiritual confusion lasted, I suspect, at least until 1927 when he published his matrix interpretation of electron spin. His solution of the anomalous Zeeman effect rests on his exclusion principle, which is a cornerstone of the new quantum mechanics, but it is not derivable from it, and that left him dissatified and confused. I believe his work on the anomalous Zeeman effect and on his exclusion principle caused him to reflect on emotional and spiritual evidence for a deeper level of reality, for an alethic reality. I believe further that his emotional and spiritual confusion that accompanied his work here lasted far longer, until he sought out Jung in 1932. I see Pauli's emotional and spiritual confusion related to three fears affecting his physics: his fear of intuition and visualization; his numerological fear of the number 2; and his fear of failure when he was unable to rationalize the physical interpretation of his work. These three fears surfaced during his work surrounding the exclusion principle. All three may be interpeted as arising from Pauli's ESTJ consciousness confronting his INFP Shadow, as if his Shadow contained forbidden ideas, like the *Pauli Verbot* as Heisenberg called Pauli's exclusion principle.[231] I like Heisenberg's term because it connotes an authoritative command, as if a Platonic god had issued an edict.

---

[227] John L. Heilbron, "The Origins of the Exclusion Principle," *Historical Studies in the Physical Sciences* **13** (1983), pp. 299-300.

[228] Abraham Pais, *The Genius of Science* (New York: Oxford University Press, 2000), p. 224. Pauli wrote this about six months after his work on the exclusion principle, on May 21, 1925.

[229] Enz, *No Time to be Brief* (ref. 1), p. 133.

[230] *Ibid.*, p. 119.

In my view, there were three phases in Pauli's route to his exclusion principle that were related to his three fears above. The first covers the period when Pauli struggled to use the methods of classical physics and the old quantum theory to analyze the problem of the anomalous Zeeman effect, exhausted these accepted methods without solving the problem, and struggled against relying on his intuition and visualization. The second covers the period when Pauli found a temporary description of the anomalous Zeeman effect, which he regarded as a "swindle" because it only described the behavior of electrons in an atom, but to use it he had to embrace the number 2 and confront the pejorative numerological significance that he associated with that number. The third covers the protracted period when Pauli struggled and failed to find a physical explanation for his *Pauli Verbot*, which represented a huge failure for his psyche, because he had not been able to complete his quest for independence from his father, to achieve professional success, and thus to prove himself worthy among his peers. He never was able to find a full explanation of the *Pauli Verbot*, but at some point later in his life that realization failed to disturb him emotionally.

**Exhausting the Old Quantum Theory**

Max Jammer, John Heilbron, Abraham Pais, and most recently and insightfully Charles Enz have discussed Pauli's route to the discovery of his exclusion principle, and Henry

---

[231] *Ibid.*, pp. 128-129. Dirac termed it the exclusion principle.

Margenau has given a penetrating discussion of its philosophical implications.[232] Pauli himself discussed the route to his discovery in his Nobel Prize lecture.[233] All of these accounts, however, focused primarily on Pauli's route as expressed by his ESTJ personality, and did not probe the INFP side of his personality, which fed his creativity. Heilbron noted that Pauli later described his style of physics as a mixture of mathematics and mysticism.[234] If there ever was a period when this characteriszation of his style was warranted, then it was during his work on the anomalous Zeeman effect and the *Pauli Verbot*.

The anomalous Zeeman effect had puzzled physicists ever since its observation by Marie Alfred Cornu in 1897. The understanding of the normal Zeeman effect followed both from classical physics (H.A. Lorentz's account of it in 1897[235]) and from Niels Bohr's *Aufbauprinzip*, governing the building up of the periodic table of the elements based upon extensions of his model of the hydrogen atom. By the early 1920s, the increasingly complicated electronic orbits resembled beautiful artwork that was striking in its detail and kernel-like aesthetic appearance, but lacked a fundamental physical interpretation.[236] It seemed that only Bohr could employ the Copenhagen methodology successfully:

> "Hafnium content" became a synonym for whatever sense there might be in the increasingly crazy theories proposed in 1923 and 1924 to extend Bohr's conceptions to the finer details of spectral and atomic structure.[237]

There were few trustworthy experimental or theoretical tools available at the time to probe the inner world of atoms.

Pauli first had to come to terms with the spectral regularities that led Alfred Landé to introduce his *g*-factor relating the orbital angular momentum of the electron *l*, the angular momentum of the atomic core [*Rumpf*] $R$, and the total angular momentum $j$, which accounted for the anomalous spectral-line splittings. Pauli was unable to do so. However, he found a phenomenological interpretation of Landé's *g*-factor. "Heisenberg thought this [Pauli's] reckoning of the *g*-values 'extraordinarily beautiful.' Pauli thought the whole work abominable."[238] That was because Pauli found that he had to assign half-integer quantum numbers to the angular momentum of the atomic core, and since half-integer quantum numbers had no place in the old quantum theory, they lacked a physical foundation. Pauli's ESTJ personality was being confronted by his INFP Shadow.

In his individuation quest, Pauli had chosen one of the most difficult problems in atomic physics for his ESTJ personality to address. As Pais marked: "In the early twenties the understanding of the Zeeman effect was marred not only by ignorance about spin; there were experimental complications as well."[239] In the normal Zeeman effect a spectral line split into three components under the influence of a magnetic field, which could be explained both classically and on the basis of the old quantum theory. The anomalous Zeeman effect defied such explanations. Pauli succeeded in explaining it phenomenologically only by assigning a "classically non-describable two-valuedness [*Zweideutigkeit*]" to the angular momentum of the atomic core. In 1923 Landé had modified his *ad hoc* formula to read as follows:

$$g = 1 + \frac{j(j+1)+R(R+1)-l(l+1)}{2j(j+1)},$$

where $j$, $l$, and $R$ are various angular momentum quantum numbers.[240] Further, Landé and Heisenberg found that the core angular momentum $R$ had to have half-integer quantum numbers for Landé's formula to reproduce the experimental spectroscopic data. Ernst Mach might have been pleased, but Pauli's operationalist psyche needed a trail back to observables, and Pauli's INFP Shadow needed more--a visualizable model.

Pauli failed to find a physical foundation for the two-valuedness, the half-integer values of the core angular momentum $R$. He knew that an understanding of the anomalous Zeeman effect would lead to deeper understanding of the behavior of electrons in complicated atoms, and that a physical understanding of the two-valuedness was pivotal, but he found that this problem was too difficult for him to solve. Nonetheless, he could not put it out of his conscious or unconscious mind. He speculated that relativity theory might need to be taken into account in addition to quantum theory, but after a complex but logical process he concluded that the behavior of the atomic core was not the source of the anomalous Zeeman effect. As he wrote in a letter to Landé:

> I might almost say that this supposed core momentum has not only half-integral values, but also only a half-reality.... A future, proper (therefore not Heisenberg) theory of the anomalous Zeeman effect will have to take into account the non-existence of the relativistic correction factor.[241]

To Pauli, the theoretically unsubstantiated but empirically required half-integer quantum numbers were preposterous. As he wrote to Bohr:

> The atomic physicists in Germany can now be divided into two classes. Some work out a given problem first with half quantum numbers, and if it doesn't agree with experiment, they do it again with integral ones. The others calculate first with integral values, and if it doesn't work, do it again with halves…. I cannot acquire a taste for this sort of theoretical physics, and am withdrawing to my heat conduction.[242]

That was in February 1924. By that fall, Pauli had reconsidered. As Pais put it:

> We have now arrived at one of those marvelous moments in science when the lessons of logic are in conflict with the lessons of the laboratory. The idea of a core angular momentum is nonsense, Pauli had shown. But the Landé formula … works, experiment had shown! Pauli found the correct way out: we need [the Landé formula], but since [it] has nothing to do with the core it must have something to do with the valence electron itself. In Pauli's words, the anomalous Zeeman effect "according to this point of view is due to a peculiar not classically describable two-valuedness (*Zweideutigkeit*) of the quantum theoretical properties of the valency electron."[243]

---

**Confronting *Zweideutigkeit***

Pauli's next step to his *Pauli Verbot* was to reinterpret Landé's scheme for the empirically known spectral-line splittings in the anomalous Zeeman effect, as if he had heard Ernst Mach asking him to provide a positivist-operationalist interpretation. Perhaps his father's voice too was in his mind as he prepared to visit his home in Vienna over the Christmas holidays of 1924. Sommerfeld had pointed out Edmund Stoner's work in experimental spectroscopy to Pauli, which gave Pauli insight. Stoner's data-classification scheme assigned the quantum number $R$ not to the atomic core, but to the electron: The electron was the source of the *Zweideutigkeit*. Pauli's ESTJ side of his personality had to accept the evidence, yet it was still unacceptable because his ESTJ side needed a physical basis for it: It was still another swindle. Further, for Stoner's scheme to work, a fourth quantum number had to be introduced that could take on only one of two values, either $+\frac{1}{2}$ or $-\frac{1}{2}$, and no two electrons could have the identical set of all four quantum numbers in the atomic "kernel."

Pauli's ESTJ consciousness was in conflict with his INFP Shadow. His INFP side required beautiful aesthetic visualization, which he had now partially achieved. His INFP side valued Sommerfled's intuitive and aesthetic methods, and awareness of them now partially entered his consciousness. Pauli's exclusion principle, his *Pauli Verbot*, appealed to his nascent sense of aesthetics and numerology, and it extended Bohr's *Aufbauprinzip,* which now could be based on four quantum numbers, but it had no theoretical explanation. Here, however, was an aesthetic element that allowed physicists to visualize Bohr's *Aufbauprinzip* as a kernel with four components. In my concluding chapter I will speculate that Pauli had a vision of just such a kernel.

In 1951 Pauli declared that introducing a fourth quantum number, going from three to four, was the most difficult step that he had to take to decipher the anamolus Zeeman effect:

> I did hit upon [Johannes] Kepler as trinitarian and [Robert] Fludd as quaternian---
> and with their polemic, I felt an inner conflict resonate with myself. I have
> certain features of both …. By the way, I wish to remark that once (in Hamburg)
> my path to the Exclusion Principle had to do precisely with the difficult transition
> from 3 to 4: namely with the necessity to attribute to the electron *a fourth* degree of freedom (which soon after was explained as "spin") instead of the *three* translations. To convince myself that, contrary to the naïve "conception" (*Anschauung*) the fourth quantum number also is a property of *one* and the same electron (on a par with the well-known *three* quantum numbers now designated as $n_r$, $l$, $m_l$)---*that was really the main work*. (I had to wrestle so much with the then accepted theories which attributed the fourth quantum number to the *atomic core* (*Rumpf*)).[244]

Pauli's ESTJ reasoning was clear, but his conclusion was troubling:

 In a puzzling, non-mechanical way, the valence electron (of an alkali atom) manages to run about in two states with the same [orbital quantum number] $k$ [earlier $l$] (but) with different [magnetic] moments.[245]

Pauli's ESTJ reasoning was in conflict with his INFP aesthetics. The inexplicable *Zweideutigkeit* of the electron was in conflict with Bohr's correspondence principle and with the assignment of only three degreees of freedom to the spatial motion of the electron. A deeper conflict also was surfacing. Pauli was beginning to form an intuition that the imagery of classical electron orbits was obsolete, but that left him with the inexplicable *Zweideutigkeit* of the four quantum numbers and transported him into the realm of the mysterious. Sommerfeld had used the term "cabalistic" to describe the numerology of the atomic quantum numbers.[246] Pauli had exerted enormous mental effort in solving the puzzle of the anomalous Zeeman effect, yet his result was still a "swindle" that forced him to revise his fundamental, rationally derived picture of the atom. His ESTJ side had reached the end of its power; his INFP style invoked pejorative connotations. He would have to reconsider heretical ideas. The *Pauli Verbot* could not be derived rationally; it simply had appeared as a law of Nature, reminiscent of Moses's commandments in the Old Testament.

Pauli has left few clues to help us understand his concerns about *Zweideutigkeit*. I can speculate on some of the difficulties he faced, however, based on his later writings. He spoke of "spiritual and human confusion" when describing his earlier work and invoked several psychological concepts, many of them Jung's. I believe his spiritual and human confusion concerned his methods of theoretical physics, his relationships to women, his inappropriate nightlife, and his spiritual grounding. He was trying to establish his style of theoretical physics, which troubled him because his intuition now was in conflict with his early training. His troubled relationships with women, which are normal for young men, were exacerbated by his lack of social skills. His use of alcohol and his nightlife conflicted with the messages of his parents on propriety and social responsibility. In his spiritual confusion over *Zweideutigkeit*, Pauli struggled with his innermost thoughts on the meaning of his life. In German the word *Zweideutigkeit* has connotations of numerological significance. We recall Heisenberg's reflections:

> This doubling of states which Pauli first had called the unmechanical doubling, was actually connected with the Lorentz group. But later on, as you know, one had found doublings which had nothing to do with the Lorentz group, say the iso-spin doubling, neutrons and protons. This doubling itself was something which Pauli liked. Therefore he was not too happy about the electronic spin. The idea that one should be forced, in such a discontinuous theory as quantum theory is, simply to double, to confront an alternative, either this or that, appealed to a very fundamental feature in Pauli's philosophy ….

> "Verdoppelung und Symmetrieverminderung. 'Das ist des Pudels Kern'." That is, "The fundamental principle from which all nature is produced is doubling of states and then, later on, reduction of symmetries." He adds, at this point, "Verdoppelung ist ein alter Zug des Teufels." In the whole medieval philosophy of the Alchemist the devil is, of course, the one who would double things. Then he adds that the devil is, of course, the one who makes doubts, hesitations, and the word "Teufel" has to do with "Zweifel," which, in the old time, meant doubling, i.e. you can do either this or that. So Pauli says that to be put in front of an alternative and to double the possibilities is an old and most fundamental feature of the devil. In this way, the devil has created the world. Pauli loved to talk about these things.[247]

Heilbron notes further:

---

[247] See Chapter 1.

> Pauli himself later fancied that the jump from three to four had been the chief, and hardest, step of all. It required, he said, a shift against the ingrained psychological prejudices not only of physicists, but of the race, the shattering of an Idol of the Tribe.[248]

A list of the emotionally laden issues in this period of Pauli's life where doubt, double trouble, and choices all somehow related to *Zweideutigkeit* might include the following: his waning Catholicism *versus* his coming to terms with his Jewishness as anti-Semitism was increasing in Germany and Austria; his emotional investment in the Catholic Trinity *versus* the Jewish cabalistic numerology associated with the number 4 or the pagan Pythagorean tetractys; his efforts to choose between rationality *versus* spirituality in various facets of his life; his concerns over his masculinity *versus* his perceived feminine traits and thoughts; his emulation of his father *versus* his mother; his following the rigorous rules of logic and reason *versus* his intuition; and his efforts to choose between his forced ESTJ personality *versus* his natural INFP personality.

This list might be extended. Pauli was encountering the deepest roots in both his physics and in his personal philosophy. Since to my knowledge he never disclosed his deepest thoughts and feelings about this period in his life, I can only speculate on them by examining his outward actions. We know that he soon withdrew from the Catholic church, and later aligned himself with Judaism. He became invested in the numerological significance of the number 4, calling his personal philosophy not trinitarian but "quaternian." He later found purely rational physics less appealing than spiritualistic ideas. He associated rationalism and empiricism with masculine qualities, those of Jung's animus; intuition and emotionalism were feminine qualities, those of Jung's anima, to which he began to align himself: He later described himself as more intuitive than rational. He was much closer emotionally to his mother than to his father; they still were married but he may have detected marital strife. After his discovery of the *Pauli Verbot* he became attracted to not repulsed by "doublings," as Heisenberg indicated. To account for all of these developments, I maintain that Pauli's personality became more openly of the INFP type, slowly eclipsing that of his earlier ESTJ type.

---

[248] Heilbron, "Origins" (ref. 9), p. 309. Note that Pauli was not under the influence of Jung until 1932, some seven years after his discovery of the exclusion principle.

This process initially caused him to experience spiritual anguish as he began to recognize the full implications of "doubling." This process soon would be disturbed by his mother's suicide.

Later, Pauli intuitively felt a sense of balance when he visualized mandalas marked by a four-part division. When he later emphasized "one" in his writings, he was referring to mandala wholes. To Pauli, the number 3 represented pure reason, empiricism, and logic, something akin to Kepler's way of doing physics where numerology was a tool but removal of nonrationality was a goal; the number 4 represented Fludd's addition of a spiritual, nonrational element into the otherwise purely physical 3 to make a quaternity whole, the "one" mandala. In physics, for example, the three spatial dimensions had to be supplemented by a time dimension involving the imaginary number $i$ to complete the wholeness of Minkowski's space-time. Prior to the discovery of his *Pauli Verbot*, he had not entertained these feelings in his physics, and thus their spiritual implications had not been pressing. After his discovery, *Zweideutigkeit* held far more meaning for him than as simply the number 2.

**Deciphering the *Pauli Verbot***

Pauli's external and internal worlds erupted soon after his work on the anomalous Zeeman effect and the discovery of his *Pauli Verbot* in 1924-1925. The later Pauli (after meeting Jung) might call that period one of psychic unsettlement and premonition. Pauli's father separated from his mother and his mother committed suicide on November 15, 1927.[249] Pauli moved from Hamburg to Zurich in 1928 in a state of despair.[250] He left the Catholic church in 1929.[251] He married on December 23, 1929, in Berlin, and the marriage fell apart soon thereafter.[252] He continued his unbridled Hamburg nightlife in Berlin and Zurich. He had difficulty coping with his emotional stress. His behavior resembled that of an addict.

If Pauli was aware of the marital strife of his parents, he would have had unsettling psychic incidents such as dreams, suspicions, and feelings that would have been difficult for him to process emotionally, and would have impinged on his ability to think rationally. The Catholic faith no longer brought him comfort. I see no indication that Pauli found emotional comfort until he visited the Jung Clinic in 1932. Prior to 1932, I thus think that Pauli might have begun to regard physics, not as concerned entirely with inert mechanisms, but as a substitute for his spiritual needs. Physics became invested with vitalism, god, and the devil; he became a reluctant and frightened Platonist after his discovery of the *Pauli Verbot*. He continued to look for a physical explanation of it from first principles, to transform it from a mystical revelation into a rational principle. He was unsettled by that quest because he found no explanation that would satisfy his requirements of rationality. He had reached an impasse. Later, he attempted to achieve deeper insight into the *Pauli Verbot* in terms of electron spin, but even here he failed to derive it from first principles. The *Pauli Verbot* was an enigma. His only recourse was to change the way in which he perceived it.

---

[249] Enz, *No Time to be Brief* (ref. 1), p. 10.
[250] *Ibid.*, p. 164.
[251] *Ibid.,* p. 211.
[252] *Ibid.,* p. 209.

Pauli discovered the *Pauli Verbot* in late 1924 before Heisenberg created the new matrix mechanics, but after Pauli had shown that the old quantum theory was inadequate. Pauli had been struggling to rid the old quantum theory of its unobservable and nonrational electronic orbits. Instead, with the *Pauli Verbot* he introduced yet another nonrational element into it: It was not derivable from known physical principles in 1924, and Pauli could not derive it completely to his satisfaction by the time he won the Nobel Prize for it in 1945. Even today a full derivation from first principles remains problematic.[253] I will treat it as I suspect Pauli viewed it: The *Pauli Verbot* stands out as a law imposed upon quantum physics; it simply *is*. The later understanding of the principle in terms of symmetric and antisymmetric wave functions replaces the original nonderivable edicts with deeper, but still problematic, ones. The significance of that view was not likely lost on the philosophically minded Pauli either in 1924 or in later years. Pauli shared with Bohr his feelings about the nonsensical nature of the *Pauli Verbot* in a letter of December 1924. Heilbron has quoted from that letter in the context of his discussion of Pauli's bewilderment:

---

[253] Jammer, *Conceptual Development* (ref. 14*)*, p. 145; Margenau, "Exclusion Principle" (ref. 14), pp. 187-208; *Nature of Physical Reality* (ref. 14), pp .427-447.

Pauli had been prepared for disappointment. He worried about squaring his rule with the correspondence principle, and about the relation between explanations by electronic duplicity and by magnetic (or relativistic) interaction. Elucidating the closure of electron groups and the lengths of Mendeleev's periods by correspondence had been the high goal of Bohr's atom building. Pauli doubted that it could be reached. "I think that everything really is much simpler; one does not have to talk about harmonic interplay [of the electrons of the subgroups]." Still the point bothered him. His paper on exclusion ends with the admission, which reads as an apology, that his results gave no hint how closure might be brought into connection with the correspondence principle.

He took with equal seriousness the cleft between his and the magnetic (or relativistic) explanations of the doublet structure. Here want of a model made the whole analysis "definitely nonsense." "But I think [Pauli wrote to Bohr] that it is no greater nonsense than the usual conception of the complex structure." Both had a bit of truth. "The physicist who succeeds in adding these two nonsenses will have the truth!" And how should one begin this odd summation? By dropping the concept of electron orbits. "I believe that energy and momenta values of the stationary orbits are something much more real than orbits. The (still unattained) goal must be to deduce these and all other physically real, observable characteristics of the stationary states from the (intergral) quantum numbers and quantum theoretical laws." So far did Pauli's renewed wrestling with atomic physics take him toward enunciating the program that led to quantum mechanics.[254]

Pauli struggled intensely and emotionally in trying to find a deeper meaning of the *Pauli Verbot*. He had reached his limit in his rationalistic physics; his ESTJ consciousness had encountered resistance and had to acknowledge the power of his latent INFP Shadow. In his Shadow, there seems to have lurked a curious mixture of religious feelings toward Catholicism and Judaism, coupled with his feelings to his idol Sommerfeld's numerology, all contributing to a nonempirical intuition that was bearing fruit. In his intuitive explorations, he could see elements of an alethic reality that held the potential

---

for a new quantum mechanics.  Experiment and positivism no longer were working for him; intuition and numerology as seen in his "mind's eye" were.  He had reached the limits not only of the old quantum theory, but of his way of doing physics.  I think that Pauli had a mystical experience that changed his view of physical reality forever, resulting in Platonic insights, but in his new view he saw demons instead of angels.

Heilbron and others have discussed the reactions of physicists to the *Pauli Verbot*, and these too carry traces of mystical connotations.  Sommerfeld felt that Pauli had found the genes of *Termbotanik* and *Spektralzoologie*.  Paul Ehrenfest regarded the *Pauli Verbot* as a blessed curse:[255] Pauli wanted it to be derivable from a new quantum mechanics, a desire that was never fulfilled.[256]  Heisenberg responded playfully on December 15, 1924:

> Swindle X  [times] swindle does not yield something correct and, therefore, two swindles can never contradict each other.  Therefore I congratulate !!!!!!!!![257]

Einstein, when he later recommended Pauli for the Nobel Prize of 1945, sent the following telegram to the Nobel Committee:

> nominate wolfgang pauli for physics-prize stop his contribution to modern quantumtheory consisting in so called pauli or exclusion principle became fundamental part of modern quantumphysics being independent from the other basic axioms of that theory stop albert einstein[258]

That Einstein acknowledged the fundamental and independent nature of the *Pauli Verbot* while still being convinced that quantum mechanics was incomplete is remarkable.  Pauli had been passed over several times earlier; now Einstein's recommendation was important in tipping the scale in his favor.[259]

In struggling to come to grips with the mystical connotations of the *Pauli Verbot*, Pauli had to admit his INFP intuitions and go beyond his ESTJ rationality.  Others have been challenged similarly by it.  Physicist-philosopher Henry Margenau discussed the broad scientific and profound philosophical implications of the Pauli exclusion principle as follows:

In a simpler form, the excluson principle (Pauli principle) requires that no two particles of the same kind--such as electrons, protons, neutrons--can be in the same state at any one time. The older quantum theories interpeted this as saying that no two particles can have the same full set of quantum numbers. The more accurate mathematical form of the principle [Margenau had discussed it in terms of symmetric and antisymmetric state functions] is shown to be equivalent to this statement under proper conditions.

There is much similarity between the present topic and the old assertion that two bodies cannot occupy the same place at the same time. On more careful investigation it turns out, however, that bodies avoid each other to the extent to which their velocities are alike. Perhaps the most spectacular application of the exclusion principle is to the "building-up" process of the elements … where it is shown that different atoms owe their characteristic features to a kind of social behavior of the electrons which may be summed up by saying: One electron knows what the others are doing and acts accordingly. And this knowledge is not conveyed by forces, or dynamic interactions, of the ordinary kind.

The exclusion principle introduces a correlation into the behavior of particles which, though its effects are similar to the effects of forces, has no explanation in dynamic terms. The resemblance with the principle of relativity is strong: relativity succeeds in accounting for certain forces (gravitational forces) by reference to the formal requirement of invariance; exclusion succeeds in accounting for other forces (*e.g.* chemical valence) by reference to the formal requirement of antisymmetry. Both are requirements imposed upon the manner in which we formulate our experience.

Leibnitz' principle of the "identity of indiscernibles" is considered from the point of view of modern physics and is shown to fail inasmuch as two electrons, which differ in no observable respects, remain nevertheless two entities.

Finally, [one of my earlier paragraphs] surveys the prospects offered by the exclusion principle--and possibly other principles of the same formal kind--for

understanding the problems of living matter.  These prospects seem very promising.[260]

If Margenau could see the principle's applicability to biological systems, perhaps Pauli could see its applicability to his father's specialty of horse-serum albumen and proteins. Historian-philosopher Max Jammer comments further:

> Thus it became clear that the principle of exclusion, like that of relativity, is not merely another theorem in physics but rather a general precept regulating the very formation of physical laws ….
>
> With the exception of a study by Henry Margenau philosophical analysis has so far shown little interest in the exclusion principle in blatant contrast to its inquisitiveness concerning the relativity principle….  Margenau's discussion on the relation between the Pauli principle, on the one hand, and the principle of causality, on the other, deserves still today serious consideration on the part of every physicist interested in the foundations of his science.[261]

When is a force not a force, and when does the law of causality apply?  Pauli might have asked himself such questions already in 1925.

Arthur Koestler, a novelist fascinated with mysticism who was originally trained in physics in Vienna, attempted to find in the exclusion principle support for his attempts to legitimize parapsychology and extrasensory perception (ESP).   Here was a noncausal requirement imposed on all of physics, and thus of fundamental importance for how material systems are formed.  The inert constituents of matter--electrons, protons, and so on--respond across the vastness of our space-time reality to its requirement.  No force, no tangible particles, no "quantities ... in principle observable" were involved, only a lofty imperative of order.  Here was an analogy to ESP.[262]  For Koestler, the exclusion principle, ESP, and synchronicity were all variations of the same phenomenon.

Physicist-philosopher F. David Peat, a colleague of David Bohm's,[263] attempted to legitimize the concept of synchronicity in a similar vein.  Peat sought support for the

concept of sychronicity, which was introduced by Carl Jung in 1948 and won Pauli's approval:[264]

> Of all Pauli's contributions to physics the best known is his exclusion principle, an addition to Heisenberg's quantum mechanics which makes an interesting resonance to the general notion of synchronicity. Synchronicity … arises out of the underlying patterns of the universe rather than through a causality of pushes and pulls that we normally associate with events in nature. For this reason synchronicity has been called by Jung an "acausal connecting principle." But an acausal connection is exactly what was proposed by Pauli in his exclusion principle ….

> So Wolfgang Pauli's most famous contribution to physics involved the discovery of an abstract pattern that lies hidden beneath the surface of atomic matter and determines its behavior in a noncausal way. It is in this sense that the Pauli principle forms a parallel to the principle of synchronicity….[265]

Koestler, in a provocative thesis, believed that Jung stole the concept of synchronicity from the Viennese Lamarckian biologist Paul Kammerer.[266] Kammerer described a very similar phenomenon already in his book, *Das Gesetz der Serie*, of 1919. He was a well-known Viennese intellectual who moved in the same circles as Pauli's father and sister, and articles about him appeared in Pauli's mother's newspaper.[267] Could it be that Pauli inadvertently passed the concept of synchronicity on to Jung? Pauli certainly helped to reinforce Jung's idea. In his Hamburg days Pauli became aware of the "Pauli effect," which he firmly believed were psychic events. Pauli may even have read Kammerer's book, since he himself held Lamarkian views of biological evolution, feeling that pure chance was insufficient to explain it. Pauli's chemist father, however, seems likely to have opposed Lamarck's views, providing yet another difference between father and son. In any case, Pauli saw in the concept of synchronicity a direct link to Arthur Schopenhauer's philosophy, which he made clear in a letter to Jung of June 28, 1949:

---

[264] C.A. Meier, ed., *Atom and Archetype:The Pauli/Jung Letters* (Princeton: Princeton University Press, 2001), p. 36.
[265] Peat, *Synchronicity* (ref. 44), pp. 16-17.
[266] Koestler, *The Roots of Coincidence* (ref. 44), pp. 82-104.
[267] Arthur Koestler, *The Case of Midwife Toad* (New York: Random House, 1973), p. 15.

> The idea of *meaningful coincidence--i.e.*, simultaneous events not causally connected--was expressed very clearly by Schopenhauer (1788-1860) in his essay…. There he postulates an "*ultimate union of necessity and chance*," which appears to us as a "force," "which links together all things, even those that are causally unconnected, and does it in such a way that they come together just at the right moment."[268]

To my knowledge, Pauli never connected his exclusion principle to sychronicity. He did talk about his exclusion principle and "action-at-a-distance," however, in a course on *Wellenmechanik* at the Federal Institute of Technology (ETH) in Zurich in 1956-1957, according to notes taken by his students:

> From a superficial consideration of the exclusion principle, it might be thought that a sort of action-at-a-distance is being postulated, as a result of which even two widely separated particles are aware of one another ("sign a contract"). However, this is not so, because the exclusion principle is valid only as long as the wave packets of the two particles overlap.[269]

Still, Pauli later called for a unification of physics and psychology, which would have entailed both his exclusion principle and the concept of synchronicity.

Pauli's Nobel Lecture of 1946 is particularly fascinating from the perspective of exploring connections between his physics and his cognitive states, especially in light of his emotional and philosophical struggles in formulating his exclusion principle. In 1924 he had reservations about the *Pauli Verbot*; by 1946 he had thought about its philosophical implications but was remarkably comfortable with these aspects of it. He recalls Sommerfeld's numerological ruminations about the wavelengths of the spectral lines of hydrogen and their connection to Kepler's and Rydberg's work. He emphasizes Sommerfeld's attempt to connect "the number 8 with the number of the corners of a cube,"[270] invoking an image of Kepler's Platonic solids. Earlier Pauli had rebelled against such unsupportable visual information; now Pauli emphasizes the significance of the word *one* and its relationship to the 4 quantum numbers of an electron:

---

> The fundamental idea can be stated in the following way: The complicated
> numbers of electrons in closed subgroups are reduced to the simple number *one* if
> the division of the groups by giving the values of the 4 quantum numbers of an
> electron is carried so far that every degeneracy is removed.[271]

Note that Pauli italicized the number word *one* but not the number 4.   The *one* became of
deep significance to him as symbol of the mandala whole, of the unity of a quaternian
physics and psychology.  Between 1924 and 1946 Pauli had made a transition from an
ESTJ to an INFP way of doing physics; he had encountered his Shadow.

In his Nobel Lecture of 1946, Pauli also discussed his attempts to find a deeper
justification for his exclusion principle.  He had explored the interconnections of spin,
statistics, and symmetry beginning in 1934.[272]  In his lecture, he discussed that particles
of spin 1/2, such as the electron, have antisymmetrical wave functions and therefore obey
Fermi-Dirac statistics, while particles of spin 1, such as the photon, have symmetrical
wave functions and obey Bose-Einstein statistics.  In his Nobel lecture, he noted a further
connection between spin and symmetry classes:

> By combination of the claims of relativistic invariance and the properties of
> field quantization, one step in the direction of an understanding of the
> connection of spin and symmetry classes could be made.[273]

Pauli regarded this as a further step in finding a deeper foundation for his exclusion
principle.  He discussed Dirac's equation of the electron in an intuitive INFP way.   Thus,
he recalled Dirac's use of rotational *Anschaulichkeit* in the hyperspace of four-
dimensional space-time, and pointed out the "two-valuedness" associated with an
electron and its antiparticle, the positron.  Dirac had revealed a new level of reality, a
manifestation of Pauli's quaternian *Hintergrundsphysik*.

Pauli now was comfortable with these developments.  His intellectual nemesis of
1924 was his intellectual glory of 1945.  At a banquet in Princeton in his honor, Einstein
praised Pauli as his "spiritual heir."[274]   By then, Pauli had consulted the art historian
Erwin Panofsky on the visual images in Kepler's works; he had been able to see the connections between his analysis of symmetry in physics and Jung's mandala symbolism. Pauli's INFP personality had been validated by 1945. Pauli thus was comfortable in calling for deeper inquiry into the alethic levels of physics. A transition had taken place. His INFP side to his personality was serving him well. He closed by calling for a fundamental explanation of the fine-structure constant, the absence of which indicated to him that quantum electrodynamics was still incomplete. Let me repeat a quotation I used earlier, now for purposes of making an additional connection. Just as he had criticized Weyl's old continuous field theory, Pauli now declared:

> I may express my critical opinion, that a correct theory [of quantum electrodynamics] should neither lead to infinite zero-point energies nor to infinite zero charges, that it should not use mathematical tricks to subtract infinities or singularities, nor should it invent a "hypothetical world" which is only a mathematical fiction before it is able to formulate the correct interpretation of the actual worlds of physics.
>
> From the point of view of logic, my report on "exclusion principle and quantum mechanics" has no conclusion. I believe that it will only be possible to write the conclusion if a theory will be established which will determine the value of the fine-structure constant and will thus explain the atomistic structure of electricity, which is such an essential quality of all atomic sources of electric fields actually occurring in Nature.[275]

Pauli's discovery of his exclusion principle, the *Pauli Verbot*, was his first truly creative act in physics. He had earned enormous respect before the discovery of his exclusion principle, stemming from the confrontation of his ESTJ and INFP sides, but respect for his creativity came with a struggle; his Nobel Prize in Physics was delayed until 1945. I believe he continued to struggle emotionally with his emerging INFP side to his personality, would not trust it, experienced another significant psychic episode when he proposed his neutrino hypothesis in 1930,[276] and continued to struggle until he sought out Jung in 1932. During the route to his discovery of the *Pauli Verbot*, elements of his philosophy of physics surfaced, which I call Pauli's Platonism. These elements are:

belief in doubling of states; attraction to quaternian symmetry in kernels; attraction to visualizable rotations of kernels in his "mind's eye"; numerological attraction to the numbers 1, 2, 4, and 137; belief in synchronicity; and belief in an alethic, Platonic reality.

In his numerology, Pauli should have included the number ½.  I have found no pertinent comments that Pauli made regarding whether or not he held the number ½ to be numerologically significant and distinct from the number 2, but we will see in the next chapter that the number ½ caused him great unrest when Goudsmit and Uhlenbeck discovered electron spin.  I believe the number ½ led Pauli to his formulation of electron spinors, which he may have started to explore before the discovery of electron spin.

## Chapter 6. Electron Spin, Spinors, and Matrix Mechanics, 1925-1927

**Intuitions, Premonitions, and Parental Tragedy**

After the publication of the *Pauli Verbot* in early 1925, Pauli's personal life was filled with intuitions and premonitions. He failed to act on most of them, relinquishing opportunities to release his creativity publicly during the formation of the new quantum mechanics. If my analysis of Pauli's psyche is correct, then I can suggest why he played only a secondary, supporting role in the creation of the new quantum mechanics. He was experiencing stress in part owing in part to his perfectionism in physics, thereby restricting his intuition. He demanded that his creative insights be both experimentally grounded and philosophically rigorous. Now, in his twenties, he was beginning to appreciate his natural INFP personality type while still holding onto his forced ESTJ type; he was struggling with his religious and philosophical identity; and he may have been aware of his parents' marital difficulties. The entire mess resulted in low self-esteem, low self-confidence, critical outbursts, frustration about his career, and hesitation to publish his creative ideas in theoretical physics. When he was hypercritical, his Shadow often was defending its turf from physicists who were working in similar areas, thus posing a threat to his hesitant creativity. Externally he was brilliant, harsh, and critical; internally he was fragile and insecure.

Carl Jung indentified Pauli's problematic behavior in 1932. I believe, however, that Pauli's behavior masked his brilliance earlier during the creation and development of the new quantum mechanics between 1925 and 1927. F. David Peat pointed out some of these features in Pauli's personality but assumed that they emerged around 1928.[277] I maintain that they emerged at least as early as 1925 and lasted until his treatment in the Jung Clinic in 1932. Jung described his patient as:

> a university man, a very one-sided intellectual. His unconscious had become troubled and activated; so it projected itself onto other men who appeared to be his enemies, and he felt terribly lonely, because everyone seemed to be against him…. [He] lived in a very one-sided intellectual way, and naturally had certain desires and needs also. But he had no chance with women at all, because he had no differentiation of feeling whatsoever.  So he made a fool of himself with women at once and of course they had no patience with him.

---

[277] F. David Peat, *Synchronicity: A Bridge between Matter and Mind* (New York: Bantam Books, 1987), pp. 17-20. Peat quotes Jung's descriptions of Pauli.

Peat has described Jung's method of treatment:

In his study of  psychological types, Jung argued that each person is the result of an equilibrim or balance between polarities.

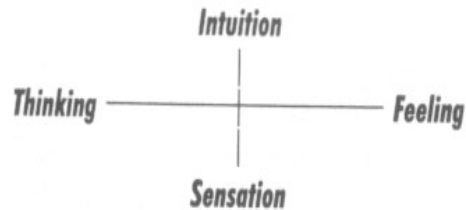

In a healthy psyche Thinking is in harmony with Feeling so that logic and reason can work in a constructive way with the emotional side of an individual.  However, in Pauli's case, thought had dominated Feeling so that the emotions were relegated to what Jung termed the Shadow side of the Ego.  In other words, Pauli's emotional and Feeling nature had never fully developed but existed in a raw and highly energized form which tended to break through in the form of irrational behavior, dreams, and neuroses.  Thought, sensing what it felt to be primitive forces at work, put the lid on even tighter so that Feeling found itself in the position of a red-hot pressure cooker with the valve jammed.  The result was Pauli's absurd marriage, his increasingly sarcastic attacks on colleagues, and his bouts of drunkenness.

According to Jung, the cure lay in bringing Feeling out of the Shadow and into the light, where it could perform its proper function and restore harmony to Pauli's whole personality.[278]

The result of Pauli's blocked Feeling, or in my terms the hidden INFP side of his personality, made him reluctant to publish his intuitive, creative ideas in theoretical physics.  Between 1925 and 1932, Pauli's creative ideas in physics were squelched by his ESTJ side, and Pauli gave them indirectly or directly to others such as Heisenberg.

---

Pauli's contributions to physics during this period were screened by his psyche, so that before he could express them publicly, they had to pass a comprehensive muster of empirical, philosophical, and logical tests. Similarly, in his personal life, Pauli led a two-sided day and night life. Externally he was viewed as a successful and composed physicist and personality; internally he struggled. His plight was quite understandable, as a suicide in a close-knit family affects all of the surviving family members in dramatic ways.

Pauli did not reside with his parents in Vienna during his twenties, but he was not estranged from them. I assume that Pauli remained close to his parents. He may have first observed their marital stress during his visit home for the Christmas holidays of 1924. They separated shortly before his mother's suicide in November 1927, when he again was active in theoretical physics, perhaps as a psychological escape mechanism. Prior to his parents' breakup, Pauli's father began a love affair with a much younger woman, the sculptress Maria Rottler, who was the same age as Pauli.[279] All of this must have been stressful to Pauli. For example, he was not having much success in his own attempts to have relationships with women at this time. I think it is quite possible that Pauli knew the gravity of his parents' situation before November 1927, which added to his stress and diverted him from trusting his intuition. He was putting many feelings into his Shadow and taking few out. I think it is remarkable that he accomplished as much as he did in theoretical physics during the these highly charged years before he met Jung. Perhaps in some strange way the intensity of his disturbed personal life contributed to his powerful creativity in theoretical physics. In my concluding chapter, I will briefly discuss Albert Rothenberg's view that creativity involves a melding of paradoxical opposites.

**Electron Spin**

Despite Pauli's reticence to follow or disclose his intuitions, he still had intuitive and creative ideas in physics between 1925 and 1927. There is good evidence that he was well in advance of others, in particular in his work on electron spin and in his contributions to the creation of quantum mechanics.  His strong requirements for the proper form of physics, his lack of trust in his intuitions and premonitions, and his unsettled personal life prevented him from publishing insights that might have brought him further public recognition.  Nevertheless, I feel that Pauli's creativity in using spin matrices that led to his spinor concept and his background role in developing quantum mechanics might have warranted a second Nobel Prize, if some historical contingencies had unfolded in slightly different ways.

Spin matrices use imaginary numbers within 2 x 2 matrices to treat rotations of objects, as in the classical-physics problem of the spinning top.  Spinors are two-valued wave-function solutions to quantum mechanical problems, and involve Pauli's spin matrices in their mathematical structure.   In the still poorly understood history of Pauli's introduction of spinors to describe the electron, it is clear that Pauli was stimulated to develop them as an extension of the mathematical power of spin matrices.

Historical discussions of Pauli's insight on *Zweideutigkeit* and the electron's fourth quantum number typically segue into a discussion of electron spin.  I did not discuss electron spin in the previous chapter because I now want to recapture Pauli's early thoughts on electron spin and quantum mechanics.  If  I am correct about his emerging Shadow and his burgeoning Platonism during this period, then Pauli's work on electron spin and quantum mechanics is related to his changing philosophical convictions and psychological makeup.  He initially opposed the concept of electron spin vehemently, apparently holding out for a deeper explanation of the two-valuedness of the electron's rotational angular momentum.  I believe that Pauli's resistance is connected to changes in his philosophical convictions and psychological makeup that began during his work on the *Pauli Verbot,* and spilled over to his subsequent work.  If one considers the discovery of electron spin and the creation of quantum mechanics as largely independent efforts, then Pauli's role in both appears to be only a secondary one.  I propose an alternative interpretation.  I believe that Pauli viewed the formulation of a new quantum mechanics and the explanation of the electron's *Zweideutigkeit* in its rotational angular momentum as constituting one and the same problem, or at least as one that could be approached using similar methods.  In this perspective, Pauli's role in both is elevated from a secondary to a primary one, since he shared his insights with Heisenberg, who adapted them to create matrix mechanics.  Additionally, Pauli stimulated, clarified, and created tools that others, notably Heisenberg and Dirac, used and reframed in their contributions to quantum mechanics.  I propose that Pauli viewed in his "mind's eye" both matrix mechanics and electron spin as rotations in alethic subdimensions, and that he shared his ideas, notably with Heisenberg, but owing to his reticence to publish he allowed Heisenberg and others to capitalize on them.  In this perspective, the mystery of Heisenberg's incorporation of matix mathematics into Hamilton-Jacobi methods can be partially explained as coming from Pauli.

Instead of focusing on Pauli's initial resistance to the concept of electron spin, I believe that he should be acknowledged for the role he played in deciphering its full complexity.  Throughout his career, in fact, Pauli continued to extend our understanding of spin as a key characteristic of elementary particles.  His contributions include the following: He conceived the concept of nuclear spin, as he recalled, before Goudsmit and Uhlenbeck introduced the concept of electron spin, but he did not describe it clearly in public.[280]  He argued for a deeper explanation for the *Zweideutigkeit* of the electron's rotational angular momentum, because he recognized that the idea of a spinning electron was too simplistic--spin had to be a quantum, not a classical characteristic.  He fully appreciated the importance of clarifying the Thomas factor 2 that had plagued the simplistic acceptance of the concept of electron spin.  He went on to brilliantly introduce spin matrices and spinors to quantum mechanics which deepened the concept of electron spin and provided the key for Dirac's ideas.  He introduced the neutrino hypothesis, which entailed spin as a conserved quantum entity, a new alethic entity without precedent, and one that provided the basis for Enrico Fermi to conceive his theory of beta decay.  And his work on spin and statistics tied these two concepts together in a new way.

According to the usual historical account, soon after Pauli explained the anamolous Zeeman effect phenomenologically by recognizing the *Zweideutigkeit* of the electron's rotational angular momentum,  Ralph de Laer Kronig interpreted it physically by assuming that the electron was spinning on its axis.  Kronig informed Pauli of his physical explanation--and was unprepared for Pauli's sharp criticism of it, so Kronig did not publish.  Kronig's calculation was off by a factor of two, which was resolved only later by Llewellyn Thomas in 1926.[281]  Goudsmit and Uhlenbeck also had a serious reservation, namely, H.A.Lorentz had pointed out to them that the spinning electron's peripheral velocity would exceed the speed of light, so they intended to withdraw their paper from publication, but Paul Ehrenfest already had submitted it without their knowledge.

Prior to Thomas's resolution of the factor of two, Pauli had resisted Kronig's and Goudsmit and Uhlenbeck's physical interpretation of electron spin.   His resistance seems confusing, since he was completely familiar with the concept of quantized angular momentum in physics, and even had introduced the concept of nuclear spin earlier in August 1924.  As Pauli recalled in his Nobel Lecture:

> I may include the historical remark that already in 1924, before the electron spin was discovered, I proposed to use the assumption of a nuclear spin to interpret the hyperfine-structure of spectral lines.  This proposal met on the one hand strong opposition from many sides but influenced on the other hand Goudsmit and Uhlenbeck in their claim of an electron spin.[282]

Goudsmit and Uhlenbeck never recalled any such influence.  Pauli thus seems to have remembered something that resided in his brain but had not emerged from it.  Pauli's memory seems to have been faulty, but since he did publish the concept of nuclear spin, the question remains why he did not see it as a physical interpretation of the *Zweideutigkeit* of the electron's rotational angular momentum.  The answer seems to be that in his work that led to his discovery of the *Pauli Verbot*, he had pushed classical physics and the old quantum theory to their limits, and now could not accept such a simplistic, classical interpretation of the electron's rotational angular momentum.   He had shown already in his doctoral thesis that the spectrum of the helium atom could not be understood on the basis of the old quantum theory.  Kronig's and Goudsmit and Uhlenbeck's classical interpretation of electron spin was simply too naïve for him to accept it.  As his close friend Heisenberg later reflected when discussing Pauli's paper on the *Pauli Verbot*:

---

I'm sure that this side of his philosophy must have played its role already in '24 when he wrote this paper on the "unmechanische Zwang." Therefore, he wasn't too happy this dissolved into a rather trivial angular momentum of an electron. In so far, he also approved of the doubling which I then tried in the iso-spin case and therefore also the doubling which occurred in the theory of elementary particles.[283]

Kronig's and Goudsmit and Uhlenbeck's classical interpretation of electron spin did not satisfy either Pauli's ESTJ needs or his emerging INFP intuitions. To Pauli, there was an insufficient logical basis for a classically spinning electron in that it was not observable in principle; there was still an insufficient rational basis for his INFP intuition of a spinning alethic reality. He would have thought that electron spin was a concept rife with entanglements about the nature of quantization.

Pauli was particularly critical of other physicists who were working along lines of research similar to his own. He later critized the ideas of Louis de Broglie, Dirac, Einstein, Res Jost, and David Bohm. Now his criticism was directed at the ideas of Kronig and Goudsmit and Uhlenbeck. Pauli had to ruminate over every detail until he achieved complete clarity in his mind, so his unconscious competitiveness tended to denigrate others and express itself as outbursts of his Shadow defending its turf, in this instance against Kronig's and Goudsmit and Uhlenbeck's concept of electron spin--a concept he likely already had considered and discounted as possessing insufficient explanatory power. He easily could have calculated that the electron's peripheral velocity exceeded the speed of light and also could have recognized that the Thomas factor of two constituted a problem. To Pauli, electron spin had to be a quantum rather than a classical concept (verified later by Bohr[284]), since Pauli's ESTJ side already had concluded--reminiscent of his analysis of the Bohr-Sommerfeld *Anschaulichkeit* models of the atom and of Hermann Weyl's unified field theory--that classical models not grounded on "quantities that are in principle observable."[285] Nor did electron spin go sufficiently far to satisfy Pauli's aesthetic needs stemming from his INFP intuitions. To

Pauli in 1925, the concept of electron spin smacked of classical models that lacked a rigorous operationalist basis,[286] as demanded by his ESTJ side. I suspect he already was speculating on a deeper explanation of it to supplant the classical naïve visualizations of electron spin, perhaps in a deeper alethic reality similar to Minkowski's relativistic space-time rotations. His earlier resistance to a fourth quantum number for the electron also was a part of this thought pattern. I think that his INFP side was already then on the trail toward his representation of electron spin that incorporated spin matrices. His INFP side saw its potential beauty as a rotation in alethic dimensions, its quaternion components, and the ubiquitous numerological sign of ½; his ESTJ side, however, required a logical and empirical basis for his spin matrix representation before he dared to publish it.

B.L. Van der Waerden has given two main reasons for Pauli's reservations about electron spin:

> From Pauli's letters to Bohr and from his Nobel prize lecture we may conclude that Pauli's main arguments against the spin hypothesis were: (1) the wrong factor 2 in the doublet splitting formula, which did not disappear when Heisenberg and Pauli made the calculations anew by means of Heisenberg's new quantum mechanics; (2) the classical mechanical character of the hypothesis of the spinning electron. Pauli's Socratic daimon, his intuition (as we call it today) told him that the 'two-valuedness' of the electron is a typical quantum effect which cannot be described in terms of classical mechanics.[287]

Van der Waerden continued by noting that electron spin indeed is a quantum phenomenon. He concluded that, "In my opinion, Pauli and Heisenberg cannot be blamed for not having encouraged Kronig to publish his hypothesis."[288]

## Spin Matrices, Spinors, and Rotations

Pauli published his spinor representation of electron spin, which incorporated his use of spin matrices, in May 1927.[289] His paper seemed to have no precedent in his earlier work, except that he adapted spin matrices to Heisenberg's and Schrödinger's quantum mechanics, which he had helped to establish.   I suggest, however, that Pauli began to think about spin matrices before Heisenberg created matrix mechanics but declined to publish his ideas until he had proof and a logical justification for them, thus delaying his publication of them.  Between 1924 and 1932, he was unsettled emotionally and dared not publish insecure ideas until he had considered and evaluated all of their aspects. I speculate that before he published his *Pauli Verbot* in early 1925, he had attempted to understand it by reconsidering Klein and Sommerfeld's mathematical treatment of a spinning top, where they had employed matrices and quaternians to analyze its rotation. Pauli was analyzing rotations already in 1924, since he recalled later that he had already considered the nucleus as a spinning body.  Soon after he published the *Pauli Verbot*, however, physicists focused their attention on Heisenberg's matrix mechanics, and Pauli's attention too was likely diverted--for example he solved the hydrogen problem using matrix mechanics, putting off his exhaustive efforts on spin matrices as applied to the electron.

I believe that in spin matrices we can see the roots of Pauli's philosophy and style of theoretical physics.   He was the first physicist to introduce them into quantum mechanics.  Spin matrices are visualizable in an abstract mathematical space involving the complex number *i*.  Pauli showed that the electron can be represented mathematically by spinors, two-valued wave functions, that use spin matrices in their formulations.  The "Pauli spin matrices" are 2 x 2 matrices that together form a kernel whole with four components.  The three Pauli spin matrices together with the unit matrix form a set of four independent matrices.  Herbert Goldstein has pointed out that, "Each of the Pauli spin matrices is therefore associated with rotation about one particular axis and may be thought of as the *unit rotator* for that axis." [290]  He notes further that:

> Characterisitic of the Cayley-Klein parameters, and of the matrices containing
> them, is the ubiquitous presence of half angles, and this feature leads to some
> peculiar properties for the *uv* space [the two-dimensional complex number space in which a transformation from one to another involves the Cayley-Klein parameters and spin matrices]. For example, a rotation in ordinary space about the $z$ axis through the angle $2\pi$ merely reproduces the original coordinate system. Thus, if in the **D** matrix …, $\varphi$ is set equal to $2\pi$, then cos $\varphi = 1$, sin $\varphi = 0$, and **D** properly reduces to the unit matrix **1** corresponding to the identity transformation. On the other hand if the same substitution is made in $\mathbf{Q}_\varphi$ … we obtain [the 2 x 2 matrices]

$$\mathbf{Q}_{2\pi} = \begin{array}{cc} e^{i\pi} & 0 \\ 0 & e^{-i\pi} \end{array} = \begin{array}{cc} -1 & 0 \\ 0 & -1 \end{array},$$

which is **-1** and not **1**. At the same time the 2 x 2 **1** matrix must also correspond to the three-dimensional identity transformation. Hence there are two **Q**-matrices, **1** and **-1**, corresponding to the 3 x 3 unit matrix. In general, if a matrix **Q** corresponds to some real orthogonal matrix then **–Q** also corrresponds to the same matrix. The isomorphism between the two sets thus involves, in this case, a one-to-one correspondence between the single 3 x 3 matrix and the *pair* of matrices (**Q**, **-Q**), and not between the individual matrices. In this sense one may say that the **Q** matrix is a *double-valued* function of the corresponding three-dimensional orthogonal matrix.[291]

This association of half angles and double-valuedness would have struck Paul as significant: quantum numbers of ½ and *Zweideutigkeit* are interconnected. Here was a mathematical connection to the electron's *Zweideutigkeit*. Spin matrices offered a potential mathematical and physical description of electron spin, but employed alethic parameters that are impossible to measure directly. They move the theoretical physicist away from the naïve visualizations of classical physics and stimulate interest in this deeper, more abstract mathematical formalism. Heisenberg emphasized Pauli's concern about *Zweideutigkeit* and his association of the Devil's handiwork in doubling. In spin matrices, the Devil's handiwork in doubling is represented mathematically: Spin matrices possess double-valuedness, and complex numbers exist in an imaginary realm, not a real one.

[291] *Ibid.*, p. 117.

On the occasion of Sommerfeld's eightieth birthday in 1948, Pauli acknowledged the influence of his mentor on his work:

> Sommerfeld's own creative scientific work forms only one part of his activity. The other part is his activity as an impressive and successful teacher. To an extent equaled by scarcely any other researcher he has given inspiration to an ever growing circle of disciples in Munich. This circle of disciples, dispersed over many countries on both sides of the Atlantic, among whom I gratefully count myself, takes care that the intellectual tradition which Sommerfeld passed on to us will be transmitted to academic youth and thereby to posterity. This tradition goes back to Sommerfeld's teacher Felix Klein, and through him also to [Bernhard] Riemann; indeed the grandly conceived work on the theory of the top, which Sommerfeld wrote with Klein, also contains the "Cayley-Klein rotation parameters" which have beome so important for the theory of spinors and hence also for Dirac's wave equations of the electron. It is not only with whole numbers that Sommerfeld's pupils will always feel at home, but also in the complex plane the use of which he, Sommerfeld, was so fond of in evaluating phase integrals and in discussing solutions of partial diferential equations.
>
> We know from the example of Kepler that the special feeling for harmony centered on the proportions of whole numbers must itself be brought into just relation as part of a larger whole, namely the advancing course of knowledge. While Kepler did not live to see the conceptual clarification brought about by Galileo's "Discorsi," and the interpretation of his own laws in Newton's "Principia," Sommerfeld has been able himself to take part in incorporating the harmonies he has discovered into the new conceptual system of quantum or wave mechanics. He may also be more fortunate than Kepler in the external circumstances of his life, in that it is now vouchsafed to him, after the completion of his eightieth year, to write up for publication, in tranquillity, his lectures delivered over a long period of years. I would not hesitate to set as superscription over Sommerfeld's works in a wider sense the title of Kepler's magnum opus— "Harmonices mundi."[292]

Pauli thus notes the significance of Sommerfeld and Klein's work for his own introduction of spin matrices and spinors and for Dirac's subsequent discovery of the relativistic equation of the electron. He also pays homage to Kepler and Sommerfeld for their Platonism, which Sommerfeld passed on to his students, including Pauli. Pauli again commented on his debt to Sommerfeld on Sommerfeld's death in 1951:

> The standard treatise on the "theory of the top," which he [Sommerfeld] wrote in conjunction with his teacher F. Klein in his early days, while he was still a "Privatdozent" in Göttingen, and in which many technical problems are discussed, possesses a significance going far beyond applied mathematics. It contains, on the basis of work by Euler and Cayley, and of Hamilton's quaternions, the essential foundations of what considerably later was called the theory of representations of the rotation group in three-dimensional space. In particular, Klein had, following Cayley, clearly worked out the relation of this group to the "covering group" of linear unitary unimodular transformations of two complex variables. Thus in this treatise, now a classic, the mathematical basis is developed for the two-component "spinors" which turned up much later in wave mechanics….[293]

Van der Waerden accepted Sommerfeld's contributions,[294] but he also emphasized Pauli's:

> [The] wave function … is equivalent to a set of $2^N$ functions of the space coordinates….
>
> In spite of these highly satisfactory results, Pauli regarded his theory as provisional and approximate only, because, as he says, his wave equation is not invariant under Lorentz transformations and does not allow the calculation of higher-order corrections to the fine structure of hydrogen terms.…
>
> It seems Pauli under-estimated the importance and the final character of his methods and results. His description of the state of $N$ electrons by a $\Psi$ function having several components which transform linearly according to a two-valued representation of a rotation group, was fundamental and final. It enabled

[Eugene] Wigner and [John] von Neumann to deduce all empirical rules of atomic term zoology without introducing any new assumption or any approximation…. Pauli's matrices …were used by Dirac to form a relativistic first-order wave equation [of the electron].  Dirac's wave equation contains matrices and is similar to Pauli's, but not to the old relativistic wave equation.  The step from one to two $\Psi$ components is large, whereas the step from two to four components is small; also, the step from vector algebra to a two-valued representation of the rotation group is large, the extension of this representation to the Lorentz group is much easier.  In all cases, it was Pauli who made the first decisive step, and in this part of his paper there is nothing provisional or approximate.[295]

I share Van der Waerden's deep appreciation of Pauli's introduction of spin matrices and spinors into quantum mechanics.  Pauli's creativity was overshadowed by Dirac's achievement.  I believe further that Pauli began to think about spin matrices before Heisenberg created matrix mechanics.  Thus, Pauli's 1924 reference to "classically nondescribable two-valuedness" of the electron I interpret as an indication that Pauli was already at that time attempting to decipher the atomic electron problem and the exclusion principle by using spin matrices.

**Matrix Mechanics**

Pauli's role in the development of matrix mechanics continues to be puzzling, and here again we gain insight by considering what was likely in his "mind's eye."  I do not believe he should be assigned a secondary role in this development.  Heisenberg's contributions to the creation of matrix mechanics were pivotal, but what was the basis of his creative leap?  Why did not Pauli create matrix mechanics instead of Heisenberg?  Pauli was adept in mathematical physics, including the use of tensors, matrices, and the Hamilton-Jacobi formalism.  I believe that Pauli's involvement with Heisenberg and others was essential to the creation of matrix mechanics.

Steven Weinberg has discussed Heisenberg's role in the creation of matrix mechanics and the mystery surrrounding it:

---

[295] *Ibid.*, p. 223.

If there is any moment that marks the birth of quantum mechanics, it would be a vacation taken by the young Werner Heisenberg in 1925…. On Helgoland Heisenberg made a fresh start. He decided that, because no one could ever directly observe the orbit of an electron in an atom, he would deal only with quantities that could be measured: specifically, with the energies of the quantum *states* in which all the atom's electrons occupy allowed orbits, and with the rates at which an atom might spontaneously make a transition from any one of these quantum states to any other state by emitting a particle of light, a photon. Heisenberg made what he called a "table" from these rates, and he introduced mathematical operations on this table that would yield new tables, one type of table for each physical quantity like the position or the velocity or the square of the velocity of an electron. (More accurately, the entities in Heisenberg's tables were what are called transition amplitudes…. Heisenberg was told [by Max Born] after he returned to Göttingen from Helgoland that his mathematical operations on these tables were already well known to mathematicians; such tables were known as matrices and the operation … matrix multiplication. This is one example of the spooky ability of mathematicians to anticipate structures that are relevant to the real world.)…

If the reader is mystified at what Heisenberg was doing, he or she is not alone. I have tried several times to read the paper that Heisenberg wrote on returning from Helgoland, and, although I think I understand quantum mechanics, I have never understood Heisenberg's motivations for the mathematical steps in his paper…. Theoretical physicists in their most successful work tend to play one of two roles: they are either sages or magicians. The sage-physicist reasons in an orderly way about physical problems on the basis of fundamental ideas of the way that nature ought to be…. Then there are the magician-physicists, who do not seem to be reasoning at all but who jump over all intermediate steps to a new insight about nature…. In this sense, Heisenberg's 1925 paper was pure magic.[296]

Weinberg thus recognized (as many others have noted) that Heisenberg was unfamiliar with matrices but had used the rules of matrix multiplication in manipulating his tables.

How did Heisenberg come to know about these rules?  I suggest that Pauli in his dialogs with Heisenberg played a key role here.  Pauli was constantly in touch with Heisenberg prior to his visit to Helgoland.  The exact nature of Pauli's role is unclear, but I can suggest some possibilities.  I suggest Pauli mentioned his interest in using spin-matrix mathematics in the atomic electron "two-valuedness" problem to Heisenberg, without naming the mathematics as such, and thus stimulated Heisenberg to apply this noncommutative mathematics in Heisenberg's own Hamilton-Jacobi analysis of dispersion theory.  Pauli was more accomplished mathematically than Heisenberg.  Pauli had used tensors and vectors, was aware of Klein and Sommerfeld's use of matrices to describe a spinning top and Lorentz transformations, and was aware of the Hamilton-Jacobi formalism to evaluate the energy values of a rotating body.  Later, in his work that led to his discovery of the exclusion principle, he had analyzed the energy and momentum states of a quantized atomic system, where the mathematics of rotations plays a key role.  Also in 1925, he thoroughly explored the old quantum theory for his encyclopedia article on quantum mechanics, the Old Testament.  I suggest that Heisenberg became aware of the rules of matrix multiplication from Pauli, and proceeded to use them in the dispersion problem with the Hamilton-Jacobi formalism to analyze the transformations of quantum states.

In general, as I now explore Pauli's role in the creation of matrix mechanics in more detail, I see an emerging pattern: Pauli's creativity often appears in the works of others and hence has not been attributed to him. Weinberg has commented on Pauli's creativity:

> The quantum-mechanical calculation of the hydrogen energy levels by Pauli was an exhibition of mathematical brilliance, a sagelike use of Heisenberg's rules and the special symmetries of the hydrogen atom.  Although Heisenberg and Dirac may have been even more creative that Pauli, no physicist alive was more clever….[297]

Abraham Pais also commented on Pauli's creativity: "That December [ of 1924] Pauli reaches a creative peak, his highest I would think [when he introduces a fourth quantum number]." [298]  Pais discussed Pauli's career at some length in 2000.  He recalled Pauli's comment: "When I was young I believed myself to be a revolutionary … [but] I was a classicist, not a revolutionary."[299]  Pais went on to note:

> As to Pauli, chutzpah never was one of his strong suits.  Had he had the temerity to publish in 1953, he would now have been remembered for his most important post-war contribution to physics, as one of the founding fathers of modern gauge theory….[300]

Pauli thus exhibited a long pattern of hesitation in publishing his insights, which led to his creativity being hidden from view.  Let us look more closely at the historical record.

Pauli wrote to Bohr on December 12, 1924, announcing his discovery of the exclusion principle and its consequence for determining the number of electrons in closed atomic shells.  On December 31, he again wrote to Bohr, commenting on the relationship of his discovery to Bohr's correspondence principle:

> I personally do not believe ... that the correspondence principle will lead to a foundation of the rule (I cannot prove this rigorously, it is only a scepticism based on my physical feeling)….  I see the promising clarification (also that of [the] coupling problem in general) rather in a physical analysis of the notions of motion and of force in the sense of the quantum theory.

Bohr no doubt shared Pauli's letter with Heisenberg,[301] so it is quite possible that some aspect of Heisenberg's train of thought that led to his creation of matrix mechanics was set into motion by Pauli.   In support of this possibility, historian John Hendry has indicated one starting point for additional historical research on the back cover of his book:

> The development of quantum mechanics is interpreted as a dynamic interaction between physical, methodological and epistemological considerations, emerging primarily as a dialogue between two profound physicist-philosophers, Niels Bohr and Wolfgang Pauli.  It is shown that Heisenberg's matrix mechanics, the quantum-mechanical transformation theory, Heisenberg's uncertainty principle and Bohr's principle of complementarity all had their roots in this central dialogue, and that the ideas characteristic of the interpretation of quantum mechanics were also essential to its creation.[302]

Hendry thus views Pauli as the central and indispensable figure in the development and interpretation of quantum mechanics, with Bohr and Heisenberg being the spokespersons for many of his ideas. Clearly, more historical research is warranted here.

The *Pauli Verbot* seemed to demand a deeper explanation for the repulsive-like inability of any two electrons in an atom to have the same set of four quantum numbers. To Pauli, the idea of electronic orbits presented a confusing array of concepts that seemed to require an explanation that went beyond continuum physics. Just as Pauli noted that Weyl's field theory could not be meaningfully applied inside an electron, so he regarded electronic orbits as meaningless for similar operationalist reasons—they could not be observed. In his letter to Bohr of December 31, 1924, after calling for "a physical analysis of the notions of motion and of force in the sense of the quantum theory," he added "in … which one will naturally have to be guided by experiment."[303] With Pauli's emphasis on experiment here, I maintain that Pauli was still clinging to his operationalist, ESTJ leanings. He felt free to speculate about methodological quantum requirements, but his rationalist ESTJ side prevented him from yielding to his INFP intuitions. Heisenberg was not similarly inhibited.

John Hendry has pointed out how the Pauli-Bohr dialog helped shape the creation of quantum mechanics, often focusing on Bohr's leadership.[304] The elder statesman Bohr posed thought-provoking questions and conceptual challenges of crucial significance to the physical foundation of quantum theory. If, however, we focus on the mathematical foundation of quantum mechanics, Pauli deserves great credit. Mara Beller recognized Pauli's significant contributions, but she focused primarily on Heisenberg in her dialogical history of quantum mechanics.[305] Pauli's role is less well documented. I believe, in fact, that Hendry's emphasis on Bohr and Beller's on Heisenberg can be traced to Pauli's hesitation in publishing his insights. Thus, to fully appreciate and

---

understand Pauli's indispensible role, we must take into account the personal restrictons that Pauli imposed upon himself.  Pauli then emerges as the key person in many of the creative dialogues that took place; he emerges as the key person in erecting a rational structure of quantum mechanics.  In my view, Bohr, Heisenberg, and others were potentially replaceable here; Pauli was indispensible.

Beller has listed the following as Pauli's contributions: Pauli stimulated Bohr and hence Heisenberg to create the matrix formulation of quantum mechanics; Pauli emphasized to Heisenberg the need for a particle kinematics; Pauli pointed out the impossibility of measuring both the momentum $p$ and the position $q$ of a particle simultaneously to absolute accuracy; Pauli considered a similar uncertainty in the simultaneous measurement of energy and time; Pauli essentially donated his insights on quantum uncertainty to Heisenberg (which suggests to me that the uncertainty principle should be named after both of them); Pauli, by translating Born's collision theory into matrix language, showed the equivalence of the matrix and wave formalisms of quantum mechanics; Pauli focused on the philosophical issues involved in the concept of causality; Pauli discussed probability and observables and hence hinted at a statistical interpretation of quantum mechanics; and Pauli laid the rational groundwork for Heisenberg's formulation of the new quantum mechanics through his enyclopedia article on the old quantum theory of 1926.[306]  Beller pointed out that Pauli later felt that he had not published many of his creative ideas owing to some restraint that he had imposed upon himself.[307]  I claim that Pauli's conflicting ESTJ-INFP personality was the source of that restraint.  His criticism of others was forthright; his criticism of himself was overwhelmingly inhibiting.

I can add to Beller's list by taking account of the work of John Hendry and Charles Enz; the sum then constitutes the heart of the efforts resulting in the creation and development of quantum mechanics between 1925 and 1930: Pauli supported Heisenberg's formulation of matrix mechanics; Pauli defended Heisenberg's physical ideas against the potential interdiction of Born's mathematics;[308] Pauli solved the problem of the hydrogen atom using Heisenberg's matrix mechanics; Pauli proved (but did not publish) the equivalence of matrix and wave mechanics;[309] Pauli recognized the importance of statistics in counting quantum states during his work on the exclusion principle, which later stimulated the recognition of the distinction between Bose-Einstein and Fermi-Dirac statistics; Pauli used the concept of symmetry in physical theorizing; Pauli clarified Bohr's ideas on the interpretation of quantum mechanics after Bohr's Como lecture in 1927; Pauli introduced spinors, which laid the basis for Dirac's relativistic theory of the electron; Pauli worked with Heisenberg on quantum field theory and charted the future development of it; and Pauli proposed the neutrino hypothesis based upon theoretical arguments decades before it was detected experimentally. This list of Pauli's contributions, including those to the creation of matrix mechanics, constitutes far more than legitimization of the creative contributions of others: Pauli's creativity undergirded the publications of others. Moreover, after 1930 he continued to make significant contributions, many of which he never published. His psychological retrictions allowed him to share his creative ideas with others, but prevented him from publishing them himself. Let us explore his hidden creativity further.

---

[306] *Ibid.* ,p. 87.
[307] *Ibid.*, p. 79.
[308] Enz,, *No Time to be Brief* (ref. 3),  p. 134.

In Heisenberg's creative leap from Hamilton-Jacobi continuum analysis to Hamilton-Jacobi matrix analysis, a possible path to his creation of matrix mechanics was to ignore the continuous and unobservable variables of the electron; focus on its initial and final states; and determine its transition frequencies. The states of the electron then form an alethic reality, and the frequencies of the emitted radiation are operationalist connections to empirical reality. Pauli was capable of formulating much of this problem mathematically, and of supplying the necessary mathematical tools of matrix mathematics to Heisenberg, who then could have adapted them to his own purposes. For example, Heisenberg might have combined Hamilton-Jacobi analysis with matrix rotational transformations to create matrix mechanics. Pauli, in contrast to Heisenberg, would not have been surprised by noncommutative matrix mathematics. If, therefore, we attribute the matrix mathematics to Pauli and the Hamilton-Jacobi analysis to Heisenberg, then the magic of Heisenberg's creative leap diminishes and the importance of Pauli's creativity increases.

---

[309] *Ibid.*, pp. 140-141; and C.Enz, "W. Pauli's Scientific Work," in Jagdish Mehra, ed., *The Physicist's Conception of Nature* (Boston: D.Reidel Publishing Co., 1973), p. 773.

Nonetheless, as in any creative leap, confusion and lack of clarity prevail, so a full explanation of Heisenberg's magic remains elusive. We might look, however, at Heisenberg's uncertainty principle,

$$\mathbf{p} \cdot \mathbf{q} - \mathbf{q} \cdot \mathbf{p} = h/2\pi i \mathbf{1} ,$$

where $\mathbf{p}$, $\mathbf{q}$ and $\mathbf{1}$ now are matrices, to attempt to gain further insight into Pauli's contributions to it. Pauli in his "mind's eye" might have viewed it as expressing the noncommutivity of matrix rotations. The dynamical variables are in an abstract mathematical space, but the imagery of a rotation remains. *Anschaulichkeit* appears in the alethic reality of momentum and position variables. The order of the rotation switches from first $\mathbf{p}$ then $\mathbf{q}$, to first $\mathbf{q}$ then $\mathbf{p}$, and Planck's constant $h$ is a measure of the minimum element of noncommutivity in this ordered process. Pauli, in addition to visualizing a kernel under rotation, also might have associated the above equation with the Pythagorean tetractys, where $\mathbf{p}$, $\mathbf{q}$, $h$, and $i$ are the four elements, and the kernel is some visualizable whole uniting all four. Thus, Pauli would not have been surprised by uncertainty; it was a consequence of a rotation-like alethic transformation. Further, Beller has pointed out that it was Pauli who provided Heisenberg with the example of the time-energy uncertainty relation.[310]

My account of how Heisenberg was led to create matrix mechanics and to propose his uncertainty principle is admittedly speculative. What have others written about potential clues to Heisenberg's route to matrices?

Max Jammer focuses on Heisenberg's substitution of quantum frequencies for classical Fourier components, which resulted in the noncommutative matrix multiplication that was unfamiliar to Heisenberg.[311] Jammer comments that:

> In connection with our present discussion of Heisenberg's matrix representation of quantum-mechanical states, the following mathematical digression seems to be of some interest. As is well known, subsequent formalisms of quantum mechanics employed--in addition to matrices and apart from complex-valued functions (Schrödinger's wave functions)--also Hilbert space vectors and, after the advent of [the] quantum mechanics of spinning particles, quaternions as well.

The striking point, now, is the fact that precisely these three kinds of mathematical entities, matrices, multidimensional vectors, and quaternions, happened to be equally involved in a historically interesting controversy concerning the priority of [Arthur Cayley or William Rowan Hamilton on] the discovery of matrices.[312]

Jammer includes the following footnote:

Quaternions had been applied to physics as early as 1867 by P.G. Tait in his *An Elementary Treatise on Quaternions*…. Their use in modern physics was revived by [Felix] Klein, who showed in 1910, in a lecture before the Göttingen Mathematical Society, that Lorentz transformations, conceived as four-dimensional rotations in Minkowski space[-time], can be conveniently expressed in terms of quaternions…. Their systematic, though never popular, use in quantum mechanics began when it was recognized that the Pauli spin matrices were essentially quaternion basis elements.[313]

Pauli was familiar with Lorentz transformations in Minkowski space-time and also with quaternians from Klein and Sommefeld's treatment of a spinning top. If he was thinking about spin matrices in early 1925, he might well have discussed rotations as Lorentz transformations with Heisenberg, then quaternians and matrices. I thus strongly suspect that Pauli provided Heisenberg with the mathematical tools necessary for the creation of matrix mechanics. Pauli himself always credited Heisenberg for creating matrix mechanics, but I believe that Pauli set the stage for Heisenberg's work and indicated a route to follow. Pauli dared not publish his intuitions owing to their lack of operationalist support, but the uninhibited Heisenberg felt no such qualms.

Pauli supported and defended Heisenberg during this critical period even though Heisenberg's bold act of creativity frustrated Pauli because it was unsupported philosophically. In 1968 Heisenberg commented on the difference between Pauli's character and his own:

Pauli's whole character was different from mine. He was much more critical, and he tried to do two things at once. I, on the other hand, generally thought that this is really too difficult, even for the best physicist.  He tried, first of all, to find inspiration in the experiments and to see, in a kind of intuitive way, how things are connected. At the same time, he tried to rationalize his intuitions and to find a rigorous mathematical scheme, so that he really could prove everything he asserted.  Now that is, I think, just too much.  Therefore, Pauli has, through his whole life, published much less than he could have done if he had adandoned one of these two postulates.  Bohr had dared to publish ideas that later turned out to be right, even though he couldn't prove them at the time.  Others have done a lot by rational methods and good mathematics. But the two things together, I think , are too much for one man.[314]

---

[312] *Ibid.,* p. 205.
[313] *Ibid.,* p. 205, see Jammer's footnote 17.

Physicist-historian Jagdish Mehra has commented further:

In his later years Pauli regretted his own "conservatism," feeling perhaps that Heisenberg's intellectual boldness (if not recklessness) reaped a richer scientific harvest and wider scientific fame.[315]

**Platonism and Alethic Reality: Rotations, Quaternians, and Kernels**

By 1927 Pauli had moved away from operationalism.  Some of his theoretical concepts, like spinors, had no connection to observables but were powerful mathematical tools.  Pauli was no longer a positivist, to the extent that he had been one; his successes encouraged him to question his father's and Mach's messages.   Sommerfeld and Bohr had made their mark on his psyche: He was beginning to value his aesthetic intuitions.  Spinors thus were more than powerful mathematical tools; they reflected some aspect of deep reality.  In his exclusion principle, Pauli had encountered the emotionally disturbing *Zweideutigkeit*, the quaternity of quantum numbers, and the need for an abstract explanation of electron spin.   In 1927 spinors satisfied these emotional needs by offering a mathematically rigorous explanation of electron spin that had been generated by his awakening intuition.  Pauli's INFP side was coming out of his Shadow.  The messages from his ESTJ father, however, were still intense, making strong demands for rationality and empiricism in his physics, as Heisenberg noted.  Thus, by 1927, Pauli's personality was a mixture of diverse components that were coming into a comfortable balance. He was opening his mind to Platonism, testing his ideas on Heisenberg, Dirac, and others, again as Heisenberg noted.[316] I see a number of components in Pauli's Platonism by 1928, some of which were present earlier. Pauli himself never elaborated on his Platonism, so I have tried to look for patterns in his physics between 1925 and 1927 that persisted later and that he discussed in his Jungian period.

---

[314] Karl von Meyenn and Engelbert Schucking, "Wolgang Pauli," *Physics Today* **54** (February 2001), p. 44.
[315] Beller, *Quantum Dialogue* (ref. 28),  p. 79

Heisenberg's new matrix mechanics and Pauli's own work using spin matrices and spinors certainly contributed to Pauli's Platonism. Its main feature was his interest in an alethic reality that made up the inner world of mathematical physics. Inside that alethic reality were kernels, quaternians, and rotations, all appearing as visualizable mathematical forms or operations. Examples of kernels that appeared in his physics between 1925 and 1927 are spin matrices and spinors, or physically the electron; and, with some unknown visualizable form, Heisenberg's uncertainty relations, or physically Planck's constant $h$ multiplied by $1/2\pi i$. Examples of quaternians are the four-component mathematical structure of spin matrices; and the four-component "primary entities" of **p**, **q**, $h$, and $i$ in Heisenberg's uncertainty relations. Examples of rotations that Pauli visualized or perceived as mathmatical operations are the spin of the electron; the mathematical operations generated by spin matrices and spinors, with their double-valuedness; and the mathematical equivalence of matrix and wave mechanics.

I will elaborate on Pauli's Platonism in my concluding chapter. By 1927, Pauli would have been growing comfortable with his physics and philosophy: He was making a name for himself as a brilliant theoretical physicist, and his released intuition was taking him productively to new levels of alethic reality. Pauli thus may have experienced intuitive visions and premonitions in his physics, but tragically, also of his mother's suicide of November 15, 1927.

---

[316] See Chapter 1 for excerpts from Thomas S. Kuhn's interview of Heisenberg.

# Chapter 7. The Neutrino Hypothesis 1928-1930

**Emotional Trauma**

The end of 1927 was a time of tremendous emotional trauma for Pauli. His mother committed suicide on November 15, 1927. He had been close to his mother, so the depth of his grief and emotional trauma must have been enormous. But his mother's suicide was not the only source of his stress. His Jewish grandmother, who it seems he knew only distantly, also died in 1927, which was a stressful event for his father and hence also for Pauli. In addition, Pauli was making plans to leave Hamburg in late 1927. To all outward appearances, his career in theoretical physics was flourishing, as is evident from his publications at this time;[317] no wonder that Jung later would comment on Pauli's heady one-sidedness. His intense intellectual activity after 1927 allowed him to escape from his emotions following his mother's suicide. He probably, in fact, had been escaping earlier from his emotions by concentrating on his theoretical physics. Jung and most psychologists agree that a person has to release emotional energy before it bursts out of the unconscious in unpredictable ways. Some of Pauli's emotional outbursts before and after 1927 pertain to his physics, even investing it with religious and moral content. As I noted earlier, Heisenberg reported to Thomas S. Kuhn:

> "Verdoppelung und Symmetrieverminderung. 'Das ist des Pudels Kern'." That is, "The fundamental principle from which all nature is produced is doubling of states and then, later on, reduction of symmetries." He adds, at this point, "Verdoppelung ist ein alter Zug des Teufels." In the whole medieval philosophy of the Alchemist the devil is, of course, the one who would double things. Then he adds that the devil is, of course, the one who makes doubts, hesitations, and the word "Teufel" has to do with "Zweifel," which, in the old time, meant doubling, i.e. you can do either this or that. So Pauli says that to be put in front of an alternative and to double the possibilities is an old and most fundamental feature of the devil. In this way, the devil has created the world. Pauli loved to talk about these things….[318]

---

[317] Ralph de Laer Kronig and Victor F.Weisskopf, ed., *Collected Scientific Papers by Wolfgang Pauli*. Vol. 1 and 2 (New York: Interscience Publishers, 1964). Pauli wrote all or major parts of eighteen articles and books published from 1927 to 1933.
[318] See Chapter 1.

By 1928, Pauli's physics thus was filled with emotional content.  His critical outbursts are well known, but the relationship of his emotional dysfunction to his creative releases has not been fully appreciated.[319]  Thus he proposed his most creative idea, the neutrino hypothesis, at the height of his emotional stress.  Pauli considered the neutrino "that foolish [*närrisch*] child of the crisis of my life (1930-1)--which also further behaved foolishly."[320]  He released this bold idea only reluctantly, first having to overcome deep inhibitions before divulging a theoretical entity that defied experimental verification.   I maintain that Pauli here again hesitated to confront his persistent psychological ESTJ restrictions and to make a transition to his emerging intuitive INFP style.  He released his neutrino hypothesis publicly only because his emotional trauma caused him to let down his guard.

We see in Pauli's career constellations of trauma and creativity.  In chronological order—which I consider to be highly significant—the most stressful events in Pauli's life during this time period were as follows: His mother committed suicide on November 15, 1927, and his Jewish paternal grandmother died sometime in 1927.[321]  He moved from Hamburg to Zurich, beginning his career at the Federal Polytechnical Institute (ETH) at the end of April 1928,[322] at which time he no longer puts "junior" after his name.[323]   His father married a much younger woman sometime in 1928.  He left the Catholic Church in May 1929 for unknown reasons.  He married Kate Deppner on December 23, 1929, divorcing her on November 26, 1930.

A few days later, on December 1, 1930, Heisenberg mentioned in a letter to Pauli "your neutrons,"[324] and only three more days later, on December 4, 1930, Pauli wrote his now-famous letter to the "Dear Radioactive Ladies and Gentlemen" in Tübingen announcing his neutrino hypothesis.[325]  These events occurred at an exceedingly emotion-laden period in Pauli's personal life.  Note especially the conjunction in time between Pauli's divorce and his neutrino hypothesis.  Note also that he shared his neutrino idea with Heisenberg before daring to release it further, and that he addressed his now-famous letter not to some scientific journal, but to a group of friendly colleagues (including especially Lise Meitner).[326]  Pauli had struggled with many difficult spiritual and emotional issues between 1928 and 1930.   For relief, he likely became absorbed in theoretical physics, and at the end of this period his extreme emotional trauma lowered his inhibitions, allowing him to divulge the most creative idea of his scientific career.

---

[319] Abraham Pais*, The Genius of Science: A Portrait Gallery of Twentieth-century Physicists* (New York: Oxford University Press, 2000), p.  230.  Pais certainly noted that Pauli's stress and his neutrino hypothesis coincided.
[320] Abraham Pais, *Inward Bound: Of Matter and Forces in the Physical World* (New York: Clarendon Press, 1986), p. 314.
[321] Charles Enz, *No Time to be Brief: A Sceintific Biography of Wolfgang Pauli* (New York: Oxford University Press, 2002), p. 8.
[322] *Ibid*., p. 193.
[323] Charles Enz and Karl von Meyenn,ed., *Wolfgang Pauli: Writings on Physics and Philosophy* (New York: Springer-Verlag, 1994), p. 18.
[324] Enz*, No Time to be Brief* (ref. 3),  p. 215.

I believe that his mother's suicide in 1927 seriously set back Pauli's individuation process and deeply affected his self-confidence.  Outwardly, he pursued his physics without interruption after his mother's suicide, but his grief must have been enormous, although I have found no record of it.  Since he probably had sensed his parent's marital discord earlier, his mother's suicide must have raised deeply disturbing issues for him.  He might have intuitively sensed something was wrong but did or could do nothing to prevent it from happening.   How should he continue his relationship with his father since his father's actions may have precipitated his mother's suicide?   How should he relate to his new stepmother, an attractive woman of his own age, whom he hated?  What in his spiritual life might provide reassurance and comfort to him when experiencing the mysteries and tragedies of life?  What would be his future relationship to women, now that his mother was dead?

In Jung's psychological metaphysics, intuition is associated with the anima, the ideal archetypal female.  Pauli was finding out what mature adult males eventually realize: The human female is not the idyllic archetypal anima.  The anima has a dark side with qualities different from idealized beauty, caring, sexuality, and creativity; the anima has a dark side of cunning, hurtfulness, rejection, and inscrutability.  He might have asked himself, "What do women want, anyway?"  In this period of transition, Pauli was psychologically engaged with his mother, his Jewish grandmother, his new stepmother, his new and soon-to-be ex-wife, and at least one woman physicist, Lise Meitner.  Pauli had problems with women and was struggling with his intuition. His mother's suicide initiated an intense period of crisis in his life involving excessive drinking, carousing in nightclubs, and a mismatched marriage that soon precipitated an affair that his wife had with another man. Jung soon would have ample material for talking about unconscious parallels between Pauli and his father. As I noted earlier, Pauli remarked after his divorce: "If it had been a bullfighter—with someone like that I could not have competed—but such an ordinary chemist!"[327] His comment is double-edged: on one level he seems to be making a joke of a difficult situation; on another, more poignant level, his father had had an affair with another woman before his own separation from his wife, and he also was a chemist. If Jung needed Shadow issues to address in his treatment sessions with Pauli, he was able to find many.

---

[325] *Ibid.*, p. 215.
[326] *Ibid.*, p. 217.

Pauli buried his feelings and intuitions more deeply into his Shadow after 1927, but its contents erupted periodically in caustic outbursts, inappropriate behavior, and (if one believes Jung) in the synchronistic events of the "Pauli effect." Another of these outbursts was the announcement of his neutrino hypothesis. Pauli managed all of these stressful circumstances from 1927 until 1932 on his own without any professional spiritual or psychological resources. Indeed, Pauli left the Catholic Church in May 1929. He might just have drifted away from it, but it seems that he intentionally and abruptly cut his ties to it. Until he met Jung in 1932, I see in the historical record no replacement resource, other than his physics, to assist Pauli in his quest for a spiritual and religious identity. I think it no accident that his two opuses on quantum theory were referred to as his Old Testament and his New Testament.

The adult's process of discovering his or her spiritual grounding can be stressful, because the process often involves confronting one's religious assumptions formed during childhood. In the process of encountering the spiritual, similar to encountering the sexual self, the adult's experiential knowledge confronts the rational. The adult's experiences of the mystical become paramount, often in confrontation with the rational messages received from parents, mentors, and the wider culture. Pauli now had to decide for himself whether to listen to his ESTJ or INFP messages: The ESTJ messages could be proven but the INFP messages could not.

---

[327] *Ibid.,* p. 211.

Adding to Pauli's stress caused by the emotional trauma that he was experiencing in his personal life, his proposal of the neutrino hypothesis took place within an atmosphere of professional controversy involving his mentor Niels Bohr. Pauli favored exact energy conservation at the level of each microscopic interaction; Bohr was willing to view energy conservation as statistical on the microscopic level. Bohr and Pauli held opposite views and values on this fundamental aspect of the nature of physical reality until Bohr conceded in 1936 that energy conservation was maintained as Pauli had believed.[328] The emotional nature of their interaction over their differing views on energy conservation is unknown, but a long debate over personal philosophical values was likely stressful to both. Bohr could be aggressive, as when he confronted Erwin Schrödinger on his wave mechanics.[329] Pauli's critical derision was likely tiring even for Bohr. Thus, Pauli's weaker INFP side was likely stressed when confronting Bohr over this intuitive value. For some inexplicable reason but perhaps related to their controversy, Bohr never nominated Pauli for a Nobel Prize.[330] Pauli was individuating from his parents and from his mentors.

To fully appreciate the boldness of Pauli's creative act, we must recall that in 1930 the proton and the electron were the only known elementary particles. The connection between spin and statistics was still unfamiliar. Bohr was challenging energy conservation and Einstein was challenging the entire edifice of quantum mechanics. The nucleus was to a large degree still *terra incognita.* Yet, despite all of these difficulties, Pauli proposed an entirely new particle on the basis of known conservation laws, for still fuzzy new conservation principles, partly for aesthetical-theoretical reasons, partly to explain the experimental evidence, in defiance of his earlier operationalism, in opposition to his fathers' and Mach's positivism, and in opposition to his mentor Niels Bohr. Occurring within Pauli's mind was a fortuitius confluence of ideas coming from both his ESTJ side and his INFP side. No wonder that it took a severe crisis in his emotional life to release his bold hypothesis, supported by his intuition but unsupported by direct experimental evidence for another quarter of a century. Pauli's creativity was finally unleashed, however hesitantly. He was able to cast off his operationalism and the ESTJ side of his personality, and to release his creativity stemming from the INFP side of his personality. This was the boldest act of creativity in Pauli's career, and marked a watershed in his philosophy of physics. It also occurred before his treatment at the Jung Clinic in 1932.

A discussion of my use of the compound term "aesthetical-theoretical" will serve to better expand on the complexity of motivations that Pauli was facing. I mean by "aesthetical-theoretical" those motivations that largely came from Pauli's INFP side, and that pertain to the intuitive visual images he saw in his "mind's eye," ones that called for internal symmetry, balance, and internal consistency of the guiding principles of theoretical physics. These constitute an aesthetical beauty. Conservation of energy, for example, is one such guiding principle of theoretical and experimental physics. Experimental physics had demonstrated time and again that energy is conserved, yet there was no way to derive this principle theoretically, as if this came from an alethic reality where deep principles unfold into reality and guide the discoveries of physics. The power of the internal beauty of this principle was such that experimental enigmas that seemed to conflict with it became of serious concern. One response would be to revise or suspend energy conservation on the microscopic scale. Another much more difficult response would be to search for a way to preserve its "aesthetical-theoretical" beauty. As I will discuss below, this was the choice that Pauli faced when confronted with the Ellis-Wooster experiment.

**The Neutrino Hypothesis**

Pauli's neutrino hypothesis stands out as an extraordinary act of his creativity, and it too was emotionally laden. The hesitancy with which he introduced this bold hypothesis, his caustic personality, and the long period of time that elapsed before it was confirmed by direct experiment may have prevented him from winning a second Nobel prize.[331] Since it was not detected experimentally much earlier, which deprived Pauli of that satisfying vindication, his psyche may have suffered yet another lingering bruise.

---

[331] Pauli's Nobel Prize-quality work, it seems to me, includes his discovery of the exclusion principle, his introduction of spinors along with his other contributions to the whole development of quantum mechanics, his introduction of the neutrino concept, his elucidation of the connection between spin and statistics, and his seminal work on the interrelationships of charge, parity, and time.

Pauli's path to the neutrino hypothesis has been discussed by Laurie Brown, Charles Enz, Allan Franklin, Karl von Meyenn, and Abraham Pais.[332] I will provide a condensed version of their interpretations, offering my own exploratory observations stemming from what I see as patterns in Pauli's personality. Pauli referred initially to the new particle as a "neutron." As noted above, he evidently first informed Heisenberg of it before December 1, 1930. His motivation to propose it appears to have stemmed from a combination of ESTJ and INFP values. Empirically, after a long controversy between Charles D. Ellis and Lise Meitner in the 1920s, Ellis and William A. Wooster proved in 1927 that the beta particles emitted from radium E are emitted with a continuous distribution of the energies.[333] The 1927 paper of Ellis and Wooster also revived the debate over energy conservation.[334] Pauli, however, likely had made up his mind on energy conservation before 1927: "The history of that foolish child ... begins with those vehement discussions about the continuous β-spectrum between [Meitner] and Ellis which at once awakened my interest."[335] Abraham Pais reported:

> It is my impression that the seriousness of the situation it [the Ellis-Wooster paper] created was not immediately appreciated. In all the literature of 1928 I found only one (uninteresting) reference to their paper. Pauli's complaint in February 1929 that Bohr was serving him up all kinds of ideas about β-decay "by appealing to the Cambridge authorities but without reference to the literature"... would indicate that not even then had he seen the Ellis-Wooster paper.[336]

The Ellis-Meitner controversy may have become sufficiently polarized among theorists by the time Ellis and Wooster published their fundamental paper that Pauli uncharacteristically may not have been paying close attention to the experimental literature.  Pauli always valued empirical information, but by this time he may have made up his mind and was listening more attentively to his own intuition.  He was searching for an intuitive model to explain the experimental results.

Pauli, stimulated also by Oscar Klein's five-dimensional relativity theory, realized that conservation of energy and momentum could be preserved in beta decay if in addition to the beta particle another particle, a "neutron," were emitted in the decay process.[337]  In addition, he was aware of the puzzling statistics of certain nuclei, as he noted in a letter to Klein on December 12, 1930:

> On the occasion of solving a school exercise about the hyperfine structure of Li+, [I] went over very thoroughly in his [my] mind once again the "wrong" statistics of the nuclei as well as the continuous β-spectrum.  Then the following possible way out occurred to me (a way out of desperation however): the nuclei could contain other elementary particles besides electrons and protons and these must be electrically neutral, obey the Fermi statistics and have the spin ½.  Let us name these particles neutrons [later renamed neutrinos by Enrico Fermi to distinguish them from neutrons].[338]

Pauli went on to explain to Klein his further reasoning:

> To start with, the electric charge is conserved in the process [of beta decay], and I do not see why the conservaton of the charge should be more fundamental than that of the energy and of the momentum….  If the conservation laws were not valid one would indeed have to conclude from these [energetic] relations that a β disintegration is always accompanied by a loss of energy, and never by a gain; this conclusion implies an irreversibility of the processes with respect to time which seems to me scarely acceptable.[339]

Pauli thus concluded that a failure of energy conservation was unlikely because, first, energy and momentum conservation is analogous to charge conservation, and second, a violation of energy conservation should also change the weight of a box filled with beta-radioactive material without affecting an external electrostatic field, and this conclusion was suspect.[340]  We see here Pauli's unique combination of experimental ESTJ and aesthetical-theoretical INFP values, except in this instance Pauli was willing to place, with ESTJ reservations, higher value on his INFP intuition.

On December 4, 1930, a few days after his divorce,  Pauli wrote his now-famous letter to his "radioactive" colleagues in Tübingen.[341]  He also issued a hesitant call to the experimentalists there (especially Lise Meitner) to give him their opinions of the reasonableness of his hypothesis.  Pauli thus was confronting his ESTJ side, which demanded experimental proof of his hypothesis.  Although aesthetically and theoretically so beautiful, his hypothesis had to be subjected to the scrutiny of experimentalists before he was willing to publish it.

In 1937 Pauli made some comments that are germane to his reasoning in postulating the neutrino, and may be interpreted as his belief in physical kernels beyond which one could not probe rationally.  At that time, Pauli treated the surface of a kernel as an impenetrable quantized entity, while Einstein saw that one could maintain the validity of continuous field equations inside it.  As Pauli noted:

> One may always choose a surface S so far away that the current density vanishes on it, then the conservation of the charge in the volume V confined by it is proven quite independently from whether the Maxwell equations are applicable to the processes occurring in the interior of this volume….  Extremely essential for this proof is the sufficient assumption that the equations of relativity theory hold at the surface S of this volume; if this is the case, then the conservation of the total energy and of the momentum in the volume V is proven quite independently of the fact whether the equations of the theory are applicable in the interior of V.  The conclusion is due to Einstein and represents a great progress.[342]

Pauli considered the conservation laws of physics to be fundamental, and he thus was willing to increase the number of conserved quantities.   The conserved quantities were primary; the number and type of particles possessing them secondary.  In the case of the neutrino, the conserved quantities were energy, charge, and spin.  In a letter to Heisenberg in 1933, Pauli wrote:

> With nuclear processes, even almost more important than the conservation laws of energy and momentum for me are the conservation laws of all the discretely quantized quantities [total angular momentum and the symmetry character of Bose-Einstein and Fermi-Dirac statistics].  For the time being I want to hold fast unconditionally to these assumptions and to pursue them in their consequences, before I modify them.[343]

In addition to Pauli's straightforward intuitive reasoning summarized above, Pauli experienced at least one "Pauli effect" in his release of his neutrino hypothesis that might have prepared him to later listen attentively to Jung's proposal of his synchronicity concept.  Pauli's belief in magic was reinforced before he met Jung.   Samuel Goudsmit recalled Pauli's sudden appearance at a conference in Rome in October 1931 to discuss his neutrino hypothesis:

> Pauli was supposed to attend the Rome meeting, but he arrived a day or so late.  In fact, he entered the lecture hall the very moment that I mentioned his name! Magic! I remarked about it and got a big laugh from the audience.[344]

### Pauli's Psychological Process

I next will discuss some of the open questions surrounding Pauli's personal motivations in his trail to the neutrino hypothesis.  For example, Alan Franklin in his book on the evidential history of the neutrino remarked:

> One does not know how seriously to take Pauli's trepidation about publishing such a radical proposal.  It [Pauli's neutrino hypothesis] was certainly novel, but no more so than the suggestion by Bohr, and others, that the strongly supported conservation laws were violated.  In any case, Pauli's hypothesis became widely known within the physics community despite its lack of publication.[345]

---

John Hendry also remarked: "It is usually impossible to connect a scientist's deepest, usually religious, concerns with his scientific work, and no attempt will be made to do so here [in Hendry's book on the Bohr-Pauli dialogue]."[346]

In contrast to Franklin and Hendry, I maintain that historians do have a unique opportunity to glean an understanding of Pauli's inner creative thoughts. His later extensive correspondence with Jung included numerous reflections on his psychological processes and their relationship to his physics. Impressed with Pauli's intellect, Jung used Pauli's dreams to reinforce his own thesis of mandalas as universal archetypes. Pauli's communications with Jung were sensitive to Pauli, and he did not wish to see them disclosed for various reasons. For example, Pauli sensed that his Platonism was not likely to be received well, as Heisenberg noted:

> But I remember that when one really started to come into these problems, I would
> say the atmosphere became so tense that it was disagreeable to continue. I could
> see that this man was so engaged in problems of that kind that it was really better if
> one did not touch it. So we started a discussion, and he could see that I could
> understand him in this plane, and from this moment on he had a strong confidence
> in me. But also in some way, it was agreed that we should not talk about it.[347]

I maintain that by 1927 Pauli's intuitive patterns in his theoretical physics were recurrent ones. He had shattered "an Idol of the Tribe" in his decipherment of the anomalous Zeeman effect with his exclusion principle, in his active role in the development of quantum mechanics, and in his brilliant use of spinors to describe the classically nondescribable double-valuedness of the electron. Rather than being an operationalist, Pauli was now a Platonist. The suicide of his mother may have affected his self-confidence in his Platonism, but he could not return to operationalism. He was using privately and successfully his intuition of kernels, *Zweideutigkeit*, and quarternians.

Jungian archetypes concern recurrent psychic forms and processes as cognitive patterns. In Jungian dream analysis, the recurrent cognitive patterns often play out with different situations, people, places, objects, and, most importantly, have different

---

meanings attached to the patterns.  To Jung, the archetypes were the cognitive patterns and not the information they carried.  He emphasized:

> Again and again I encounter the mistaken notion that an archetype is determined in regard to its content, in other words that it is a kind of unconscious idea (if such an expression is admissible).  It is necessary to point out once more that archetypes are not determined as regards their content, but only as regards their form and then only to a very limited degree.[348]

Analogous to the situation of modern digital computers where ones and zeroes constitute data that might represent different information, the recurrent cognitive patterns can carry different meanings.  In a digital computer, data consisting of three ones might code for the number seven or it might code for a particular color in a pixel.  Perhaps the brain uses electrochemicals as data to carry information.  Whether the archetypes and cognitive patterns are related to particular brain electrochemicals is an open question, but the analogy between computer data with attached information, and cognitive patterns with attached meanings, is a useful one.  In Pauli's dreams, the patterns' meanings were multi-layered.  He dreamed of egg-splitting processes, mandalas, and the number four, but at another level the meanings of Pauli's dreams were also about physics.  These are the cognitive patterns I will now explore in Pauli's development of his neutrino hypothesis.

In the development of his neutrino hypothesis, I believe that Pauli was seeing another kernel in his "mind's eye," a kernel whole that was connected to the alethic physics behind beta decay.  The kernel that Pauli might have envisioned when pursuing the path to his neutrino hypothesis was an abstract one of energy, charge conservation, and the still confusing connection between spin and statistics.  These are placed on an equal footing and constitute the endpoints of three axes of symmetry.  To  Pauli, charge conservation was to Maxwell's electrodynamics as energy conservation was to Einstein's general relativity, and the connection between spin and statistics was another still fuzzy corollary.  Thus, neglecting spin and statistics for awhile, since charge is conserved in

electrodynamics and inside the nucleus, energy must be conserved there as well.[349]
Pauli's "mind's eye" neutrino kernel may have appeared as follows:

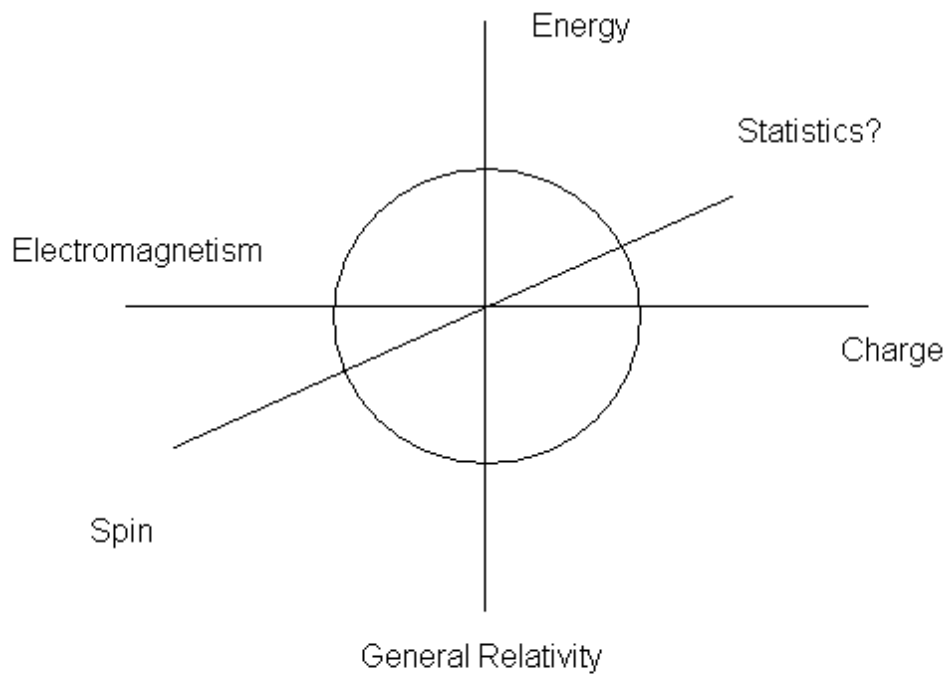

Figure 11:  Author's representation of Pauli's neutrino kernel with three
orthogonal abstract dimensions.

Pauli's neutrino kernel thus united charge conservation and Maxwell's electrodynamics, energy conservation and general relativity.  Angular momentum (spin) and statistics might have formed a third axis of symmetry within the kernel, although here even Pauli's powerful intuition needed the clarification that was to come only later with the discoveries of the neutron and positron, with the contributions of other physicists such as Victor Weisskopf, and finally, in Pauli's own seminal work on spin and statistics published in 1940.[350]   To Pauli, energy conservation was only one aesthetic value that had to be conserved in the reaction in which beta rays (electrons) were ejected from the nucleus.  Other conservation laws held as well, namely conservation of charge and angular momentum (spin).  These aesthetical-theoretical values in confluence with the experimental Ellis-Wooster constraint required an entirely new, perhaps undetectable particle.  Pauli's strict adherence on the microscopic level to the conservation laws was in opposition to Bohr's questioning of their validity inside the nucleus.  Pauli was willing to place energy conservation on the same fundamental level as charge conservation, and by extension to insist that spin had to be conserved as well.

Another point of exploration concerns Pauli's hesitation to clarify whether he felt the neutrino existed inside the nucleus or was created at the moment of beta decay.  He vacillated between these two positions as if he did not see a significant difference between them.  Laurie Brown writes:

> Had Pauli proposed in 1930 that neutrinos were created (like photons) in transitions between nuclear states, and that they were otherwise not present in the nucleus, he would have anticipated by three years an important feature of [Enrico] Fermi's theory of beta decay.  Pauli did not claim to have had this idea when he wrote the Tübingen letter, but he did say (in his Zürich lecture) that by the time he was ready to speak openly of his new particle, at a meeting of The American Physical Society in Pasadena, held in June of 1931, he *no longer* considered his neutrons to be nuclear constituents.  It is for this reason, he says, that he no longer referred to them as "neutrons"; indeed, that he made use of no special name for them.  However, there is evidence ... that Pauli's recollections are incorrect; that at Pasadena the particles *were* called neutrons and *were* regarded as constituents of the nucleus.

> ...a short note in *Time*, 29 June 1931, headed "Neutrons?", says that Pauli wants to add a fourth to the "three unresolvable basic units of the universe" (proton, electron and photon); adding, "He calls it the *neutron*."[351]

Continuing my exploration of Pauli's recurrent cognitive patterns, Pauli reported to Jung in 1948 about his dreams that involved quaternity symbolism and *Zweideutigkeit* splitting:

> Seven pictures [appear] in a row.  No words are spoken until right at the end and I am the one speaking.

> *Picture 1.* A woman comes with a bird, which lays a large egg.

---

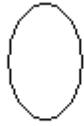

*Picture 2.* This egg divides itself into two:

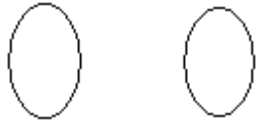

*Picture 3.* I go closer and notice that I have in my hand another egg, with a blue shell.

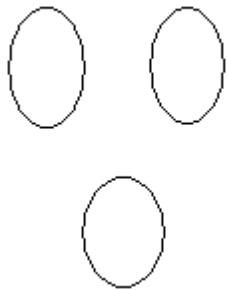

*Picture 4.* I divide this last egg into two. Miraculously, they remain whole, and I now have two eggs with blue shells.

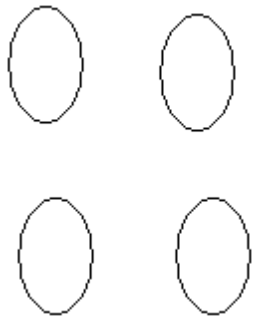

*Picture 5.* The four eggs change into the following mathematical expressions

cos δ/2          sin δ/2

cos δ/2            sin δ/2

*Picture 6.*  This gives the formula

cos δ/2      +      *i* sin δ/2

cos δ/2      -      *i* sin δ/2

*Picture 7.*  I say, "The whole thing gives $e^{i\delta}$, and that is the circle."[352]

Note the quaternities, the imaginary number *i*,  and the factor of ½.   Pauli goes on in this lengthy letter to discuss his hypothesis of the connection between physics and psyche, and does not mention the neutrino.  I maintain, however, that his cognitive pattern is the key and not the attached meaning of splitting eggs or mathematical formulas.  This is how Pauli thought.  Heisenberg recalled Pauli's repeated admonition: "Verdoppelung und Symmetrieverminderung.  'Das ist des Pudels Kern'."  That is, "The fundamental principle from which all nature is produced is doubling of states and then, later on, reduction of symmetries."[353]  In his development of the neutrino hypothesis, the splitting and doubling process within beta decay produced a holistic quaternity of four particles-- the proton, the electron, the photon, and now the neutrino.

Another example of Pauli's recurrent cognitive patterns appeared in a letter of February 27, 1952, to Jung.  Pauli refers to the diagram as a "quaternio" involving synchronicity, causality, energy, and time. His quaternio is similar to what I have termed a kernel.  Note especially in his diagram, below, that the endpoints of the energy-time axis split into three additional components. Here is an additional example, I maintain, of Pauli's conception of *Zweideutigkeit* splitting.

---

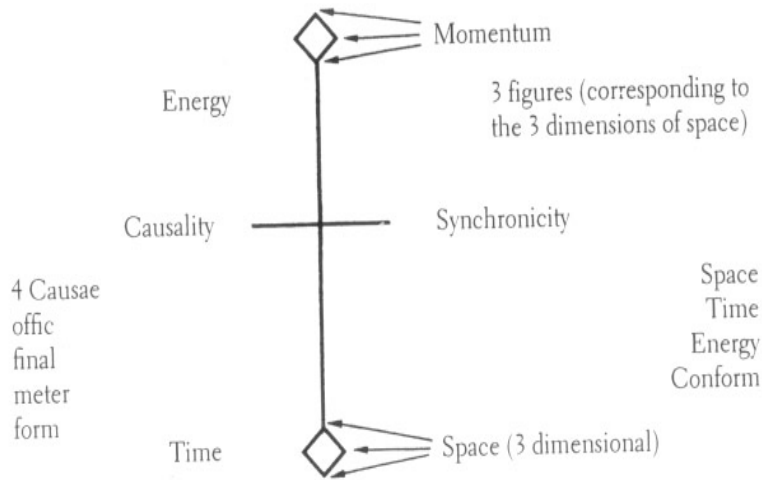

Figure 12. Pauli's quaternio of 1952 splitting into many components.[354]

       The above examples of Pauli's recurrent cognitive patterns do not prove he thought similarly in his path to the neutrino, but do raise interesting speculations that seem relevant to my exploration of his reasoning.  Four-part symmetry and *Zweideutigkeit* splitting were part of his recurrent cognitive patterns, and therefore may have occurred to him along his path to the neutrino.  Thus, the neutrino was essential to the nuclear kernel's wholeness and the total symmetry of the nucleus; in his mind the neutrino existed both inside the nucleus at some alethic level and was produced and made a separate entity at the moment of beta decay.

**Platonism and Alethic Reality**

The impact of Pauli's neutrino hypothesis on elementary-particle physics was monumental, yet one of delayed gratification.  Pauli gave talks to his colleagues on his hypothesis before publishing his ideas.  Enrico Fermi in 1934 built his theory of beta decay around Pauli's neutrino, which supported Pauli's hypothesis, yet there was a logical circularity.  As noted by Franklin, "The conservation laws support the neutrino, but the neutrino saves the conservation laws."[355]  The neutrino was for the time being unsupported by direct empirical evidence but needed for Pauli's, and now other physicists' intuitive theories.  Nuclear physicists exploring the workings of the nucleus

---

were led more by intuitive aesthetical-theoretical values rather than empiricism and operationalism.

After 1930, Pauli should have been able to renew his individuation process, and release his psyche from the voices that had been within his mind since childhood. He now could see the fruitfulness of trusting his intuition, of following his INFP side. This, however, remained a difficult process for him. His emotional crisis, heavy drinking, and the like, continued until he began treatment at the Jung clinic in 1932. I thus see his psychological struggles continuing beyond 1930, but I also see little change in his style of Platonism. He continued to be reserved in disclosing his creativity. His Shadow continued to defend its turf with biting criticism even into his final years. For example, there is an indication he may have preceded Chen Ning Yang and Robert Lawrence Mills in their work of 1954 on new fields involving isospin, but Pauli could not publish such speculative ideas based only on his intuition. His Shadow, however, could defend its claim:

> Yang and Mills knew the charge and the isospin of the new field particles, but they had no idea of their masses, and they recognized this as a weakness in their theory. When Yang presented the theory in a seminar at Princeton in February 1954, he found himself under attack from no less a person than Pauli. As soon as Yang had written on the blackboard an expression involving the new field, Pauli asked, "What is the mass of this field?" When Yang explained that it was a complicated problem and that he and Mills had come to no definite conclusions, the acerbic Pauli charged, "That is not sufficient excuse."[356]

Pauli's physics after 1930 involved psychological *Anschaulichkeit* where empiricism and intuition competed. He still needed both before he felt sufficiently comfortable to publish. His cognitive processes, however, freely but privately flowed along intuitive lines. His Platonism involved the mathematical forms and physical processes that he had seen in his "mind's eye." The forms and processes that Pauli saw in his physics were more than inert mathematical or physical concepts, more than positivistic or descriptive tools. They were a spiritual resource for him. I think that Pauli felt that they were filled with life.  Pauli, however, needed validation from Jung before he could be comfortable with Platonism.

---

[356] Christine Sutton, "Hidden Symmetry: The Yang-Mills Equation," in Graham Farmelo, ed., *It Must Be Beautiful: Great Equations of Modern Science* (New York: Greta Books, 2003), p. 243.

# Chapter 8.  Conclusions

## Pauli's Psychological Makeup

Repeating my introductory comments, to understand the history of quantum mechanics, it is essential for the historian to understand Pauli's role in that history through his unique contributions to its physics.  That requires the historian to understand Pauli's personality and philosophy.  That, in turn, requires the historian to understand Pauli's receptivity to Carl Jung's psychological philosophy.  When all of these factors are taken into account, Pauli's creativity and significance emerges with new clarity.  Pauli was consistently at the nexus of developments in quantum physics; moreover, he was the creative nexus that united these developments.  His perfectionism restricted his publishing, and his extreme rational and empirical self-testing, his critical outbursts toward others who were working on ideas similar to his own, and his silence about his intuitions, kept his creativity well hidden.  I have attempted to illuminate these aspects of his creativity, but a full understanding of them remains elusive.  Pauli was a theoretical physicist of genius whose human qualities were a product of his times, heritage, parents, and teachers.  Like everyone else, he struggled during adolescence and young adulthood, seeking independence from his parents and from others; he needed to become his own man.  His worldview changed during his process of individuation and it affected his physics.  When difficulties arose, he reacted in understandable ways.  By exploring his life, his personality, and his physics up to 1930, I have attempted to understand his complex character and his unique contributions to physics.

Pauli's inner turmoil appears to have continued until he underwent therapy at the Jung Clinic in 1932, and diminished greatly thereafter.  He second marriage to Franca Bertram in 1934 was successful and lasted until his death in 1958, although he evidently strayed at times.[357]  He continued to work in elementary-particle physics and quantum-field theory, and won the Nobel Prize in 1945, supported by Einstein's nomination, for his discovery of the exclusion principle.  He continued his patterns of secretiveness, hesitancy, criticism, and perfectionism.   He remained fascinated with Jung's archetypal psychology, and his relationship with Carl Jung deepened, although shortly before his death it and his relationships to Heisenberg and others became strained.  He died of pancreatic cancer, stilling a voice that profoundly influenced the course of twentieth-century physics.

---

[357] Abraham Pais, *The Genius of Science* (New York: Oxford University Press, 2000), p. 241.

I have suggested that Pauli saw the following Jungian elements in his physics prior to his treatment in the Jung Clinic in 1932: He believed that physics operated in an alethic reality; kernels existed in the alethic reality of physics; kernels often possessed quaternian structure, and hence the number 4 was especially significant numerologically; rotations in his "mind's eye" corresponded to mathematical transformations between coordinate systems within kernels; within kernels, *Zweideutigkeit*, doubling of physical parameters, was common and represented some aspect of the Devil's handiwork; synchronistic events were an aspect of alethic reality; and numerology was an important methodology.  I claim that Pauli had these elements in his mind as a curious and fuzzy mixture of concepts, emotions, and intuitions.  I further claim that Pauli subordinated his creative INFP side in his work on the exclusion principle, electron spin, and spinors, and in his contributions to the development of matrix mechanics, in proving the equivalence of matrix and wave mechanics, in explaining beta decay, and in searching for a new quantum theory.  He did not publish many of these contributions, however, and hence they were overshadowed by those of Heisenberg, Schrödinger, Dirac, Fermi, Bohr, and Einstein.

**Pauli's Treatment  at the Jung Clinic**

The years 1930-1932 were not Pauli's best.  Grief over his mother's suicide and the turmoil of his first marriage remained in his mind, and his drinking alone might have led him to seek Jung's help; he broke his arm in a drinking episode in 1931.[358]  In light of these and other factors, one can only speculate about the contents of Pauli's Shadow before he met Jung.  Pauli's father recommended that Pauli seek treatment in the Jung Clinic.[359]  Pauli was receptive to his father's advice, and Pauli attended Jung's lectures and read Jung's materials before deciding to seek treatment.  He felt that his relationships to women constituted his main psychological problem.

---

[358] Charles Enz, *No Time to be Brief: A Scientific Biography of Wolfgang Pauli* (New York: Oxford University Press, 2002), p. 225.
[359] C. Enz and K. von Meyenn, ed., *Wolfgang Pauli: Writings on Physics and Philosophy* (New York: Springer-Verlag, 1994), p. 18.

The details of Pauli's psychological analysis at the Jung Clinic will remain unknown until his records are located and made available for study.   We do know that in January 1932 Jung recommended to Pauli that he see a female analyst, rather than himself, because of his problems with women.  Pauli thus wrote to the Jungian analyst Erna Rosenbaum on February 3, 1932, seeking her help.[360]  Thus, during the eight months of his treatment,[361] problems with women no doubt were the focus of part of his treatment.  I suggest that Rosenbaum, with Jung in the background, expanded her analysis to include the following areas: an appreciation of the anima, the female muse of intuition and creativity; an understanding of personality typology and thus Pauli's relationship to his father and mother; conveying empathy for Pauli's grief over his mother's suicide and his own divorce; an understanding of his own dormant INFP personality with its intuition and creativity, which was masked by his ESTJ side with its demands for rationality and empirical proof; an appreciation of Pauli's Shadow; using the Shadow concept to interpret his hesitancy, perfectionism, and critical outbursts; the significance of his dreams, and archetypal psychology in general; conveying empathy for Pauli's spiritual struggles, which could be addressed through Jung's perspective of the universality of religions and the archetypal character of Catholicism and the Jewish Cabbala; and endorsing the archetypal roots of Pauli's *Anschaulichkeit* in physics.  Since Pauli was in treatment for only eight months, I believe that Rosenbaum and Jung primarily offered him validation for his perspectives on the above elements and areas: If he had had chronic or severe problems, then these would have required a longer period of treatment.

Pauli's physical concepts were closely related to those of Jung's archetypal psychology.  Alethic reality in quantum physics translated into the unknowable collective unconscious.  Pauli's physical kernels translated into Jung's mandalas, where mandalas had added vitalistic and nonrational features.  Mandalas thus were knowable through visual impressions, had a holistic character greater than the sum of their parts, and added philosophical-religious features to Pauli's inert physical symbols.  Jung adopted the Sanskrit word *mandala* to symbolize a "magic circle" that captures in the "mind's eye" the psychic center of a transformational process**.**  The idea defies precise definition, since the person seeing a mandala image is supposed to experience a mystical feeling of appreciation going beyond verbal definition, seeing a whole greater than the sum of its parts. The viewer can extract the mandala's components, but notably they did not always have four-part symmetry. To Jung, quaternian forms were more significant than trinitarian forms. According to the Jungian psychologist Mary Ann Mattoon:

> Jung's unorthodox conclusion is that the Christian Trinity is incomplete; wholeness is a quaternity. The fourth principle was posited variously by Jung as evil, the earth, matter, the body, and the feminine. Jung seemed to see these manifestations of the fourth as various ways of describing concrete reality, which contrasts with the entirely spiritual (and presumably male) Trinity…. Evil is a force in itself, he insisted; it is part of the collective unconscious and empirically verifiable…. It is no accident that the knowledge that good and evil are opposites is a metaphor for the emergence of consciousnes. A crucial factor in individuation is the encounter with the dark side of the personality and with the evil in the world.[362]

Among Jung's methodologies of treatment was his use of personality typing, the precursor of Myers and Briggs's typology, which I have used because it is easier to understand than Jung's. Pauli, however, was likely subjected to Jung's simpler one during his treatment and no doubt was receptive to it because of its clear visual model, as shown below:[363]

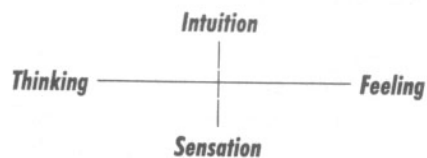

Figure 13: Peat's representation of Jung's mandala of psychological types.

Pauli, armed with his new appreciation for intuitive, visual knowledge, would have gleaned a quaternian mandala for the complete personality type.

---

[362] Mary Ann Mattoon, *Jungian Psychology in Perspective* (New York: The Free Press, 1981), pp. 201-202.
[363] F. David Peat, *Synchronicity: The Bridge between Matter and Mind* (New York: Bantam Books, 1987), p. 18.

Pauli's attraction to rotations, which were intimately connected to his ideas of kernels, translations, and symmetry principles in physics, were related to Jung's changing perspectives of a mandala's components or his different interpretations of a dream. The physical process that Pauli saw in *Zweideutigkeit* was nearly identical to the operation of the archetypal "tricksters" that Jung saw operating in the unconscious. Jung had not yet published his concept of synchronicity, which Pauli likely helped Jung to refine. Pauli also likely felt more comfortable with the concept of time symmetry and acausal connections in physics, as in the *Pauli Verbot*, following his discussions with Rosenbaum and later Jung. Rosenbaum and Jung helped Pauli appreciate nonrational thinking.

Armed then with Jung's metaphysical, archetypal psychology, Pauli began to pursue an avid interest in it. He assumed it was valid in part because it brought him inner peace following his therapy. His INFP side went on to secretly develop his mystical philosophical interests which were closely related to Jung's ideas, and which I see as a kind of Platonism. Moreover, during the course of Rosenbaum's treatment, Pauli not only found validation for his ideas, he may even have experienced a religious conversion. Jung reported one of Pauli's dreams—one of some 1300 that Rosenbaum recorded—that has become particularly famous.[364] Jung described it at several times in several places. It was Pauli's dream of a "world clock," as shown below.[365]

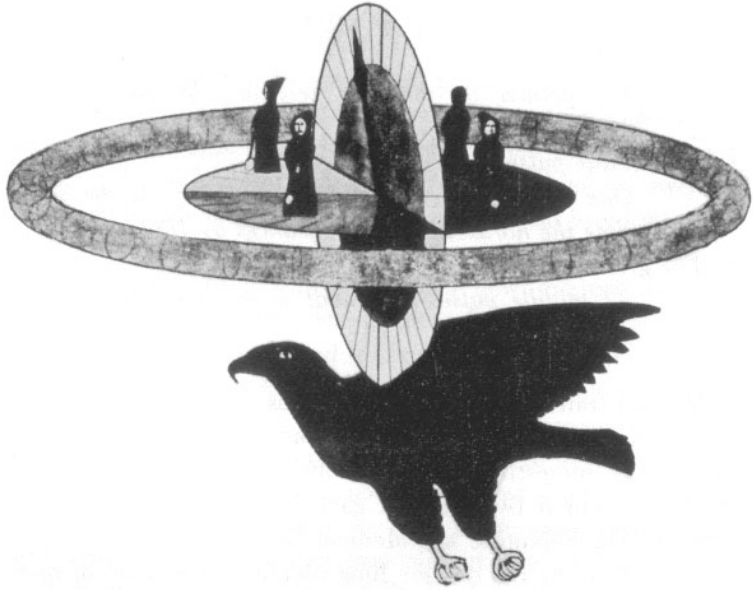

Figure 14. W. Byers-Brown's interpretation of Pauli's "world clock" dream.

Rosenbaum recorded Pauli's dream image, and noted that it generated a mystical experience in him.  Then, in 1935, Jung published a collection of Pauli's dreams, carefully hiding Pauli's identity, to illustrate the power of the mandala during psychological analysis.  Jung described Pauli's complex mandala dream as follows:

> There is a vertical and a horizontal circle, having a common centre.  This is the World Clock.  It  is supported by the black bird.
>
> The vertical circle is a blue disc with a white border divided into 4 X 8 = 32 partitions.  A pointer rotates upon it.
>
> The horizontal circle consists of four colours.  On it stand four little men with pendulums, and round about it is laid the ring that was once dark and is now golden (formerly carried by children).
>
> The "clock" has three rhythms or pulses:
>
> 1. The small pulse: the pointer on the blue vertical disc advances by 1/32.
> 2. The middle pulse: one complete revolution of the pointer.  At the same time the horizontal circle advances by 1/32.
> 3. The great pulse: 32 middle pulses are equal to one revolution of the golden ring.

This remarkable vision made a deep and lasting impression on the dreamer, an impression of "the most sublime harmony," as he himself puts it.

... a three dimensional mandala ... in bodily form signifying realization. (Unfortunately medical discretion prevents my giving the biographical details. It must suffice to say that this realization did actually take place.) [366]

Jung goes on to interpret Pauli's mandala dream as that it:

... aspires to the most complete union of opposites that is possible, including that of the masculine trinity and the feminine quaternity on the analogy of the alchemical hermaphrodite.

Since the figure has a cosmic aspect--world clock--we must suppose it to be a small-scale model or perhaps even a source of space-time, or at any rate an abstract of it and therefore, mathematically speaking, four-dimensional in nature although only visible in a three-dimensional projection. I do not wish to labour this argument, for such an interpretation lies beyond my powers of proof.[367]

**Pauli's Creativity**

I believe that Pauli's dream helped him to appreciate the value of visualization, *Anschaulichkeit*, as his preferred but secret source of knowledge. His "world-clock" dream contains symbolism resembling the kernels that he had recognized earlier in his mathematical physics, abstract interacting components in abstract rotations, that he may have seen in his "mind's eye." Thus, for example, the symbolism is reminiscent of the Bohr-Sommerfeld atom and Bohr's *Aufbauprinzip*, with the Pauli exclusion principle adding a fourth quantum number for the electron to build up the periodic table of elements; or possibly, but more tenuously, the symbolism points to a relationship that Pauli elucidated later, the Charge-Parity-Time (CPT) three-component symmetry of quantum-field theory. Pauli mentioned his dream in a letter to Jung in 1934:

---

[366] Enz, *No Time to be Brief* (ref. 2), pp. 246-247.
[367] Sir Herbert Read, Micheal Fordham, and Gerhard Adler, ed., *The Collected Works of C.G. Jung*. Vol. 12. *Psychology and Alchemy* (London: Routledge & Kegan Paul Publishers, 1953), p. 196.

The relationship of the Trinity is perfectly understandable to me, for I once had a dream in which the Trinity is turned into the 3 rhythms (the "world clock"). And the interplay of the latter is said to involve danger at a certain point in time.[368]

Pauli was attracted to Kepler later in part because Kepler was a bridge between "trinitarian" and "quaternian" thinking, as he indicated in his analysis of the Kepler-Fludd polemic, which he related to the interpretation of modern quantum physics:

Modern quantum physics has come closer to the quaternary point of view, which was so violently opposed to the natural science that was germinating in the 17th century, to the extent that it takes into consideration the role of the observer in physics than is the case in classical physics.[369]

Pauli's appreciation for the value of visualization was greatly strengthened by his treatment in the Jung Clinic. I will repeat some previously used quotations to better connect Pauli's views to Jung's. When Jung published his sessions with the unidentified Pauli, he opened his book by explaining why he included many illustrations in it:

...the wealth of illustrations ... is justified by the fact that symbolical images belong to the very essence of the alchemist's mentality. What the written word could express only imperfectly, or not at all, the alchemist compressed into his images; and strange as these are, they often speak a more intelligible language than is found in his clumsy philosophical concepts. Between such images and those spontaneously produced by patients undergoing psychological treatment there is, for the expert, a striking similarity both in form and in content....[370]

Pauli agreed with Jung's views, since his treatment was based on dream analysis, which depended upon his endorsement.

By 1948 Pauli assigned a high status and significance to visual knowledge and symbols:

Not only alchemy but also the heliocentric idea furnishes instructive examples of the problem as to how the process of knowing is connected with the religious experience of transmutation undergone by him who acquires knowledge (*Wandlungserlebnis des Erkennenden*); it transcends natural science and can be

---

apprehended only through symbols, which both express the emotional, feeling aspect of the experience and stand in vital relationship to the sum total of contemporary knowledge and the actual process of cognition. Precisely because in our times the possibility of such symbolism has become an alien idea, it may be considered especially interesting to examine another age to which the concepts of what is now called classical scientific mechanics were foreign, but which permits us to prove the existence of symbols that had simultaneously a religious and scientific function.[371]

Since Jung viewed dreams as consisting of many levels, he or Rosenbaum may have told Pauli that his "world clock" could symbolize both the Trinity and physical concepts. Pauli had left the Catholic church in 1929 and regarded the Trinity with some apprehension, as he mentioned in his letter to Jung in 1934. Similarly, in his discovery of the exclusion principle, he had to go beyond three and assign four quantum numbers to the electron, which was filled with emotion for him because it could not be supported on rational grounds, only on intuitive and aesthetic ones. He had to shatter "an Idol of the Tribe."[372] To fully grasp the significance of this event for Pauli, we must look more closely at the meaning of a mystical or religious conversion experience. Both Pauli and Jung were aware of William James's book, *The Varieties of Religious Experience*,[373] where James delineated four characteristics of a mystical or religious conversion experience: ineffability--the experience defies verbal expression; noetic quality--the experience seems to be another state of knowing, of deep insight; transiency--the experience cannot be sustained nor remembered except imperfectly; and passivity--the person feels his or her own will is in abeyance, as if held by a superior power.[374] James wrote that the result was "that deepened sense of the significance of a maxim or formula which occasionally sweeps over one."[375] In Pauli's case, one result of his therapy was that his self-confidence returned, and he soon opened up a long correspondence with Jung. Pauli's experiences in treatment in the Jung Clinic were quick, dramatic, effective,

---

and lasting, as if they were as James had described them.  Rosenbaum and Jung, however, may only have helped Pauli process earlier events in his life.

I find it remarkable that Pauli allegedly had 1300 dreams within his short eight months of treatment.  I suspect, therefore, that he had been recording his dreams for several years before his treatment with Rosenbaum and Jung.  If that is the case, then his "world clock" dream may have occurred much earlier than 1932 and gains added significance, because it may have been the source of both his inspiration and his anxiety over his discovery of the exclusion principle in 1924.  Similar to August von Kekulé's dream of a snake that seized its own tail and inspired him to visualize the molecular ring structure of benzene,[376] Pauli's dream may have been of Bohr's *Aufbauprinzip*.  Pauli's dream has a Jungian interpretation of many layers: the clock ticks off the ordered structure of the periodic table of the elements as protons are added to the nucleus and electrons to the atomic shells; the four colors symbolize the four quantum numbers of the electron; the four "sages" symbolize Sommerfeld, Bohr, Mach, and Pauli's father, all in heated debate over the methods of science; the three rings symbolize the Trinity; and Pauli's sense of danger ignites over the quaternian kernel intuitive visualization that came to him without rational cause.  If Pauli simultaneously experienced a religious conversion with this dream, no wonder he entered into "a brief period of spiritual and human confusion, caused by a provisional restriction to *Anschaulichkeit*."[377]

To further understand Pauli's psychological roots and his style, we must examine not only what he did but also what he did not do.  Thus, Pauli could have made his mark in solid-state physics--he had worked in this field already in Hamburg but he chose not to pursue it.   To Pauli, solid-state physics was dirty, messy; he referred to it as *Dreck* (dirt) and *Schweinerei* (filth),  also commenting that "I don't like this solid-state physics … I initiated it though."[378]   His Shadow was expressing an important message.
He was at root an INFP personality; his ESTJ side would have been comfortable with the empirical messiness of solid-state physics.  Thus, Pauli's comments often appear contradictory, and we have to be aware of which Pauli personality is speaking.  While his unusual dual personality typology of ESTJ and INFP caused him stress at times, I believe it also contributed to his high creativity.

Albert Rothenberg has analyzed the process of creative thinking and has interpreted the creative act as a union of contrary psychological forces. In support for his thesis, he quotes historian of physics Gerald Holton as writing:

> Not far below the surface, there have coexisted in science, in almost every period since Thales and Pythagoras, sets of two or more antithetical systems or attitudes, for example, one reductionistic and the other holistic, or one mechanical and the other vitalistic, or one positivistic and the other teleological. In addition, there has always existed another set of antitheses or polarities, even though, to be sure, one or the other was at a given time more prominent—namely, between the Galilean (or more properly, Archimedean) attempt at precision and measurement that purged public, "objective" science of those qualitative elements that interfere with reaching reasonable "objective" agreement among fellow investigators, and, on the other hand, the intuitions, glimpses, daydreams, and *a priori* commitments that make up half the world of science in the form of a personal, private, "subjective" activity.

> Science has always been propelled and buffeted by such contrary or antithetical forces. Like vessels with draught deep enough to catch more than merely the surface current, scientists of genius are those who are doomed, or privileged to experience these deeper currents in their complexity. It is precisely their special sensitivity to contraries that has made it possible for them to do so, and it is an inner necessity that has made them demand nothing less from themselves.[379]

To Rothenberg such psychological opposition is relative and abstract and takes place in the mind of the creative person. I believe the psychological union of opposites that Pauli found so powerful and creative was that between his rationalism and empiricism and his intuitive religiosity and mysticism. Alternately, we can say that Pauli had a combined ESTJ and INFP personality type, and that combination resulted in his high creativity. I believe that this perspective on Pauli also explains his hesitancy to publish, his critical nature, and his tendency to share his creative but empirically-questionable ideas only with close friends.

## Insights into Pauli's Later Physics and Philosophy

I believe that an understanding of Pauli's personality offers insights into his physics and philosophy. Pauli's Platonism can be seen in his physics after 1930 when his use of symmetry principles became his hallmark. His work on spin and statistics, CPT symmetry, and isospin reflected his predilection for mandalas, quaternities, and rotations. In contrast to the *Zweideutigkeit* of the electron, he took T.D. Lee and C.N. Yang's 1956 discovery of the nonconservation of parity in stride, as if he could easily incorporate it into his thinking.[380] After his death two years later, quantum physics continued to unfold along quaternian lines with the identification of the four fundamental forces of nature, the gravitational, electromagnetic, weak, and strong forces, and the continuing attempts to unify them. Pauli would have been pleased to see the appearance of Stephen Adler's book, *Quaternionic Quantum Mechanics and Quantum Fields*.[381] In 1955 Pauli displayed the extent to which his reflections on the nature of reality had taken him, calling for unification of rationalism and mysticism, physics and psychology:

> At the present time a point has again been reached at which the rationalist outlook has passed its zenith, and is found to be too narrow. Externally all contrasts appear to be extraordinarily accentuated. On the one hand the rational way of thought leads to the assumption of a reality which cannot be directly apprehended by the senses, but which is comprehensible by means of mathematical or other symbols, as for instance the atom or the unconscious. But on the other hand the visible effects of this abstract reality are as concrete as atomic explosions, and are by no means necessarily good, indeed sometimes the extreme opposite. A flight from the merely rational, in which will to power is never quite absent as a background, to its opposite, for example to a Christian or Buddhist mysticism is obvious and is emotionally understandable. Yet I believe that there is no other course for anyone for whom narrow rationalism has lost its force of conviction, and for whom also the magic of a mystical attitude, experiencing the external world in its crowding multiplicity as illusory, is not effective enough, than to expose himself in one way or another to these accentuated contrasts and their conflicts. It is precisely by this means that the scientist can more or less consciously tread a path of inner salvation. Slowly then develop inner images, fantasies or ideas, compensatory to the external situation, which indicates the possibility of a mutual approach of poles in the pairs of opposites. Taking warning from the failure throughout the history of thought of all premature endeavors to achieve a unity I shall not venture to make predictions about the future. As against the strict division of the activities of the human spirit into separate departments since the seventeenth century, I still regard the conceptual aim of overcoming the contrasts, an aim which includes a synthesis embracing the rational understanding as well as the mystic experience of one-ness, as the expressed unspoken mythos of our own present age.[382]

---

[380] Enz, *No Time to be Brief* (ref. 2), pp. 518-519.
[381] Stephen Adler, *Quaternionic Quantum Mechanics and Quantum Fields* (New York: Oxford University Press, 1995).

There was a common thread in Pauli's physics throughout his life, as Ralph de Laer Kronig and Victor Weisskopf pointed out:

The tendency towards invariant formulations of physical laws, initiated by Einstein, has become the style of theoretical physics of our days, upheld and developed by Pauli during all his life by example, stimulation, and criticism. For Pauli, the invariants in physics were the symbols of ultimate truth which must be attained by penetrating through the accidental details of things. The search for symmetry and general validity transcended the limits of physics in Pauli's work; it penetrated his thinking and striving throughout all phases of his life, in all fields of philosophy and psychology. Some of his writings dating from his last years reflect his thoughts in this direction.[383]

In the last years of his life, Pauli collaborated with Heisenberg in an attempt to discover an all-encompassing World Formula (*Weltformel*), but in response to severe criticism he

---

[382] Enz and von Meyenn, *Wolfgang Pauli* (ref. 3), pp. 147-148.
[383] Ralph de Laer Kronig and Victor Weisskopf, ed., *Collected Scientific Papers by Wolfgang Pauli*. Vol.1 (New York: Interscience Publishers, 1964), p. vii.

refused to join Heisenberg in publishing their work. Charles Enz has commented on their collaboration:

> Again later in 1957 it was the neutrino that motivated Pauli to collaborate with Heisenberg on his nonlinear spinor equation.... Towards the end of the year this collaboration developed into a real euphoria, which ended when Pauli met severe criticism at a special meeting convened at the end of January 1958 at Columbia University, New York.... The result of this collaboration was a manuscript entitled *On the Isospin Group in the Theory of the Elementary Particles* which, however, was not published until recently.[384]

After the seminar in New York in which Pauli spoke of his and Heisenberg's new theory, Pauli said to Bohr: "You may well think that all this is crazy." Bohr remarked pithily: "Yes, but unfortunately it is not crazy enough."[385] Mara Beller seems to suggest that Pauli's mysticism played a negative role in his extensive dialogue with Bohr when she commented: "He [Pauli] advanced imaginative, metaphorical ideas of wholeness and acausality to a mystical abyss that Bohr felt no inclination to approach."[386] Pauli's philosophical ideas are indeed more obscure than Bohr's and Heisenberg's and need to be explored further by historians and philosophers of science, but they were not dismissed by his colleagues.

Pauli's life, work, and thought will be studied long into the future. Two areas seem to me to be especially significant for future research: Pauli's role in the creation of matrix mechanics, and Pauli's increasing Platonism after 1932 in comparison to the Copenhagen hegemony in the interpretation of quantum mechanics. We already know that Pauli in letters to, and conversations with Bohr and Heisenberg seems to have contributed ideas to Heisenberg for his creation of matrix mechanics, defended Heisenberg against Born, revealed the power of matrix mechanics by deriving the spectral formula for the hydrogen atom, proved the equivalence of matrix and wave mechanics, and provided clues to Heisenberg on the uncertainty principle--although he did not publish much of this work. Pauli also pointed Dirac toward his relativistic equation of the electron, and he

---

refined Bohr's Como lecture on the principle of complementarity. We thus see Pauli as a brilliant theoretical physicist who delayed publishing many of his highly creative ideas but who shared them freely and willingly with others: He was too self-critical to follow his creativity through to publication.

As I mentioned in Chapter 6, historian John Hendry has provided one starting point for additional historical research on the back cover of his book:

> The development of quantum mechanics is interpreted as a dynamic interaction between physical, methodological and epistemological considerations, emerging primarily as a dialogue between two profound physicist-philosophers, Niels Bohr and Wolfgang Pauli. It is shown that Heisenberg's matrix mechanics, the quantum-mechanical transformation theory, Heisenberg's uncertainty principle and Bohr's principle of complementarity all had their roots in this central dialogue, and that the ideas characteristic of the interpretation of quantum mechanics were also essential to its creation.[387]

Hendry terms Pauli's philosophy of physics "operationalism," by which he seems to mean that Pauli believed that the foundations of quantum mechanics should be based on sound physical concepts from which direct experimental consequences can be calculated.[388] Still, Pauli's philosophical position remains obscure. Most physicists and historians have shied away from its mystical aspects as revealed in Pauli's attraction to Jungian psychology, and thus have dismissed their stimulating role in Pauli's philosophy.

Pauli's later Platonism seems to me to be almost indecipherable without considering its Jungian roots. Pauli believed in what I have called an alethic reality involving both physics and psychology. He believed that the fine-structure constant required explanation, that a Lamarckian view of biological evolution was valid, and that quantum-field theory was not aesthetically pleasing; the seventeenth-century scientific revolution had gone too far in excluding mind from matter. To Hendry, the creation of quantum mechanics embodies "competing demands of visualisation on the one hand and physical operationalism on the other towards a common recognition that neither was obtainable."[389]  Heisenberg commented on his obituary of Pauli to Thomas S. Kuhn:

> I like this article on Pauli's philosophical views.  I think that I had succeeded in describing very accurately how Pauli's mind was constructed.  I also hoped that I had made it clear to many people that I liked this kind of mind, and that my own mind is not so very different from that of Pauli.[390]

Heisenberg too was a Platonist.  We might also include Sommerfeld here.  Pauli's interactions with Einstein also imply that Einstein appreciated Platonism.[391]

Pauli had numerous discussions with Einstein during the war years, even collaborating on an article on relativity theory, and Einstein called Pauli his "spiritual heir in physics" at a dinner in Princeton celebrating Pauli's Nobel Prize.[392]  Einstein's remark, too, invites further exploration, since it implies that they both had serious reservations about the completeness of quantum theory.  Pauli and  Einstein later attempted to clarify their differences on quantum mechanics.  Pauli recognized, as Einstein did, that quantum mechanics was a provisional theory, but the two disagreed about the nature of its incompleteness.  Pauli sought an explanation for the fine-structure constant, that kernel of fundamental constants, without which quantum mechanics was incomplete.  Both Pauli and Einstein seemed to want a formulation of quantum mechanics that addressed the alethic nature of Schrödinger's wave function.  Pauli, however, remained skeptical of Einstein's views, as he commented to Max Born:

> As O. Stern said recently, one should no more rack one's brain about the problem of whether something one cannot know anything about exists all the same, than about the ancient question of how many angels are able to sit on the point of a needle.  But it seems to me that Einstein's questions [on quantum theory] are ultimately always of this kind.[393]

What is striking is that Pauli and Einstein remained respectful of each other despite their diverse talents, opinions, and philosophical positions.  Pauli's philosophical position appears to have been one that would support both Born's and Einstein's: Pauli supported Born's statistical interpretation of quantum mechanics, and he also supported Einstein's quest for a unified field theory.  Still, Pauli and Einstein disgreed on some fundamental issues.  As Pauli recalled:

> What is more, on the occasion of my farewell visit to him [sometime in 1954?] he told me what we quantum mechanists would have to say to make our logic unassailable (but which does *not* coincide with what he himself believes): "Although the description of physical systems by quantum mechanics is inconsistent, there would be no point in completing it, as the complete description would not agree with the laws of nature."  I am not, however, altogether satisfied with this formulation, as it seems to me to be one of those metaphysical formulations of the "angels on the point of a needle" type (whether something exists which nobody can know anything about).

Pauli soon clarified Einstein's position on the incompleteness of quantum mechanics further:

> He is only interested in his assertion that the characterisation of a state by a wave-function ("pure case" after v. Neumann) is also "incomplete," as the "true objective state of reality" always has a quasi-sharp location (even when the wave-function does not have it).[394]

Pauli wrote about his views of ultimate reality in 1954, presenting a picture that appears to be composed of Jung's psychological space consisting of both consciousness and unconsciousness with physical reality inextricably entwined.[395]  Although he refers to Platonism, he insists on not being associated with any philosophical school.  To Pauli, a physicist's mind, when trying to ascertain the nature of physical reality, is centered only in its conscious psychological space but is controlled and affected by its unconscious psychological space.  Hence, theoretical ideas determine what is experimentally measureable, and the concept of reality changes both with conscious ideas and experiment.  Similar to the divide between observer and observed in a quantum system, there is a divide between the conscious and unconscious that changes constantly in each

---

individual and in physics itself as it is refined.  Thus, the totality of reality is inaccessible to any conscious human mind at any time in history, but a deep totality of reality nevertheless exists, what I have called an alethic reality.

Pauli continued to be fascinated with the fine-structure constant because it implied numerologically that an alethic reality was behind it.  In an article of 1954 he again pointed out the need for a rational understanding of it:

> We may take a third, more fundamental, example [of physics being in a state of development].  One of the most assured empirical results of physics is the atomistic structure of electric charge.  Charge values are integral multiples of a fundamental unit, the electric elementary quantum, from which, along with the quantum of action and velocity of light, one can form a dimensionless number, 137.04.  To reach this result one requires a considerable part of the classical theory of electricity.  In the 17th century, for instance, when it was not known how to measure electric charges and how they are defined quantitatively, this empirical result could never have been obtained and formulated.  But we are unable to understand or explain the above number [today in 1954].[396]

Throughout his life, Pauli often closed his major works with a call for explanation of the fine-structure constant.  In honor of Pauli and his traditional call, I too will close by citing a recent announcement that surely would have attracted Pauli's attention and aroused his curiosity:

> In 1999 a group of scientists at the University of New South Wales in Australia reported some positive evidence that [the fine-structure constant] alpha was not staying the same.  The evidence for a changing alpha at the level of a part in 100,000, according to a new report issued by the same group, consists of the spacing of pairs of absorption lines of metal atoms in gas clouds in front of quasars at various redshifts….  The new observations suggest that alpha is growing bigger.[397]

---

[396] *Ibid.*,  p. 135.
[397]  Phillip F. Schewe, Ben P. Stein, and James R. Riordon, ed., *Physics News in 2001*: A Supplement to APS News, Media and Government Relations Division, American Institute of Physics (College Park, Maryland).  The article summarizes the article by J.K. Webb, M.T. Murphy, V.V. Flambaum, V.A. Dzuba, J.D. Barrow, C.W. Churchill, J.X. Prochaska, A.M. Wolfe. *"Further Evidence for Cosmological Evolution of the Fine Structure Constant,"* Physical Review Letters **87** (2001), 091301.

I wonder how Pauli, with his sure intuition aroused by empirical evidence, would have responded to this provocative possibility.

---

**Appendix: Pauli Timeline**

This timeline stops in 1935, and thus covers only the portion of Pauli's life that is relevant to my dissertation. I considered Charles Enz's book, *No Time to be Brief,* [398] to be the most accurate for Pauli's life events, and Abraham Pais's book, *Inward Bound,* [399] to have the best timelime of developments in modern physics.

-------------------------------------------------------------------------------------------------

**1869**   Wolfgang Joseph Pauli (nee Pascheles), father of Wolfgang Pauli, is born in Prague, September 11

Grandfather Jacob W. Pascheles inherited from his father Wolf a bookstore in Prague; acquired a house that had been a former convent of the congregation of the Paulans; was a respected member of Jewish community of Prague, as elder in "Gypsy's Synagogue" where he presided over the bar mitzwah of Franz Kafka, whose family also lived on the Old Town Square.

**1871**   The wavelengths of three lines in the hydrogen spectrum are found to have simple ratios. The ratios are $H_\alpha : H_\beta : H_\gamma = 1/20 : 1/27 : 1/32$

**1878**   Birth of Pauli's mother, Bertha Camilla Schütz on November 29, in Vienna

**1879**   Birth of Einstein

Death of Maxwell

**1885**   Balmer's formula for the hydrogen spectrum

Birth of Bohr

**1887**   Birth of Schrödinger

**1892**   First detection of fine structure in the spectral lines of hydrogen

Pauli's father studied medicine at Charles University in Prague with Ernst Mach's son Ludwig, and obtained his doctor's degree

**1897**   Grandfather Pascheles dies

**1898**   Pauli's father receives permission to change his name to Pauli in July

**1899**   Pauli's father converts from Jewish to Roman Catholic in March, shortly before his marriage to Camilla Schütz on May 2

**1900** Birth of Wolfgang Ernst Friedrich Pauli on April 25

Pauli is born in Vienna to father Wolfgang Josef Pauli, Jewish physician and later professor and director of a biological institute in the faculty of medicine of the University of Vienna, and to mother Berta Camilla Pauli, nee Schütz, daughter of a Viennese writer. Father had converted to the Catholic faith, which also was his wife's religion.

Planck discovers the quantum

Paul Villard discovers of γ -rays

**1901** Births of Heisenberg and Fermi

**1902** Birth of Dirac

**1905** **March:** Einstein postulates the light quantum

**June:** Einstein's first paper on special relativity

**September:** His second paper on special relativity, with $E = mc^2$

**1906** Pauli's younger sister Hertha Ernestina Pauli is born September 4. She later became a well-known novelist and biographer. Her birth shocks Pauli, who becomes jealous.

Rutherford discovers α-particle scattering

**1909** Einstein analyzes energy fluctuations in blackbody radiation and gives first statement about wave-particle duality

**1910** Pauli enrolls in the Döbling Gymnasium, takes courses in the classics, and pursues the natural sciences under the direction of his father.

Arthur Erich Haas makes first attempt to link atomic structure to Planck's constant

**1911** Parents Wolfgang and Bertha Pauli both leave Catholic church, for unknown reasons

Rutherford proposes the nuclear model of the atom.

First statement that Planck's law is related to the indistinguishability of light quanta

First Solvay Congress

**1912** First attempt to link Planck's constant to angular momentum.

Paschen-Back effect: Paschen and Back first observed the "normal" Zeeman triplet in a very strong magnetic field. In their paper the term "anomalous Zeeman effect" appears for the first time.

**1913** Bohr's trilogy on the constitution of atoms and molecules appears

First statement that β-decay is a nuclear process appears

First recognition that $A$ and $Z$ are independent nuclear parameters

Proton-electron model of the nucleus emerges

Moseley's experiments lead to the definitive interpretation of the periodic table

**1914** First detection of the continuous β-spectrum

Pauli visits Mach for the last time

Great War erupts

**1915** Field equations of gravitation from Einstein appear

Bohr is appointed professor at Copenhagen

Fritz Hasenöhrl is killed by a grenade in Great War

**1916** Pauli learns that he is of Jewish descent. Ernst Mach dies

Sommerfeld announces his fine structure constant, based on relativistic modifications to the orbit of electrons in the hydrogen atom

Einstein introduces the $A$ and $B$ coefficients for radiative emission and absorption

**1918** Pauli graduates from the Döbling Gymnasium, in the "class of geniuses." That fall, Paul enrolls at Munich under Sommerfeld. Pauli submits his first paper for publication in September, dealing with a topic in general relativity and unified field theory, before he leaves Vienna.

First selection rules appear for atomic spectra, restricting electron transitions between certain orbits

Emmy Noether's theorem on symmetry and energy conservation

**1919** Heisenberg enrolls at Munich and meets Pauli

Versailles Peace Treaty is signed June 28, initiating economic hardships in Germany and Austria

**1921** Pauli's encyclopedia article, which becomes his book *The Theory of Relativity*, is published

Pauli graduates from Ludwig-Maximilian University in Munich *summa cum laude*.  Pauli's doctoral thesis under Sommerfeld contains the first quantitative proof that the old quantum theory of Bohr and Sommerfeld had reached its limits. His thesis is on the limitations of the old quantum theory of Bohr and Sommerfeld when applied to the hydrogen molecular ion.

Pauli becomes Max Born's assistant at Göttingen for the winter term 1921-1922; Pauli and Born use astronomical perturbation theory in atomic problems.  Pauli meets Niels Bohr who invites him to come to Copenhagen.

Lande´ introduces half-integral magnetic quantum numbers

Einstein receives Nobel Prize for his equation on the photoelectric effect

**1922** After a summer in Hamburg as assistant to William Lenz, with Niels Bohr spends the academic year 1922-1923 in Copenhagen, Pauli works on the anomalous Zeeman effect, while helping Bohr on numerous other matters.

Bohr receives Nobel Prize for his atomic model.

**1923** Pauli accepts position at Hamburg, where he stays until going to Zurich in 1928.

Arthur Holly Compton discovers "Compton effect"

Louis de Broglie introduces particle-wave duality for matter

The core model for the anomalous Zeeman effect is developed

Bohr publishes review aricle on *unmechaniker zwang*

Pauli retreats from the anomalous Zeeman effect problem because it is too stressful

"Pauli effect" originates while Pauli is in Hamburg, perhaps as early as 1923

**1924** Pauli begins to write his Old Testament on the old quantum theory and suffers frustration with theoretical physics.

**January**: Bohr-Kramers-Slater paper on theory of radiative processes

**February**: Laporte discovers a new selection rule in the iron spectrum.  This is referred to by Pauli and others for awhile as "signature" and only later as "parity."

**July**: Bose introduces a new statistics for light quanta.

**July:** Einstein applies Bose statistics to matter and proposes a theory of the ideal gas

**August:** Pauli's theory of hyperfine structure

Stoner's rule for the periodic system is published

**November:** Pauli writes Landé that the complexity of fine structure does not lie in the nucleus of the atom. Pauli soon publishes his reasoning, which appears in early 1925.

**December**: Pauli recognizes a two-valuedness of the quantum properties of valence electrons. Pauli writes to Bohr on December 12 giving his exclusion principle reasoning and conclusions. Heisenberg reads the letter about the *Pauli Verbot* in which Pauli advocates a new quantum mechanics with energy and momentum being more fundamental than orbits, which contributes to Heisenberg's thoughts on matrix mechanics.

Pauli visits his parents in Vienna over the Christmas holidays.

1925    **January:** Pauli's two papers, the first on two-valuedness, the second on his exclusion principle are published. Kronig thinks of electron spin and discusses the concept with Pauli, but Pauli discourages Kronig from publishing the idea.

**May:** Pauli is in a state of frustration, and writes to a colleague that he "wishes he were a comedian" instead of being a theoretical physicist.

**June:** Bothe and Geiger's experimental demonstration of energy-momentum conservation in individual elementary processes

**July:** Heisenberg's first paper on quantum mechanics

**August**: Uhlenbeck and Goudsmit introduce half-integral quantum numbers for the hydrogen atom

**September**: Born and Jordan recognize that Heisenberg's theory is a matrix mechanics and that there is a need for a matrix electrodynamics

**October:** Goudsmit and Uhlenbeck's proposal of electron spin. Pauli is frustrated in working on his lengthy summary of the principles of the old quantum theory, his Old Testament.

**November:** Quantum algebra is intoduced by Dirac. Born, Heisenberg, and Jordan give a comprehensive treatment of matrix mechanics and recognize the concept of second quantization.

1926    **January:** Pauli's derivation of the discrete hydrogen spectrum by matrix methods. Schödinger's first paper on wave mechanics

**February:** Fermi statistics. Discovery of the Thomas factor of 2

**June:** Born's first paper on the probability interpetation of quantum mechanics

**July:** Pauli's Old Testament is published. Pauli finally accepts electron spin after the Thomas factor of 2 is resolved

**August:** Dirac relates symmetric-antisymmetric wave functions to Bose-Einstein and Fermi-Dirac statistics and derives Planck's law from first principles

**October:** G.N. Lewis introduces the name "photon," which he did not regard as Einstein's light quantum

**November:** Wigner introduces group theory into quantum mechanics

**December**: Dirac's first paper on quantum electrodynamics

From 1926 to 1929 the proton-electron model of the nucleus leads to a series of paradoxes.

1927   **March**: Heisenberg uncertainty relations. Davisson and Germer detect electron diffraction

**May:** Pauli introduces two-component "spinor" non-relativistic wavefunctions into quantum mechanics to address electron spin.

**August**: Ellis and Wooster's paper on their calorimeter experiment proves that the β particles from radium E are emitted with a continuous distribution of energies

**September:** Bohr proposes the concept of complementarity at the Volta Conference in Como

**October:** Jordan-Klein matrices. Fifth Solvay Congress in Brussels

**15 November:** Pauli's mother commits suicide

**December:** Jordan and Pauli introduce covariant commutation relations for free electromagnetic fields

1928   **January:** The Dirac equation for the electron. Jordan-Wigner matrices

**February:** Wigner introduces parity. Death of Lorentz

**April:** Pauli moves to the ETH in Zurich

**October:** Relationship of odd and even number of fermions to Fermi-Dirac and Bose-Einstein statistics is found. Klein-Nishina formula for Compton scattering

**December:** Klein paradox

Pauli's father remarries sometime in 1928

**1929**  **February:** First observation of cosmic-ray showers.

**March:** Heisenberg and Pauli give the Lagrangian formulation of quantum-field theory

**May 6:** Pauli leaves Catholic Church for unknown reasons

**September:** First systematic treatment of quantum electrodynamics in the Coulomb gauge and proof of its covariance

**December 23:** Pauli marries Kate Deppner. The marriage soon ends in divorce

From **1929 to 1936**, Bohr considers the possibility that energy is not conserved in β-decay

**1930**  **November 26**: Pauli's divorce from Kate Deppner

Sixth Solvay Congress is held in Brussels. Pauli attends

**December 1:** In a letter to Pauli, Heisenberg refers to Pauli's "neutron" postulate

**December4:** Pauli publicly announces his neutrino hypothesis

**1931**  Pauli breaks his arm in a drinking episode. G. Beck, H. Bethe, and W. Riezler write famous spoof on the fine-structure constant

**1932**  **January:** Carl Jung recommends Pauli to see a woman analyst, rather than Jung himself, because of the nature of Pauli's issues of relationships to women. Pauli's first contact with his analyst Erna Rosenbaum of the Jung Clinic was in the form of a letter dated February 3.

In skit at the Bohr institute, Pauli plays Mephistopheles and Gretchen represents the neutrino.

**1933**  Pauli meets Franca Bertram, who soon becomes his second wife. Pauli takes his bride-to-be to his father's and stepmother's home in Vienna at Christmas.

**1934**  **April 4**: Pauli marries Franca Bertram.

**1935**  Jung publishes a collection of Pauli's dreams without identifying their source

## Articles